\documentclass[notitlepage,a4paper,aps,prd,onecolumn,superscriptaddress,nofootinbib,longbibliography,20pt]{revtex4}
\usepackage{amsthm}
\usepackage{makeidx}
\usepackage{graphicx}
\usepackage{microtype}
\usepackage{amsmath}
\usepackage{amssymb}
\usepackage{subfigure}
\usepackage{hyperref}
\usepackage{url}
\usepackage{adjustbox}
\usepackage{xcolor}
\usepackage{color}
\usepackage{mathrsfs}
\usepackage{calrsfs}
\usepackage{amsfonts}
\usepackage{lipsum}
\usepackage{eufrak}
\usepackage{tabularx}
\usepackage{eucal}
\usepackage{float}
\usepackage{latexsym}
\usepackage{ragged2e}
\usepackage{epsfig}
\usepackage{textcomp}\usepackage{multirow}
\usepackage{booktabs}
\usepackage{caption}
\usepackage{orcidlink}
\usepackage{hyperref} 
\usepackage{xcolor}

\DeclareCaptionJustification{justified}{\leftskip=0pt \rightskip=0pt \parfillskip=0pt plus 1fil}
\usepackage{calligra}
\DeclareMathAlphabet{\mathcalligra}{T1}{calligra}{m}{n}
\DeclareFontShape{T1}{calligra}{m}{n}{<->s*[2.2]callig15}{}

\definecolor{vividviolet}{rgb}{0.62, 0.0, 1.0}
\definecolor{amaranth}{rgb}{0.9, 0.17, 0.31}
\definecolor{palatinateblue}{rgb}{0.15, 0.23, 0.89}
\definecolor{brightpink}{rgb}{1.0, 0.0, 0.5}
\definecolor{cornflowerblue}{rgb}{0.39, 0.58, 0.93}
\definecolor{deepcarminepink}{rgb}{0.94, 0.19, 0.22}
\definecolor{radicalred}{rgb}{1.0, 0.21, 0.37}
\colorlet{Mycolor1}{green!10!orange}

\hypersetup{ linktoc=all,
	colorlinks, linkcolor={palatinateblue},
	citecolor={brightpink}, urlcolor={amaranth}
}

\graphicspath{{Images/}}

\def\sideremark#1{\ifvmode\leavevmode\fi\vadjust{\vbox to0pt{\vss
			\hbox to 0pt{\hskip\hsize\hskip1em
				\vbox{\hsize1.3cm\tiny\raggedright\pretolerance10000
					\noindent #1\hfill}\hss}\vbox to8pt{\vfil}\vss}}}%
\def\beq{\begin{equation}}
	\def\eeq{\end{equation}}
\newcommand{\be}{\begin{equation}}
	\newcommand{\ee}{\end{equation}}
\newcommand{\ba}{\begin{eqnarray}}
	\newcommand{\ea}{\end{eqnarray}}

\begin{document}

	\title{A Unified Dynamical Systems Framework for Cosmology in $f(Q)$ Gravity: Generic Features Across the Connection Branches}

	\author{Jibitesh Dutta\orcidlink{0000-0002-6097-454X}}
	\email{jibitesh@nehu.ac.in}
	\affiliation{Mathematics Division, Department of Basic Sciences and Social Sciences, North Eastern Hill University, Shillong, Meghalaya 793022, India}
	\affiliation{Visiting Associate, Inter University Centre for Astronomy and Astrophysics, Pune 411 007, India}
	
	\author{Wompherdeiki Khyllep\orcidlink{0000-0003-3930-4231}}
	\email{sjwomkhyllep@gmail.com}
	\affiliation{Department of Mathematics, St. Anthony’s College, Shillong, Meghalaya 793001, India}
	
	\author{Saikat Chakraborty\orcidlink{0000-0002-5472-304X}}
	\email{saikat.c@chula.ac.th,\,saikat.chakraborty@nwu.ac.za}
	\thanks{Corresponding author}
	\affiliation{High Energy Physics Theory Group, Department of Physics,\\
		Faculty of Science, Chulalongkorn University, Bangkok 10330, Thailand}
	\affiliation{Center for Space Research, North-West University, Potchefstroom 2520, South Africa}
	
	\author{Daniele Gregoris\orcidlink{0000-0002-0448-3447}}
	\email{danielegregoris@libero.it}
	\affiliation{School of Science, Jiangsu University of Science and Technology, Zhenjiang 212100, China}
	
	\author{Khamphee Karwan\orcidlink{0009-0004-4993-4733}}
	\email{khampheek@nu.ac.th}
	\affiliation{The Institute for Fundamental Study, Naresuan University, Phitsanulok 65000, Thailand}

	\begin{abstract}
		We develop a unified dynamical systems framework for spatially flat FLRW cosmology in $f(Q)$ gravity, covering all three connection branches using a single set of Hubble-normalised variables without fixing the function $f(Q)$ \textit{a priori}. This model-independent and connection-agnostic approach enables direct comparison across connection choices and uncovers structural features of the cosmological dynamics not visible in connection-specific formulations. While the existing works narrowly focus only on fixed point analysis, in our work we put special efforts to identify the invariant submanifolds, model-independent trajectories and physically viable regions of phase space across connections. For a broad class of viable $f(Q)$ models, we establish the generic existence of de Sitter attractors and matter-dominated fixed points in the two branches of connection, offering a robust route to late-time acceleration without fine-tuning. We further identify an invariant submanifold that yields $\Lambda$CDM-like background evolution despite underlying dynamics distinct from General Relativity, providing a geometric origin for cosmic acceleration distinguishable only at the perturbation level. We also derive a first integral on this submanifold, allowing analytic reconstruction of the dynamical connection and uncovering hidden conservation laws.
		Another key feature we found is that while trivial connections exhibit strong parameter dependence, the nontrivial branches often feature parameter-independent behaviours. We also study the variation of the effective gravitational coupling, $k_{\text{eff}}$, across branches and show how this can be constrained using astrophysical observations, which bridges theoretical viability with observational consistency in a novel way. Applying our framework to the illustrative model $f(Q)=\alpha Q + \beta(-Q)^n$, we find late-time acceleration and $\Lambda$CDM-like behaviour without vacuum energy. Finally, we propose a general route for extending dynamical systems analysis to broader classes of $f(Q)$ models using the $m_i$-hierarchy method. This framework enables closure for models previously inaccessible to standard approaches and offers a diagnostic tool for identifying structurally viable cosmologies within modified gravity theories.
	\end{abstract}
	
	\maketitle
	
	\section{Introduction}
	
	Recent cosmological observations, such as results from DESI~\cite{DESI:2025fii,DESI:2025zgx}, suggest that the standard $\Lambda$CDM model may be incomplete.  The cosmological tensions, such as the Hubble discrepancy and anomalies in the growth of structure, indicate that General Relativity (GR) might require modifications to explain cosmic evolution fully.
	A promising direction is offered by the so-called \emph{geometric trinity of gravity}, which presents GR in three equivalent formulations: based on curvature ($R$), torsion ($T$), or nonmetricity ($Q$). However, their modifications are not equivalent, and among these, \textit{\( f(Q) \) gravity} has recently emerged as a promising avenue for modifying GR while preserving second-order field equations~\cite{BeltranJimenez:2019tme}.   The theory operates within the framework of \textit{symmetric teleparallel gravity}, where both curvature and torsion are absent, and gravitation arises solely from nonmetricity.  $f(Q)$ gravity thus offers a flexible yet straightforward framework to explain late-time cosmic acceleration without invoking a cosmological constant or dark energy. Its analytical structure and geometric clarity have made it a popular candidate in modified gravity studies~\cite{Dimakis:2021gby, Bahamonde:2022zgj, Paliathanasis:2023pqp, Jarv:2023sbp, Narawade:2023rip}. Moreover,  it belongs to the broader metric-affine gravity class, where the metric and connection are treated independently. It also introduces new degrees of freedom, leading to numerous interesting gravitational solutions.

	In $f(Q)$ gravity, the underlying flat and torsionless geometry allows a three-parameter family of affine connections. These give rise to three distinct connection branches, $\Gamma_1$, $\Gamma_2$, and $\Gamma_3$, each characterised by a free function $\gamma(t)$. In the $\Gamma_1$ branch, however, $\gamma(t)$ enters the connection definition but does not influence the cosmological evolution~\cite{Guzman:2024cwa}. Notably, the three connection branches are not only mere mathematical constructs but also of physically relevant quantities. Despite all being flat and torsionless, numerous studies have shown distinctive signatures of the connection branches on the Friedmann-Lema\^itre-Robertson-Walker (FLRW) dynamics. For instance, some cosmological singularities that arise in one connection branch may correspond to smooth de Sitter epochs in another for the same $f(Q)$ function~\cite{Ayuso:2025vkc}. Likewise, it was found in \cite{Shi:2023kvu} that for a given $f(Q)$, the scenarios based on the three different branches perform differently in accounting for the same observational dataset SNe+CC+BAO+QSO. In particular, it was found that a specific choice of the free gauge parameter defining the affine connection can lead to phantom crossing of the dark energy equation of state, which the authors have dubbed as `gauge-induced phantom crossing' in \cite{Shi:2023kvu}. Another recent work \cite{Basilakos:2025olm} shows that the connection branch $\Gamma_2$, in the absence of matter, is equivalent to a quintom model on the FLRW minisuperspace. This provides a framework in which the phantom field is well-behaved, with a lower bound on its energy. The latter work also provides a realisation of the phantom field from first principles. Phantom fields are generally needed to cross the effective divide line $\omega=-1$, a phenomenon predicted by recent DESI dataset analysis \cite{DESI:2025fii}. In light of these recent works, a deeper understanding of the dynamics under the non-trivial connections is warranted.

	In spite of the recent surge of interest in $f(Q)$ gravity as a novel framework for modifying gravity, a study has  highlighted the potential presence of ghost and strong coupling issues in $f(Q)$ gravity~\cite{Gomes:2023tur}. In fact, the strong coupling issue also arises in the study of $f(T)$ gravity, where certain degrees of freedom lose their kinetic terms on FLRW backgrounds~\cite{BeltranJimenez:2020fvy,Bahamonde:2021gfp}. However, Hu et al.~\cite{Hu:2023gui} argue that the ghost degree of freedom is non-propagating. It was pointed out in~\cite{Heisenberg:2023wgk} that the inclusion of hypermomentum or a nonminimal matter-geometry coupling might resolve these issues, potentially leading to a healthy theory. Ref.~\cite{Bello-Morales:2024vqk} constructs a class of ghost-free symmetric teleparallel modified gravity theories that break full diffeomorphism invariance but retain transverse diffeomorphisms. 
	
	The question regarding the actual number and nature of propagating gravitational degrees of freedom remains an important area of research and is expected to shed light on the theoretical viability of $f(Q)$ theories. Lastly, although ghosts—i.e., propagating degrees of freedom with negative kinetic terms and consequently a Hamiltonian unbounded from below—are usually associated with instabilities, this need not always be the case. In~\cite{Deffayet:2021nnt}, the authors demonstrate analytically and numerically that mechanical systems with Hamiltonians unbounded from both below and above can exhibit stable dynamics for any initial conditions. Further detailed computations showing integrability and global Lyapunov stability (not just local linear stability) have been presented in a later work by the same authors~\cite{Deffayet:2023wdg}. Additionally, a ghost may be considered \textit{benign} if the associated divergence does not occur in finite time~\cite{ErrastiDiez:2024hfq}.

	Our primary focus in this paper is on the phenomenological viability of $f(Q)$ gravity,  while we understand that theoretical consistency is also important.  In particular, we aim to address whether viable cosmological scenarios, such as a matter-dominated era followed by late-time acceleration, can arise generically within this framework. This question is most effectively tackled using the tools of dynamical systems analysis. This approach provides an effective framework for studying the qualitative evolution of cosmological models in modified gravity, which are governed by nonlinear equations; for details, see the well-known review~\cite{Bahamonde:2017ize}. In this formalism, cosmological evolution is reformulated as a flow in a phase space of Hubble-normalised, dimensionless variables. Fixed points in this space represent possible cosmological epochs. A realistic modified gravity theory should exhibit a heteroclinic trajectory connecting a matter-dominated saddle in the past to a stable accelerated attractor in the future.
	However, obtaining a generic, model-independent dynamical system is often obstructed by the difficulty of closing the system, especially in higher-order theories like $f(R)$, $f(T)$, $f(G)$, or $f(Q)$~\cite{Amendola:2006we, Carloni:2007br}. A major goal of this work is to address these challenges within a unified framework that applies to all connection branches of $f(Q)$ gravity.

	Several studies have developed dynamical system (DS) formulations for $f(Q)$ cosmology, primarily within the connection branch $\Gamma_1$ ~\cite{BeltranJimenez:2019tme, Khyllep:2021pcu, Khyllep:2022spx, Boehmer:2022wln, Boehmer:2023knj}. In this branch, Boehmer et al. showed that the phase space of a spatially flat FLRW universe with a single fluid can be reduced to a one-dimensional system due to the nondynamical nature of the connection.
	Alternative DS formulations have introduced different choices of dimensionless variables. The earliest works used energy-density-like variables, while later approaches proposed refinements tailored to specific subclasses of $f(Q)$ models. However, most of these methods are either model-dependent or limited in their scope.
	Beyond $\Gamma_1$, fewer studies have tackled dynamical system formulations. Paliathanasis~\cite{Paliathanasis:2023nkb, Paliathanasis:2023hqq} explored specific models in the $\Gamma_2$ and $\Gamma_3$  branches. Shabani et al. \cite{Shabani:2023nvm} proposed a generic dynamical system formulation for the connection $\Gamma_2$, which has very recently been extended to Bianchi-I models \cite{Murtaza:2025gme}. However, the analysis remains restricted by simplifying assumptions and does not explore model-independent or geometric features such as invariant submanifolds. To date, no general DS framework exists for the $\Gamma_3$ branch.

	In dynamical system analysis of $f(Q)$ gravity,  parameters play an important role. While fixed points in $\Gamma_1$ exhibit sensitivity to model parameters, the  $\Gamma_2$ and $\Gamma_3$ connection  branches often reveal parameter-independent behaviour under generic conditions. This difference has direct implications for observational constraints on viable $f(Q)$ models,  particularly in explaining cosmic acceleration without invoking a cosmological constant or exotic dark energy components.
	Moreover, recent studies suggest that the connection parameter $\gamma(t)$, which distinguishes different connection branches, can lead to cosmological acceleration as an apparent effect of non-inertial motion~\cite{Shi:2023kvu}. In this interpretation, dark energy and even aspects of dark matter can be viewed as manifestations of non-inertial effects rather than fundamental sources of energy-momentum. This perspective avoids the cosmological constant problem altogether and predicts the existence of an inertial frame in which cosmological observations remain consistent without dark energy. A similar idea has also been advocated in the context of teleparallel gravity~\cite{Ulhoa:2011aa}, where non-inertial effects have been shown to mimic the influence of dark energy. 
	
	Recent observational studies indicate that certain functional forms of $f(Q)$ may alleviate current cosmological tensions. Specifically, exponential forms of $f(Q)$ gravity can jointly tame the Hubble and $S_8$ observational tensions by supporting a higher than $\Lambda$CDM value of $H_0$ together with a lower value of the effective gravitational constant $G_{\rm eff}$ \cite{Boiza:2025xpn}. Furthermore, these models in the $\Gamma_{1}$ branch can also realise a quintom-like redshift evolution, where the effective equation of state parameter crosses the phantom divide, thus being consistent with DESI DR2 BAO, SNe, and CMB combined datasets. \cite{Yang:2025mws}.  Thus, exponential forms of $f(Q)$ appear to provide more realistic astrophysical description than power laws which are challenged by GRB datasets \cite{Paliathanasis:2025hjw}.  Interestingly, Ref.~\cite{Dialektopoulos:2025ihc} has shown via a minisuperspace analysis that powerlaw $f(Q)$ models are the only ones admitting Noether point symmetries, with the corresponding conservation laws constructed in that work. The minisuperspace method has similarly been applied to reconstruct $f(Q)$ models in the $\Gamma_{2}$ and $\Gamma_{3}$ branches, describing both early and late cosmic evolution \cite{De:2025swf}.  Moreover, certain interactions between non-metricity and matter allow cosmic birefringence in the CMB and the baryon–antibaryon asymmetry to be interpreted as a single phenomenon \cite{Yu:2025job}. On smaller, galactic scales, a tetrad–spin formulation of $f(Q)$ gravity has enabled some black hole solutions to be reinterpreted in the framework of torsionful $f(T)$ gravity, establishing a link between otherwise distinct geometrical descriptions \cite{Wu:2024vcr}.
		
As discussed, all existing results depend on specific functional choices of $f(Q)$, raising the question of how general such behaviours are. Given the theoretical and phenomenological motivations outlined above, constructing a \emph{unified dynamical systems framework}, and a subsequent \emph{comprehensive analysis across all three spatially flat FLRW connection branches} of $f(Q)$ gravity, is both timely and necessary. This challenging construction is the central objective of the present work. As mentioned earlier, most cosmological analyses have focused exclusively on the $\Gamma_1$ branch, typically assuming specific $f(Q)$ forms from the outset. In contrast, our approach enables the exploration of broader families of $f(Q)$ models by imposing only the generic existence of matter-dominated and accelerating attractors. Furthermore, this framework also opens the door to broader and model-independent comparisons with recent DESI observations and the ongoing cosmic tension data.

As a first step toward identifying such generic cosmological features, we perform a \emph{dynamical system analysis} of a spatially flat, isotropic, and homogeneous universe for each of the three connection branches, identifying a de Sitter epoch and studying the dynamics in its vicinity for the non-trivial branches, whose analysis has remained mathematically challenging~\cite{Yang:2025mws}. To respect earlier attempts and maintain a standardised definition of dynamical variables, we utilise the same variables introduced in~\cite{Shabani:2023nvm}. As it turns out, this set of variables has the added advantage of being strikingly similar to the standard dynamical variables used in $f(R)$ dynamical systems (e.g., see~\cite{Amendola:2006we}), a point we briefly mention at the end of Sec.~\ref {Sect:Conclusion}. With this formalism, we recover a 1-dimensional phase space for the connection branch  $\Gamma_1$, consistent with Boehmer et al.~\cite{Boehmer:2022wln, Boehmer:2023knj}. In contrast, the  connection branches $\Gamma_2$ and $\Gamma_3$ yield 4-dimensional phase spaces, in agreement with the earlier discovery that on the FLRW minisuperspace the dynamics can be effectively attributed to two scalar fields, each having its own kinetic and potential contribution \cite{Paliathanasis:2023pqp}. While most of the dynamical system works in the literature content themselves with the fixed point analysis, particular effort has been put into this paper to single out the physically viable region of the phase space, and to identify possible invariant submanifolds and fixed points that exists \emph{(almost) generically} irrespective of any particular model under consideration. The strength of our work lies in the detailed, generic analysis conducted for each connection branch, which enhances the general applicability and theoretical robustness of our framework.

To better understand the general analysis, we further consider a specific illustrative model, $f(Q) = \alpha Q + \beta(-Q)^n$ ($\alpha,\beta\neq0,\,n\neq0,1$) applied uniformly across all three branches. This example captures a wide range of cosmological behaviours, including transitions from matter-dominated to accelerated epochs, $\Lambda$CDM-mimicking solutions. Moreover, we provide a dedicated discussion on the behaviour of the effective gravitational coupling $k_{\text{eff}}$, which becomes a dynamical quantity in $f(Q)$ gravity, in contrast to its fixed value in GR. By systematically analysing $ k_{\text{eff}} $ across all three connection branches, our framework provides deeper insight into how different geometric choices in $f(Q)$ gravity can lead to distinct observational signatures. Finally, we outline a framework that offers a possible route for the dynamical systems analysis of a broader class of $f(Q)$ gravity theories.

	The paper is organised as follows: Section~\ref{sec:background} introduces the general setup of $f(Q)$ gravity and the affine connection branches, while Section~\ref{sec:DSA_intro} presents the unified dynamical system framework and the conditions required for its closure. Sections~\ref{sec:Gamma1}, \ref{sec:Gamma2}, and \ref{sec:Gamma3} contain the detailed dynamical analyses for the branches $\Gamma_1$, $\Gamma_2$, and $\Gamma_3$, respectively. We apply our framework for each branch to the illustrative model $f(Q) = \alpha Q + \beta (-Q)^n$. In Section~\ref{app: broader_class}, we outline a generalisation of the framework that enables the dynamical systems analysis of a broader class of $f(Q)$ gravity theories. Finally, Section~\ref{Sect:Conclusion} summarises our key findings, discusses implications for observational cosmology, and outlines directions for future research. A structural comparison between $f(Q)$ and $f(R)$ gravity is provided in ~\ref{Appendix:f(R)_comp}, highlighting key differences in phase space topology and attractor structure.
	
	
	\section{$f(Q)$ gravity field equations in a spatially flat Friedmann universe}
		\label{sec:background}
		In order to study the cosmological dynamics of $f(Q)$ gravity in the FLRW metric given by 
		\begin{align}\label{FLRW}
			ds^2 &= -dt^2 + a(t)^2 \left(dr^2 + r^2 d\theta^2 + r^2 \sin^2 \theta d\phi^{2} \right) \,,
		\end{align}
		where $ r, \theta, \phi$ are the spherical polar coordinates and $a(t)$ is the scale factor, we demand the Lie derivatives of the affine connection to be zero with respect to Killing vectors corresponding to spatial rotations and translations. As a result, the following three branches of connection, which determine the form of the nonmetricity scalar $Q$ compatible with homogeneity and isotropy, are obtained \cite{Hohmann:2021ast,Guzman:2024cwa}:

		\begin{equation}
			\label{Qbranches}
			\Gamma_1:\, Q=-6H^2,\qquad
			\Gamma_2:\, Q=-6H^2+9\gamma H+3\dot{\gamma},\qquad
			\Gamma_3:\, Q=-6H^2+\frac{3\gamma H}{a^2}+\frac{3\dot{\gamma}}{a^2},
		\end{equation}
		where $H=\dot{a}/a$ is the Hubble parameter and $\gamma(t)$ is a time-dependent function characterizing the dynamical connection; an overdot denotes differentiation with respect to cosmic time. For a detailed decomposition of the affine connection into its symmetric, antisymmetric, and nonmetric parts, as well as the explicit derivation of $Q$ under FLRW symmetry, see Refs.~\cite{BeltranJimenez:2019tme,Hohmann:2021ast}.
		
		The corresponding Friedmann and Raychaudhuri equations can be written in a unified form as
		
		\begin{eqnarray}
			3H^2 f_Q + \tfrac{1}{2}(f - Qf_Q) + \Xi_F& =& \rho_m,\label{fried-rev}\\ 
			\qquad
			-2\frac{d}{dt}(H f_Q) - 3H^2 f_Q - \tfrac{1}{2}(f - Qf_Q) + \Xi_R &=& 0,\label{raych-rev}
		\end{eqnarray}
		where $\rho_m$ denotes the energy density of pressureless dust, $\Xi_F$ and $\Xi_R$ encode the connection-dependent corrections:
		\begin{equation}
			\Xi_F=
			\begin{cases}
				0, &  {\rm for}~~ \Gamma_1,\\[4pt]
				\dfrac{3\gamma\dot Q f_{QQ}}{2}, &  {\rm for}~~  \Gamma_2,\\[8pt]
				-\dfrac{3\gamma\dot Q f_{QQ}}{2a^{2}}, &  {\rm for}~~  \Gamma_3. 
			\end{cases}
		\end{equation}
		
		In these equations, $\Xi_F$ and $\Xi_R$ coincide for the $\Gamma_1$ and $\Gamma_2$ branches,	while for $\Gamma_3$ they are related by
		\begin{align}
			\Xi_R = -\tfrac{1}{3}\,\Xi_F .
		\end{align}

		For $\Gamma_{1}$ case, we can combine the  equations \eqref{fried-rev} and \eqref{raych-rev} as
		\begin{equation}\label{frd_ray-1_rev}
			2 \frac{d}{dt}(Hf_Q)=-\rho_m\,.
		\end{equation}
		
		Additionally for $\Gamma_2$ and $\Gamma_3$ branches, the evolution of the nonmetricity scalar, ensuring connection compatibility, satisfies the unified relation
		\begin{equation}\label{Q-dot_rev}
			\dot Q^{2} f_{QQQ} + \left[\ddot Q + \Xi_H \dot Q \right] f_{QQ} = 0,
		\end{equation}
		where the effective Hubble-like coefficient $\Xi_H$ depends on the connection branch as 
		\begin{equation}
			\Xi_H =
			\begin{cases}
				3H, &  {\rm for}~~  \Gamma_2, \\[4pt]
				H + 2\dot{\gamma}/\gamma, &  {\rm for}~~  \Gamma_3.
			\end{cases}
		\end{equation}
		For $\Gamma_1$, no such relation appears since $Q=-6H^2$ depends only on $H$, and $\dot Q$ is algebraically fixed by $\dot H$.

		For completeness, we note that under the redefinition $\gamma \to a^2 \gamma$ (as in \cite{Chakraborty:2025qlv}), the $\Gamma_3$ field equations can be written as
		\begin{subequations}
			\begin{eqnarray}
				&& Q = -6H^{2} + 9\gamma H + 3\dot{\gamma}\,, \label{Q-3_redef_rev}
				\\
				&& 3H^{2}f_Q + \frac{1}{2}\left(f - Qf_Q\right) - \frac{3\gamma\dot{Q}f_{QQ}}{2} = \rho_{m}\,, \label{fried-3_redef_rev}
				\\
				&& -2\frac{d}{dt}\left(H f_Q\right) - 3H^{2}f_Q - \frac{1}{2}\left(f - Q f_Q\right) + \frac{\gamma\dot{Q}f_{QQ}}{2} = 0\,, \label{raych-3_redef_rev}
				\\
				&& \dot{Q}^{2}f_{QQQ} + \left[\ddot{Q} + \left(5H + \frac{2\dot\gamma}{\gamma}\right)\dot{Q}\right]f_{QQ} = 0\,. \label{Qdot-3_redef_rev}
			\end{eqnarray}
		\end{subequations}

		In our analysis, we employ these redefined equations to construct the dynamical system, as they provide a consistent  formulation for the phase-space variables.
		
		The matter sector is assumed to be pressureless and minimally coupled,
		\begin{equation}\label{mat_cons_rev}
			\dot{\rho}_m + 3H\rho_m = 0,
		\end{equation}
		while the effective gravitational coupling is given by $G_{\rm eff}=1/f_Q$, which must satisfy $f_Q>0$ to preserve an attractive gravitational interaction \cite{Guzman:2024cwa}.  
		
		The above equations fully characterize the background cosmology of $f(Q)$ gravity under the three admissible connections and serve as the starting point for the autonomous dynamical system formulation developed in the next section.
		At the mathematical level, the metric field equations of the branch $\Gamma_2$ reduce to those of $\Gamma_1$ in the limit $\gamma \to 0$, which is not the case for those of $\Gamma_3$. Our dynamical system formulations (\ref{eq:ds_2}) and (\ref{eq:ds_3}) will preserve such behaviours in terms of the dynamical variable $x_4$ defined in (\ref{dynvar_def}). This formulation is, however, purely a mathematical characteristic of the corresponding field equations, and not any inherent physical aspects of the gravitational theory; in the so-called  $\Gamma_2$ and $\Gamma_3$ branches, in fact the dynamical connection cannot be switched off at the level of the affine connection \cite{Hohmann:2021ast}, and $\gamma$ is necessarily nonzero. 
		
		The dynamics described by the above cosmological equations in each branch are complicated to determine. Therefore, in what follows, we investigate the generic evolution of the three branches of connections using dynamical system techniques.

	
	\section{Dynamical System in $f(Q)$ gravity: Formulation}\label{sec:DSA_intro}
	
	In this work, we utilise the variables that Shabani \cite{Shabani:2023nvm} used to present a generic approach to closing the system in $f(Q)$ gravity across all three connection branches. In contrast to most of the previous studies, which assume particular functional forms of $f(Q)$, such as power laws, logarithmic modifications, or polynomial corrections, our formulation provides a systematic framework that can be applicable to a broad class of $f(Q)$ theories. This generality is achieved by introducing an auxiliary quantity that governs the closure mechanism and determines how nonmetricity influences the cosmic evolution. We define a common set of dimensionless variables that capture the key physical quantities to set up our dynamical system. Following \cite{Shabani:2023nvm}, we introduce  
	\begin{equation}\label{dynvar_def}
		x_1=-\frac{f}{6H^2 f_Q},\,~x_2=\frac{ Q}{6H^2},\,~x_3=-\frac{\dot{f_Q}}{H f_Q},\,~x_4=\frac{\gamma}{2H},\, \Omega=\frac{\rho_m}{3H^2 f_Q}.
	\end{equation}
	These variables allow us to analyse the system across different connection choices, providing a unified framework for comparing the cosmic evolution in different formulations of $f(Q)$ gravity. Notably, our dynamical system formulation relies on five variables, as for the quintom cosmology \cite[Eq.(4.19)]{Cai:2009zp}, consistently with the correspondence established in \cite{Basilakos:2025olm}.

	A crucial ingredient in our analysis is the role of the auxiliary function
	\begin{equation}
		\label{defofm}
		m(Q) \equiv Q \frac{f_{QQ}}{f_Q} = \frac{d \ln f_Q}{d \ln Q}\,.
	\end{equation}
	Note that $m(Q) \to 0$ recovers GR\footnote{Apart from the standard limits $f_{QQ}\!\to\!0$ or $Q=\mathrm{const}$, there also exist nontrivial cases where the background equations reduce exactly to those of GR even for genuinely nonlinear $f(Q)$. This structural reduction has been shown in \cite[(Eqs. (48), (63), (69)]{Guzman:2024cwa}.}. Physically, the quantity $m(Q)$ characterises the deviation from GR. One must express $m(Q)$ in terms of the other dynamical variables to close the system. We achieve this as follows. Notice that the quantity $\frac{x_1}{x_2}$ is a function of $Q$ only
	\begin{equation}\label{invertibility}
		-\frac{x_2}{x_1} = \frac{Q f_Q}{f} \equiv r(Q)\,.
	\end{equation}
	If and when the relation \eqref{invertibility} is invertible to obtain $Q\left(x_1,x_2\right)$, it can be substituted back into the expression for $m(Q)$ to get $m=m(x_1,x_2)$, hence closing the system.
	
	Some important observations are worth mentioning here.
	\begin{enumerate}
		\item For monomial theories like $f(Q)=\alpha Q^n$, which include the case of STEGR without a cosmological constant ($n=1$) and the limiting case of polynomial $f(Q)$s at the high-$Q$ limit (for $n>1$) or low-$Q$ limit (for $n<1$), the phase space dimensionality is reduced by one, because in this case one has $r(Q)=n$, so that an additional constraint $x_2 + n x_1 = 0$ holds true. The same happens for the dynamical system formulation for $f(R)$ gravity, as mentioned in \cite[Sec.5]{Carloni:2007br}. 
		\item For the connection branch $\Gamma_1$, the $x_2$ dimension freezes ($x_2=-1$), so that one can write $m=m(x_1)$. The connection function $\gamma(t)$, and hence $x_4$, does not play any role in the cosmological dynamics, so that one must be able to write a closed system without $x_4$. Lastly, an additional hidden constraint exists on the phase space, which can be revealed as follows. Eq.\eqref{frd_ray-1_rev} provides
		\begin{equation}
			\frac{\dot{H}}{H^2} = -\frac{3}{2}\Omega + x_3\,,
		\end{equation}
		using which, the very definition of the dynamical variable $x_3$, after some straightforward manipulations, give the following constraint
		\begin{equation}
			x_3 = \left(\frac{3m(x_1)}{1+2m(x_1)}\right)\Omega\,.
		\end{equation}
		Ultimately, one is left with $x_1$ and $\Omega$, which will be related via the Friedmann constraint. Therefore, cosmological phase space dimensionality in the $\Gamma_{1}$ connection branch reduces to one.
		\item It is interesting to seek for possible physical interpretations that can be given to the the dynamical variables defined in \eqref{dynvar_def}. They are as follows:
		\begin{itemize}
			\item Since the $x_2$ dimension freezes ($x_2=-1$) in the $\Gamma_{1}$ connection, a dynamics along the $x_2$ direction characterizes a deviation from the $\Gamma_{1}$ gauge  dynamics.
			\item In the absence of a cosmological constant term, the quantity $(x_1+x_2)$ characterises a deviation from GR.
			\item The variable $x_3$ can be related to the rate of change of the effective gravitational coupling
			\begin{equation}
				x_3 = -(1+z)\frac{\kappa_{\rm eff}'(z)}{\kappa_{\rm eff}(z)}\,.
			\end{equation}
			From the observations, it is possible to provide an upper bound on the time variation of $\kappa_{\rm eff}$ \cite{Ray:2005ia}, and therefore, to $x_3$. 
			\item The variable $\Omega$ is related straightforwardly to the standard matter abundance parameter $\Omega_m$ whose present day value is treated as a free parameter during model constraining
			\begin{equation}
				\Omega = \kappa_{\rm eff}\Omega_m\,.
			\end{equation}
		\end{itemize} 
	\end{enumerate}
	
	The evolution of the system depends on the specific functional form of $f(Q)$, which influences the behaviour of the auxiliary quantity $m(Q)$ and the dynamical equations. To gain deeper insight into the underlying dynamical aspects like the fixed point structure, invariant submanifolds, emergence of parameter (in)dependence, etc, it is necessary to consider explicitly at least a broad class of $f(Q)$ models. Special care must be taken to single out the physically viable region of the phase space. The condition of positivity of the effective gravitational coupling $\kappa_{\rm eff}=\frac{1}{f_Q}>0$ already constrains the phase space to the region $\Omega>0$ for all three connection branches. Particular connection branches can provide specific constraints on the availability of the physically allowed region of the phase space.
	
	The example we have chosen to work with in this paper is $f(Q)=\alpha Q + \beta(-Q)^n$ with $\alpha>0,\,n\neq0,1$. For this theory, one can calculate that
	\begin{equation}
		r(Q) = \frac{Q f_Q}{f} = \frac{\alpha - n\beta(-Q)^{n-1}}{\alpha - \beta(-Q)^{n-1}}\,, \qquad m(Q) = \frac{Q f_{QQ}}{f_Q} = - \frac{n(n-1)\beta(-Q)^{n-1}}{\alpha - n\beta(-Q)^{n-1}}\,.
	\end{equation}
	
	Inverting $r(Q)$,  one gets
	\begin{equation}
		(-Q)^{n-1} = \frac{\alpha}{\beta}\left(\frac{1-r}{n-r}\right)\,,
	\end{equation}
	which, when substituted into the expression for $f(Q)$, gives
	\begin{equation}
		m(r) = n\left(1 - \frac{1}{r}\right) \quad \text{or,} \quad m(x_1,x_2) = n\left(1 + \frac{x_2}{x_1}\right)\,.
	\end{equation}

	\section{DYNAMICAL SYSTEM  ANALYSIS  FOR  $\Gamma_1$}\label{sec:Gamma1}
	
	In the connection branch $\Gamma_1$, the connection is chosen to vanish, simplifying the system by attributing all gravitational effects to nonmetricity. This gauge provides the simplest setup in symmetric teleparallelism and allows us to explore the influence of nonmetricity on cosmic evolution. We formulate the dynamical system using only the following dimensionless variables in the case :
	
	\begin{equation}
		\Omega \equiv \frac{\rho_m}{3 H^2 f_Q}\,, \quad x_1 \equiv -\frac{f}{6 H^2 f_Q}=\frac{f}{Qf_Q}\,.
	\end{equation}
	In terms of these variables, Eq.(\ref{fried-rev}) yields the Friedman constraint 
	\begin{equation}\label{Gamma1_fried_constr}
		x_1 + \Omega = 2\,.
	\end{equation}
	Using the Friedmann constraint $\Omega= 2 -x_1  $, the system of eqs.\eqref{fried-rev}, \eqref{raych-rev}, \eqref{mat_cons_rev}  reduces to a one-dimensional dynamical system given by
	\begin{equation}\label{eq:dx1_c}
		\Omega' = 3\Omega \left[\Omega\left(\frac{1 + m(Q)}{1 + 2m(Q)}\right) - 1\right]\,.
	\end{equation}
	Notice that $x_1$ is a function of $Q$ only:
	\begin{equation} \label{eq:ds_conn1}
		x_1(Q) = \frac{d\ln Q}{d\ln f} \,.
	\end{equation}
	For a given $f(Q)$, if this relation is invertible, we can express $Q=Q(x_1)$, leading to $m(Q)=m(Q(x_1))$. Along with the Friedmann constraint $x_1 = 2 - \Omega$, the invertibility condition is crucial for closing the system and obtaining a self-consistent evolution.
	
	It was pointed out by Boehmer et.al. \cite{Boehmer:2022wln} that the $f(Q)$ dynamical system for the connection branch $\Gamma_1$ in the presence of a single fluid is 1-dimensional. Our result is consistent with that. In fact, our dynamical system is the same as that obtained in that paper (see \cite[Eq.(14)]{Boehmer:2022wln}).
	
	From \eqref{frd_ray-1_rev}, we can trade between $q$ and $m(Q)$ as
	\begin{equation}
		\label{trade}
		q=-1+\frac{3 \Omega}{2(1 + 2m(Q))}\,.
	\end{equation}
	This relation helps us to  find the cosmology corresponding to a particular fixed point.

	\subsection{Generic considerations for $\Gamma_1$}\label{subsec:Gamma1_generic}
	
	First, without specifying a particular $f(Q)$ model, we list the general features that can be extracted from the dynamical system. We can frame these considerations as theory-space constraints, which will allow us to extract model-independent insights.
	
	\begin{enumerate}
		
		\item \textbf{Singularity at $m = -\frac{1}{2}$:}
		
		The dynamical system \eqref{eq:dx1_c} becomes singular for $m(Q) =- \frac{1}{2}$, corresponding to the theory $f(Q) = C_1 + C_2\sqrt{-Q}$. This case is exceptional: the corresponding minisuperspace action consists solely of a constant and a total derivative term, implying the theory lacks genuine cosmological dynamics (see \cite{Paliathanasis:2023pqp}). 
		
		\item \textbf{GR Limit:}
		
		The GR limit is recovered when $m(Q) \to 0$. In this case, Eq.~\eqref{eq:dx1_c} reduces to the familiar GR form. This limit is smooth in the $\Gamma_1$ branch.
		
		\item \textbf{Existence of a vacuum fixed point $\mathcal{P}_1$:} 
		
		A fundamental requirement for any physically reasonable $f(Q)$ cosmological model is that $\Omega=0$ (vacuum-dominated epoch) must be an \emph{invariant submanifold}, which is nothing but a fixed point in a 1-dimensional dynamical system. This condition ensures that once the system reaches a state where matter density vanishes, it remains there, aligning with the standard cosmological paradigm where no classical mechanism exists for spontaneous particle creation from vacuum. For a vacuum fixed point to exist, in the vicinity of $\Omega=0$, the following relationship must hold
		\begin{equation}\label{cond_Omega_c1}
			\Omega \left(\frac{1+m(Q)}{1+2m(Q)} \right) -1 \sim -\frac{C}{\Omega^p}, \qquad p<1 ~ \text{near} ~ \Omega \gtrsim 0.
		\end{equation} 
		The vacuum fixed point will be stable or unstable depending on whether $C>0$ or $C<0$. For constant $m$, this condition is always satisfied, and the corresponding  2-parameter class of theories  $f(Q) = -2\Lambda + (-Q)^{1+m}$ admits a stable vacuum point. 
		
		\item \textbf{de Sitter Nature of the Vacuum Point:}
		
		In the last item, we established the conditions to be satisfied in the theory space by an $f(Q)$ theory such that there is a vacuum fixed point. The relationship \eqref{trade} reveals that, whenever a vacuum fixed point exists, the corresponding cosmological solution must necessarily be a De-Sitter solution $q=-1$. In other words, the vacuum solutions admitted by any $f(Q)$ theory with the connection branch $\Gamma_1$ must necessarily be De-Sitter solutions \footnote{Although the condition $q=-1$ is widely associated with De-Sitter solutions, this does not necessarily have to be true. In certain rare situations $q\to-1$ can also correspond to a super-inflationary solutions with $a(t)\sim e^{t^2}$ (see \cite[Sec. 3.5]{Barrow:2006xb})}.
		
		\item \textbf{Matter-Dominated Fixed Point $\mathcal{P}_2$:}
		
		For cosmological viability, an $ f(Q) $ theory must admit a matter-dominated fixed point \(\mathcal{P}_2\) at:
		\begin{equation}
			\Omega = \frac{2m + 1}{m + 1}, 
		\end{equation}
		for constant $m$ satisfying  $m < -1$ or $m > -\frac{1}{2}$.  The corresponding deceleration parameter is:
		\begin{equation}
			q = \frac{1 - 2m}{2(1 + m)}.
		\end{equation}
		For instance, $m = 0$ yields $\Omega = 1$ and $q = \frac{1}{2}$, mimicking a matter-dominated universe. The 2-parameter class of theories  $f(Q) = -2\Lambda + (-Q)^{1+m}$   satisfy this when $m < -1$ or $m > -\frac{1}{2}$.   For non-constant $ m(Q) $, $\mathcal{P}_2$ exists if  the equation $ \Omega  =  \frac{2m(\Omega) + 1}{m(\Omega) + 1}$ yields a positive root, depending on specific $ f(Q) $ (see Section \ref{sectG1example}).
		
		\item \textbf{Stability of fixed points:}
		
		The stability of the fixed points is established by analysing the eigenvalues of the linearised system. The Jacobian of the 1-dimensional dynamical system \eqref{eq:dx1_c} is
		\begin{equation}\label{sta_coin_recns}
			J = \left[-3 \left( m'(\Omega)\right) \Omega^{
				2} + 6\left(  \left( \Omega -1\right) m(\Omega) + \Omega -\frac{1}{2} \right) \left( 1 + 2m(\Omega) \right) \right] \,.
		\end{equation}
		In the above $m'(\Omega)\equiv\frac{dm}{d\Omega}$, which can only be calculated once a theory is specified. A fixed point is stable/unstable according to whether $J<0$ or $J>0$.
		For the vacuum point $\mathcal{P}_1$, since $\Omega = 0$, the stability condition  reduces to:
		\begin{equation}\label{P-1stability}
			m'(0)\cdot 0^2+ (2m(0) + 1)^2 > 0,
		\end{equation}
		implying stability whenever $m$ is constant. The matter point $\mathcal{P}_2$ is stable or unstable depending on the form of $m(\Omega)$; for constant $m$, it is always a repeller. Thus, the stability and existence of fixed points depend critically on whether $ m$ is constant or non-constant, as summarised in Table~\ref{tab:fixedpoint-summary_c1}.
		
	\end{enumerate}
	\begin{table}[h!]
		\begin{center}
			\renewcommand\arraystretch{1.3}
			\begin{tabular}{|c|c|c|c|}
				\hline
				{\bf Fixed Point} & $m$-{\bf type} & {\bf Existence} & {\bf Stability}
				\\ \hline
				\multirow{2}*{$\mathcal{P}_1 (\Omega=0)$}
				& Constant    & Condition (\ref{cond_Omega_c1})  & Attractor   \\
				& Non-constant & Condition (\ref{cond_Omega_c1})   & Condition (\ref{P-1stability})  \\
				\hline
				\multirow{2}*{$\mathcal{P}_2 \left(\Omega=\frac{2m+1}{m+1}\right)$}
				& Constant    & $m\neq -1$  & Repeller   \\
				& Non-constant & Model-dependent   & Repeller if $J>0$  \\
				\hline
			\end{tabular}
		\end{center}
		\caption{Existence and stability of fixed points in the $\Gamma_1$ dynamical system for $f(Q)$ gravity, categorized by constant and non-constant $m(\Omega)$.} \label{tab:fixedpoint-summary_c1}
	\end{table}

	\subsection{Physically Viable Regions in Phase Space}
	
	The structural features identified above—such as the de Sitter attractor and matter-dominated fixed point—highlight the potential of $f(Q)$ gravity to reproduce key epochs of cosmic evolution. However, to ensure that such models are not just mathematically interesting but also physically viable, additional constraints must be imposed on the phase space. These include the positivity of effective energy, ghost-freedom ($f_Q > 0$), and well-defined behaviour of $Q$. In this subsection, we outline the key restrictions required for a consistent and physically meaningful $f(Q)$ cosmology.
	
	We have already mentioned that because of the physical viability condition $f_Q>0$, the dynamical variable $\Omega$ is restricted to the range $\Omega\geq0$. This condition, together with the Friedmann constraint \eqref{Gamma1_fried_constr} yields 
	\begin{equation}
		x_1 \leq 2\,.
	\end{equation}
	
	For a specific $f(Q)$ model, provided that the relation \eqref{eq:ds_conn1} can be inverted to express $Q = Q(\Omega)$, the following physical restrictions on the range of ($\Omega $ must be imposed in the $\Gamma_1$ connection branch:
	\begin{itemize}
		\item The requirement $f_Q>0$ implies $\Omega>0$, which is actually true for all the connection branches.
		\item Since for $\Gamma_1$, $Q=-6H^2$, this puts another restriction $Q(\Omega)<0$ on the range of $\Omega$. Notice that here we assume that a phase with $H=0$, such as a cosmological turnaround, cannot appear in the late time context. Since we do not expect a late-time cosmology that is drastically different from the General Relativistic cosmology, and a phase with $H=0$ cannot appear in a General Relativistic late-time cosmology, $Q(\Omega)<0$ is a valid requirement for cosmic viability.
		\item Since the quantity $f_Q$ is a function of $Q$, which can be expressed as a function of $\Omega$, the condition $f_Q>0$ itself can be recast as a restriction on $\Omega$.
		\item From the modified Friedmann equation (\ref{fried-rev}) one gets a further constraint
		\begin{align}
			f - 2Q f_Q = 2\rho_m \geq 0.
		\end{align}
		The above conditions serve as non-trivial viability filters. For specific $f(Q)$ models,  they may further restrict the admissible range of $\Omega $, or place bounds on model parameters.
	\end{itemize}
	
	In this work, we examine power-law modifications of the form $f(Q) = \alpha Q + \beta (-Q)^n$, where $n$ may be fractional (e.g., $n=1/4, 3/2$). In order to confirm the mathematical consistency of such models, we require $-Q > 0$, so that real-valued powers remain well-defined. This condition is naturally satisfied in the connection $\Gamma_{1}$ ($Q=-6H^2$), but may require careful handling in the  $\Gamma_2$ and $\Gamma_3$ gauges. Notably, power-law models of the form
	$f(Q) = Q^\alpha$ was initially studied in the $\Gamma_{1}$ connection and later extended to other branches. However, when adopting different connection choices,  special attention must be given to the sign of $Q$ to ensure that the formulation remains mathematically consistent. In particular, the choice of dynamical variables should reflect this requirement to avoid ambiguities in the phase space analysis. 
	
	By analysing this class of $f(Q)$ theories, we aim better to understand the parameter dependence of the dynamical system, the role of $m(Q)$, and the viability of $f(Q)$ models in explaining cosmic evolution. 
	
	Thus, as a general caution,  we must ensure the proper sign conventions in $f(Q)$ models when extending results across different connections. This convention guarantees the robustness of the dynamical system approach and avoids potential misinterpretations arising from implicit assumptions about the behaviour of $Q$. 
	
	\subsection{Example : $f(Q) = \alpha Q + \beta (-Q)^n$, $\alpha>0,\,\beta\neq0,\,n\neq0,1$}\label{sectG1example}
	
	In this work, we consider a power-law deformation of the STEGR, defined by
	\begin{equation}
		f(Q)=\alpha Q+\beta (-Q)^n\,,
	\end{equation}
	with $\alpha>0,\,\beta\neq0,\,n\neq0,1$ where $\alpha, n$ are dimensionless parameters and $ \beta$ is parameter of suitable dimension. This form generalises $\Lambda$CDM by incorporating nonlinear modifications of the nonmetricity scalar $Q$, allowing for richer cosmological dynamics. 
	For $n>1$, the nonlinear term becomes dominant at early times, a potential description of inflationary behaviour \cite{Anagnostopoulos:2022gej}, whereas $0 < n < 1$ modifications naturally support late-time cosmic acceleration \cite{BeltranJimenez:2019tme}.
	
	Power-law models of this type have been extensively studied both at the background and perturbative levels~\cite{BeltranJimenez:2019tme, Khyllep:2021pcu, Khyllep:2022spx}, and have been constrained using a variety of astrophysical datasets such as Type Ia Supernovae (SNe Ia), Baryon Acoustic Oscillations (BAO), and Hubble rate measurements~\cite{Koussour:2023rly, Mandal:2023cag, Wang:2024eai}.
	Such observational constraints are key to checking whether $f(Q)$ gravity models can match cosmological data, especially when compared to the standard $\Lambda$CDM CDM model. They help determine if the theory can reproduce the observed expansion history and structure formation, while also placing constraints on the model parameters.

	For this particular model, the auxiliary quantity $m(Q)$ introduced earlier
	can be expressed in terms of the dynamical variable 
	$\Omega$  as
	\begin{equation} \label{m_dynvar_c1}
		m=n(\Omega-1)\,,
	\end{equation}
	which follows directly from the definition of $m$,  $f$ and the  Friedman constraint \eqref{Gamma1_fried_constr}.
	
	Substituting the above $m$ into the general dynamical equation (\ref{eq:dx1_c}) for $\Gamma_1$, we obtain a  first-order differential equation governing the evolution of $\Omega(N)$:
	\begin{equation}\label{eq:dx2_c}
		\Omega' = \frac{3\Omega (\Omega-1)(\Omega n-2n+1)}{(2\Omega n-2n+1)}\,.
	\end{equation}
	In addition, the deceleration parameter $q$ given by (\ref{trade}) can be rewritten as 
	\begin{equation}
		\label{eq:q_Gamma1}
		q=-1+\frac{3 \Omega}{2[1 + 2n(\Omega-1)]}\,.
	\end{equation}
	Thus, the model's entire phase-space dynamics can be analysed through the evolution of $\Omega(N)$ and the corresponding behaviour of $q(N)$.

	\subsubsection*{Physically viable region of the phase space}
	
	\noindent We must impose additional constraints beyond the dynamical evolution to ensure physical viability. These constraints ensure that the gravitational theory remains physically meaningful. Specifically, for the model $f(Q) = \alpha Q + \beta (-Q)^n$ with $\alpha>0$, by virtue of the relation \eqref{eq:Q_n_c1}, let us list how these four constraints restrict the available range for $\Omega$ and/or the available parameter space $\{\alpha,\beta,n\}$. 
	
	Using the definition of the auxiliary quantity $m(Q)$ and Eq.~\eqref{m_dynvar_c1}, we can express $Q$ in terms of $\Omega$:
	
	\begin{equation}\label{eq:Q_n_c1}
		(-Q)^{n-1}=\frac{\alpha (\Omega-1)}{\beta (\Omega n-2n+1)}\,.
	\end{equation}
	In the subsection \ref{subsec:Gamma1_generic}, we listed four conditions that put some requirements on the range of the dynamical variable $\Omega$. Each physical condition translates into specific restrictions on $\Omega$ and the model parameters $(\alpha,\beta,n)$ by virtue of the relation \eqref{eq:Q_n_c1}, as discussed below.
	
	\begin{itemize}
		\item Firstly, we should demand non-negative matter density, i.e. $\Omega\geq 0$.
		\item Condition  $Q<0$: From Eq.~\eqref{eq:Q_n_c1}, the requirement $Q<0$ implies 
		\beq 
		\frac{(\Omega-1)}{\beta (\Omega n-2n+1)} > 0 \,,
		\eeq
		given that $\alpha>0$. The sign of the above fraction depends on $\beta$ and $n$. The detailed allowed ranges for $\Omega$ are summarised in the following Table:
		$$
		\begin{array}{c|c|c}
			&\beta>0 & \beta<0 \\\hline & \\
			n<0 &
			1< \Omega <\frac{2n-1}{n}  & \Omega >\frac{2n-1}{n}~ \text{or}~ \Omega<1\\[1.5ex]
			0<n<1  & \Omega>1~\text{or}~ \Omega<\frac{2n-1}{n} & \frac{2n-1}{n}<\Omega<1   \\[1.5ex]
			n>1& 	\Omega<1~\text{or}~ \Omega>\frac{2n-1}{n} & 1<\Omega<\frac{2n-1}{n}  
		\end{array}
		$$
		\item Condition $f_Q>0$: The condition $f_Q>0$ implies 
		\beq \label{fQ1_phys_c1}
		\frac{(1-n)}{\Omega n-2n+1}>0\,,
		\eeq
		where again $\alpha>0$ has been used. 
		\item Condition $f(Q)-2Qf_Q\geq 0$: Similarly, this condition leads to the same inequality as \eqref{fQ1_phys_c1}:
		\beq  \label{fQ2_phys_c1}
		\frac{(1-n)}{\Omega n-2n+1}\geq 0\,.
		\eeq 
		Thus, the conditions $f_Q>0$ and $f(Q) - 2Qf_Q\geq 0$ coincide in this model.
		\item Taking into account that $\alpha>0$ and $\Omega \geq 0$, the condition \eqref{fQ1_phys_c1} for this model leads to the following restrictions on $\Omega$:
		$$
		\begin{array}{ccc}
			
			0\leq \Omega <\frac{2n-1}{n} & \text{if}& n<0\\[1.5ex]
			\Omega> \frac{2n-1}{n}   & \text{if}&0<n<1 \\[1.5ex]
			0\leq \Omega<\frac{2n-1}{n}   & \text{if}& n>1
		\end{array}
		$$
		\item Taking into consideration all of the above requirements, the physical phase space of fixed points of the dynamical equation \eqref{eq:dx2_c} is given in Table \ref{psp_c1} for different choices of $\beta$ and $n$.
		\begin{table}[H]
			\centering
			\begin{tabular}{c|c|c}
				& $\beta>0$ & $\beta<0$ \\\hline & \\
				$n<0$ & $\left\lbrace \Omega \in \mathbb{R} \mid 1< \Omega <\frac{2n-1}{n}  \right\rbrace$
				&  $\left\lbrace \Omega \in \mathbb{R} \mid 0 \leq \Omega <1  \right\rbrace$ \\ [1.5ex] \hline &\\
				$0<n<1$  & $\left\lbrace \Omega \in \mathbb{R} \mid \Omega > 1\right\rbrace$
				& $\left\lbrace \Omega \in \mathbb{R} \mid  \frac{2n-1}{n}  < \Omega <1  \right\rbrace$ \\ [1.5ex] \hline &\\
				$n>1$ & $\left\lbrace \Omega \in \mathbb{R} \mid   0 \leq \Omega <1 \right\rbrace$
				& $\left\lbrace \Omega \in \mathbb{R} \mid 1< \Omega <\frac{2n-1}{n}   \right\rbrace$ \\ [1.5ex] \hline
			\end{tabular}
			\caption{The physically allowed region of the phase space of the dynamical equation \eqref{eq:dx2_c}}
			\label{psp_c1}
		\end{table}
	\end{itemize} 
	The physically allowed range for the dynamical variable $\Omega$ is shown in the $n$-$\Omega$ plane in Figs.\ref{fig:phys_reg_plot_c1} and \ref{fig:phys_reg_plot_c12} below. Evolution of cosmological parameters along a cosmologically relevant trajectory is shown in  Fig.\ref{fig:qcon1}.
	
	\begin{figure}[H]
		\centering
		\subfigure[]{%
			\includegraphics[width=8cm,height=6cm]{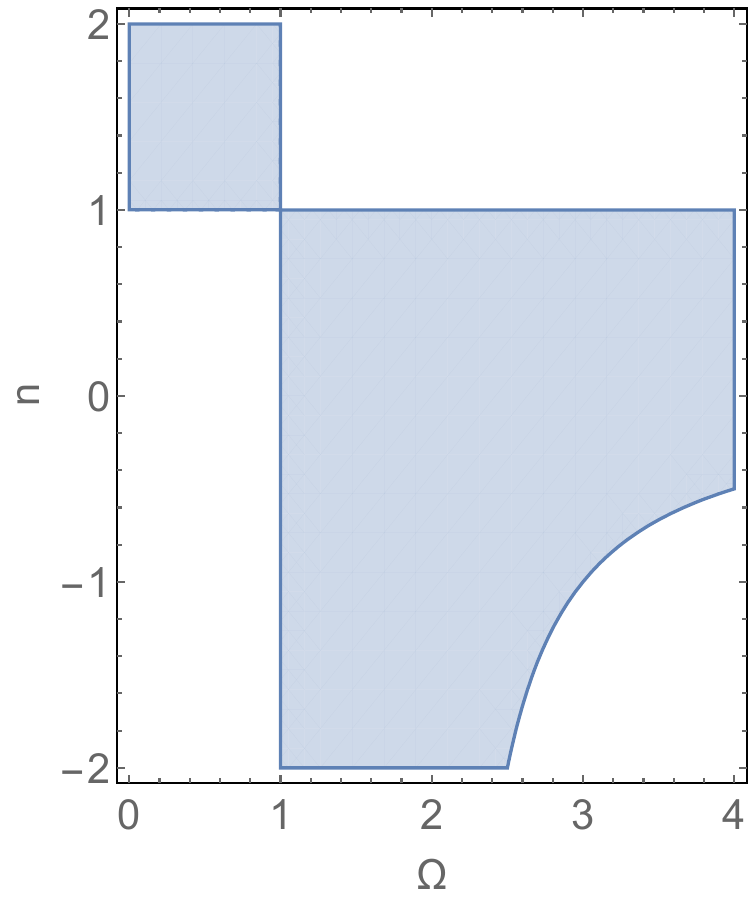}\label{fig:phys_reg_plot_c1}}
		\quad
		\subfigure[]{%
			\includegraphics[width=8cm,height=6cm]{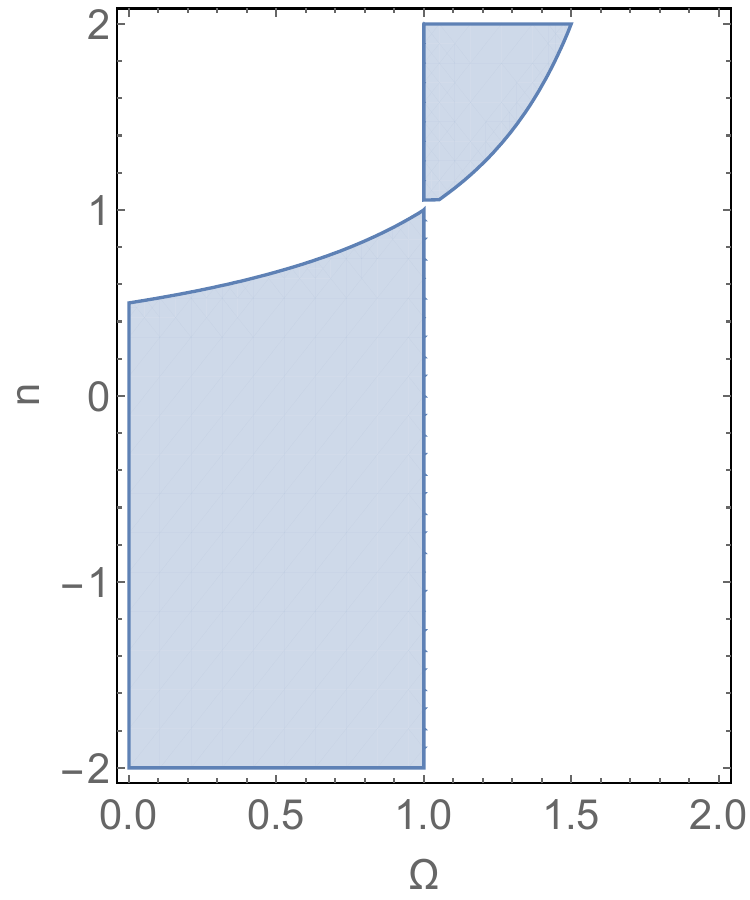}\label{fig:phys_reg_plot_c12}}
		
		\quad
		\subfigure[]{%
			\includegraphics[width=8cm,height=6cm]{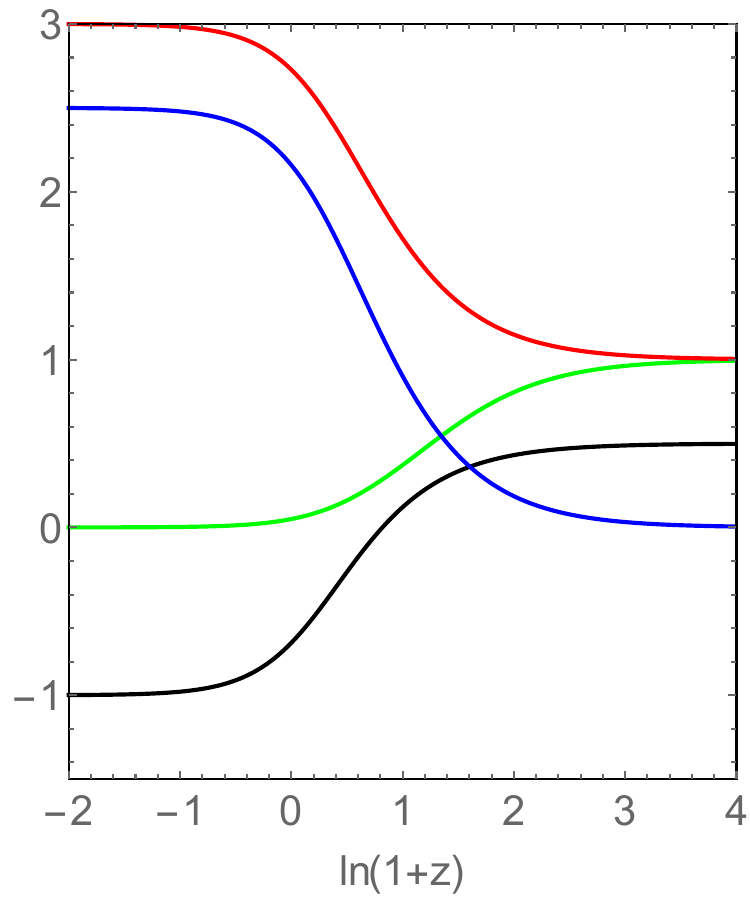}\label{fig:qcon1}}
		\caption{{\it Upper panel:}~Physically viable regions in the $n$-$\Omega$ plane for $\Gamma_1$ with $f(Q) = \alpha Q + \beta (-Q)^n$, satisfying $\Omega \geq 0$, $Q < 0$, $f_Q > 0$ and $f-2Qf_Q\geq0$ for $\alpha = 1$ and $\beta>0$ in (a) and $\beta<0$ in (b). {\it Lower panel:}~Evolution of the deceleration parameter $q$ (black curve), matter density parameter $\Omega$ (green curve), $\frac{\alpha (1 - n)}{\Omega n - 2n + 1}$ (red curve) and  $\frac{(\Omega-1)}{\beta (\Omega n-2n+1)}$ (blue curve) for $n = 0.4$, $\alpha = 1$, $\beta = 1$.}
	\end{figure}

	\subsubsection*{Fixed points}
	
	We now present the fixed points for the dynamical system (Eq.~\eqref{eq:dx1_c}) governing this specific model in the following Table~\ref{tab:C1_2}.

	\begin{center}
		\begin{table}[H]
			\centering
				\renewcommand\arraystretch{1.5}
			\begin{tabular}{|c|c|c|c|c|c|}
				\hline 
				Fixed point& $\Omega$ & Stability  & $q$& Cosmological solution & Physical viability \\ 
				\hline 
				$A_1 $ & 0 & Attractor & $-1$ for $n\neq \frac{1}{2}$ & $a(t)=a_0 e^{H_0 t}$ (deSitter) & $(\beta<0 ~\& ~n<0) ~\text{or}$~\\&&&&& $(\beta<0 ~\& ~n<\frac{1}{2})  ~\text{or}~(\beta>0 ~\& ~n>1)  $\\ 
				\hline
				$A_2$ &  $1$ &   Repeller for $n<1$ & $\frac{1}{2}$  & $a=a_0 t^{\frac{2}{3}}$ & Nonviable \\
				& & Attractor for $n>1$ &  &  & \\
				\hline 
				$A_{3}$ & ${\frac {2n-1}{n}}$ & Attractor if $0<n<1$ & $\frac{3-2n}{2n}$ for $n\neq\frac{1}{2}$ & $a=a_0 t^{\frac{2n}{3}}$ for $n\neq\frac{1}{2}$ & Nonviable \\
				\hline
			\end{tabular} 
			\caption{Fixed points of the dynamical system \eqref{eq:dx1_c} for the model $f(Q) = \alpha Q + \beta (-Q)^n$ in the $\Gamma_1$ connection, along with their stability, deceleration parameter $q$, cosmological evolution, and physical viability (as assessed in Table \ref{psp_c1}). Note that while $A_3$ becomes degenerate with $A_1$ for $n = \frac{1}{2}$ in terms of $\Omega$,  their corresponding scale factor evolutions do not match. Hence,  we should treat this case separately.}
			\label{tab:C1_2}
		\end{table}
	\end{center}
	We note here that point $A_1$  corresponds to point $\mathcal{P}_1$ of the general case,  while points $A_2, A_{3}$ are fixed points associated with the general fixed point type $\mathcal{P}_2$. The stability analysis, based on the Jacobian of Eq.~\eqref{eq:dx1_c}, reveals that point $A_1$  is always stable. Point $A_2$ is unstable for $n<1$ and stable otherwise, while point $A_3$ is stable for $0<n<1$. 
	
	It is worth highlighting that $n=\frac{1}{2}$ is a special case. One can see from Eq.\eqref{eq:q_Gamma1}, the value of the deceleration parameter freezes to $q=2$ (which corresponds to $a(t)\sim t^{2/3}$). In this case, direct integration of Eq.~\eqref{eq:dx2_c} gives
	\begin{equation}
		\Omega(N) = \frac{\Omega_0}{\Omega_0 + (1 - \Omega_0) e^{3 (N - N_0)/2}},
	\end{equation}
	so that $\Omega$ is monotonically decreasing from $\Omega=1$ in the asymptotic past to $\Omega=0$ in the asymptotic future. However, despite this smooth evolution in $\Omega$, there is no transition from a decelerated to an accelerated phase since $q$ remains fixed at $\tfrac{1}{2}$ throughout. 
	
	Thus, the theory $f(Q) = \alpha Q + \beta \sqrt{-Q}$ (i.e., $n = \tfrac{1}{2}$) cannot account for late-time acceleration without an additional cosmological constant term. This conclusion is consistent with Ref.~\cite{Chakraborty:2025qlv}, which found that the square-root $f(Q)$ model requires a cosmological constant to achieve cosmic acceleration in the $\Gamma_1$ connection.

	Another solvable case is $ n = 2 $, where the solution\footnote{Note that the arbitrary constant of integration is the same for both branches.}:
	\begin{equation}
		\Omega(N) = \frac{C e^{3N} - 12 \pm \sqrt{(C e^{3N})^2 + 12 C e^{3N}}}{2 (C e^{3N} - 4)}, \quad C = \frac{(2 \Omega_0 - 3)^2 e^{-3N_0} }{\Omega_0 (\Omega_0 - 1)},
	\end{equation}
	which yields $\Omega \to \frac{3}{2}$ as $N \to -\infty$ in both branches and independently of the initial condition $\Omega_0$. Hence, keeping in mind the last row in Table \ref{psp_c1}, this point is not viable\footnote{We can  identify that also the future asymptotic limit $N \to \infty$ is ill-defined because it depends on the choice of the branch for the square root term, either reading $\Omega \to 1$ or $\Omega \to 0$ independently on the initial value $\Omega_0$.}.

	Notably, only the de Sitter point $A_1$ consistently resides within the physically allowed region of the phase space for certain combinations of the model parameters. In particular, from Table~\ref{psp_c1}, we see that $A_1$ is viable when
	$$
	(\beta < 0 ~\&~ n < 0), \quad \text{or} \quad (\beta < 0 ~\&~ n < \tfrac{1}{2}), \quad \text{or} \quad (\beta > 0 ~\&~ n > 1).
	$$
	These regimes satisfy the conditions $Q < 0$ and $f_Q > 0$, while keeping $\Omega \geq 0$ at the fixed point. In contrast, $A_2 $, which is the sole candidate for a matter-dominated past attractor, as well as $A_3 $ lie outside the viable phase space for all \( n \). That the fixed point $A_2$ falls in a physically nonviable region of the phase space can be shown explicitly from Eq.\eqref{eq:Q_n_c1}, according to which $\Omega=1$ makes $Q=0$, i.e. $H=0$. This conclusion is inconsistent with the fact that  $q=\frac{1}{2}$ at this fixed point, which requires a non-zero Hubble rate.

	The particular form of $f(Q)=\alpha Q+\beta(-Q)^n$, within the $\Gamma_{1}$ connection, has a long history of study in the context of cosmology, starting from one of the earliest papers on cosmology in $f(Q)$ gravity \cite{BeltranJimenez:2019tme}. An extensive phase space analysis considering all possible physical viability requirements was missing, which we perform here. Surely, such an $f(Q)$ form may give an adequate fit with data near $z\sim0$. But suppose one truly wants to investigate this as a candidate for the late-time cosmology starting from the matter-dominated epoch until $z\sim0$. In that case, our investigation proves that there is a problem. The absence of a physically viable matter-dominated era renders the $\Gamma_{1}$ connection branch cosmologically inadequate for the model $f(Q) = \alpha Q + \beta (-Q)^n $. Thus, merely inspecting the properties of the functional form of the theory Lagrangian density $f(Q)$, as done in \cite{Anagnostopoulos:2022gej}, may provide a first qualitative taste of the kinematics. However, in this paper, we delve into a much deeper understanding of the dynamics of the model. A given model may exhibit different features at the kinematics and dynamical levels as pointed out in \cite[Sect.IV]{Chakraborty:2021jku}. We, however, point out that inclusion of a nonvanishing cosmological term may alter the situation, but that is not investigated here. 
	
	Further, as pointed out in \cite{Jarv:2018bgs}, the cosmological equations of the $\Gamma_{1}$ connection coincide with those of $f(T)$ gravity. It is therefore interesting to compare our results with $f(T)$ gravity \cite{Hohmann:2017jao}. Unlike \cite{Hohmann:2017jao}, we restrict to $H \neq 0$ and focus on late-time evolution. Consistently with \cite{Hohmann:2017jao}, we find no phantom-divide crossing and obtain a physically viable late-time de Sitter attractor.

	\subsubsection*{Variation of the effective gravitational coupling}
	
	Despite the absence of a viable matter-dominated epoch in the $\Gamma_1$ branch for the model $f(Q) = \alpha Q + \beta (-Q)^n$, the robust late-time de Sitter attractor (point  $A_1$) motivates further investigation into the behavior of the effective gravitational coupling $G_{\text{eff}} \propto 1/f_Q$. First of all, we note that the present-time condition $G_{\rm eff}(z=0)<G$ \footnote{We write the condition as appeared in \cite{Boiza:2025xpn}. In our convention $G=1$, so that the condition just reduces to $f_Q>1$.},  which helps to alleviate  the $S_8$ tension (see {\it e.g.} \cite{Boiza:2025xpn}), in our scenario provides the upper/lower bound $H_0 \gtrless \sqrt{\frac{1}{6}\left( \frac{1-\alpha}{n\beta}\right)^{1/(n-1)}}$ on the Hubble constant\footnote{We restrict to parameter choices satisfying $\frac{1-\alpha}{n\beta}\geq 0$, ensuring that the analytic expression for the root remains real.} for $n \gtrless  1$; this constraint is tighter than those obtained in a model-independent way from DESI data when assessing modified gravity frameworks which are still consistent with $G_{\rm eff}(z=0) \approx G$ \cite{Ishak:2024jhs}. After discussing these observational constraints on the effective coupling itself, we will now inspect those on its relative time variation.

	Specifically, observational constraints on the time variation of $G_{\text{eff}}$ force stringent bounds that any modified gravity theory must respect. Studying the variation of $G_{\text{eff}}$ near the de Sitter point thus provides critical insights into the physical viability of the theory at late times and informs the construction of more refined $f(Q)$ models.
	
	Starting from Eq.~(\ref{frd_ray-1_rev}), for the selected model $f(Q) = \alpha Q + \beta (-Q)^n$, the evolution equation for the Hubble parameter $H(z)$ can be recast as
	\beq \label{for_k_eff_gam1_1}
	\left[6^n \beta n \left(n-\frac{1}{2}\right) H^{2n-1}(z)-3\alpha H(z)\right] \frac{dH(z)}{dz}-\frac{3 \rho_0 (1+z)^2}{2}=0\,,
	\eeq
	where $\rho_0$ is the present-day matter density. 
	
	Equation~(\ref{for_k_eff_gam1_1}) can be implicitly integrated to yield
	\beq 
	\rho_0 (1+z)^3 + 3\alpha H^2 + 6^n \beta \left( \frac{1}{2}-n \right)H^{2n}+C=0\,,
	\eeq
	where $C$ is an integration constant.
	
	This solution allows us to investigate the astrophysical constraint on the relative variation of the effective gravitational coupling. Observational bounds require that
	\begin{equation}
		\left| \frac{\dot{f}_Q}{f_Q} \right| < D,
	\end{equation}
	where $D$ is a small numerical constant obtained from astrophysical data (e.g., Ref.~\cite{Ray:2005ia}).
	
	Using Eqs.~(\ref{frd_ray-1_rev}) and (\ref{for_k_eff_gam1_1}), the quantity $\dot{f}_Q/f_Q$ can be expressed as
	
	\begin{equation}\label{for_k_eff_gam1_2}
		\left| \frac{\dot f_Q}{f_Q} \right| = \left| \frac{2 \cdot 6^n n (n-1) \beta (1+q)H^{1+2n} }{6\alpha H^2-6^n n \beta H^{2n} } \right|=\left| \frac{2n(n-1)(1+q)\{[(1+z)^3 +3C] \rho_0 + 3\alpha H^2 \} H}{3\alpha H^2 (n-1) - n[(1+z)^3 +3C]\rho_0} \right| <D \,.
	\end{equation}

	Rearranging Eq.~(\ref{for_k_eff_gam1_2}) yields a constraint on $1+q$ as
	\begin{equation}
		\label{finalvariation}
		\qquad |1+q|<D \left|\frac{3\alpha H^2 (n-1) - n[(1+z)^3 +3C]\rho_0}{2n(n-1)\{[(1+z)^3 +3C] \rho_0 + 3\alpha H^2 \} H}\right|.
	\end{equation}
	This inequality sets an upper limit on the deceleration function $q$ as a function of the Hubble parameter $H$ and the redshift $z$.
	
\begin{figure}[H]
	\centering
	\subfigure[]{%
		\includegraphics[width=8cm,height=6cm]{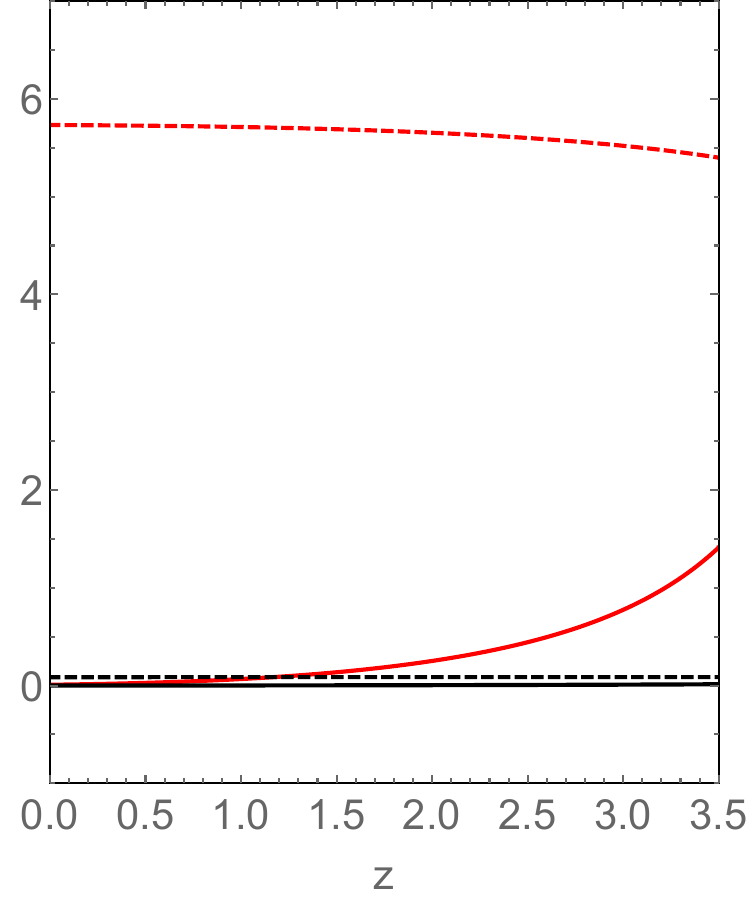}\label{fig:q_const_c1}}
	\quad
	\subfigure[]{%
		\includegraphics[width=8cm,height=6cm]{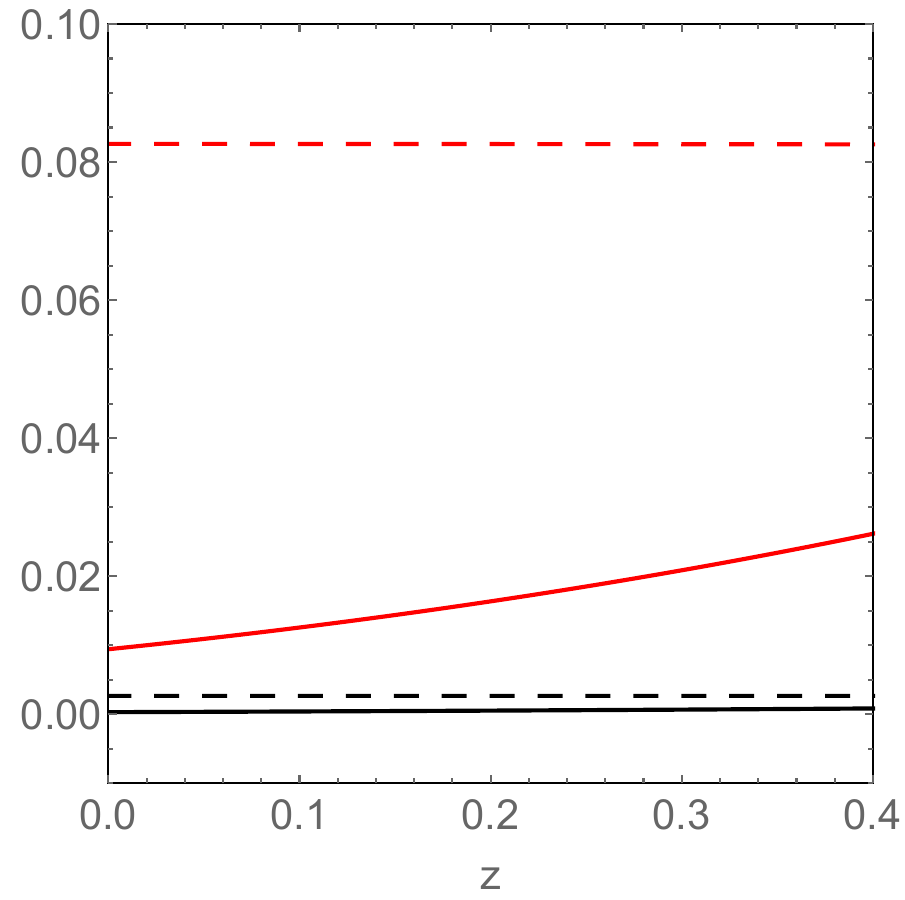}\label{fig:q_const_c1_lunar}}
	
	\quad
	\subfigure[]{%
		\includegraphics[width=8cm,height=6cm]{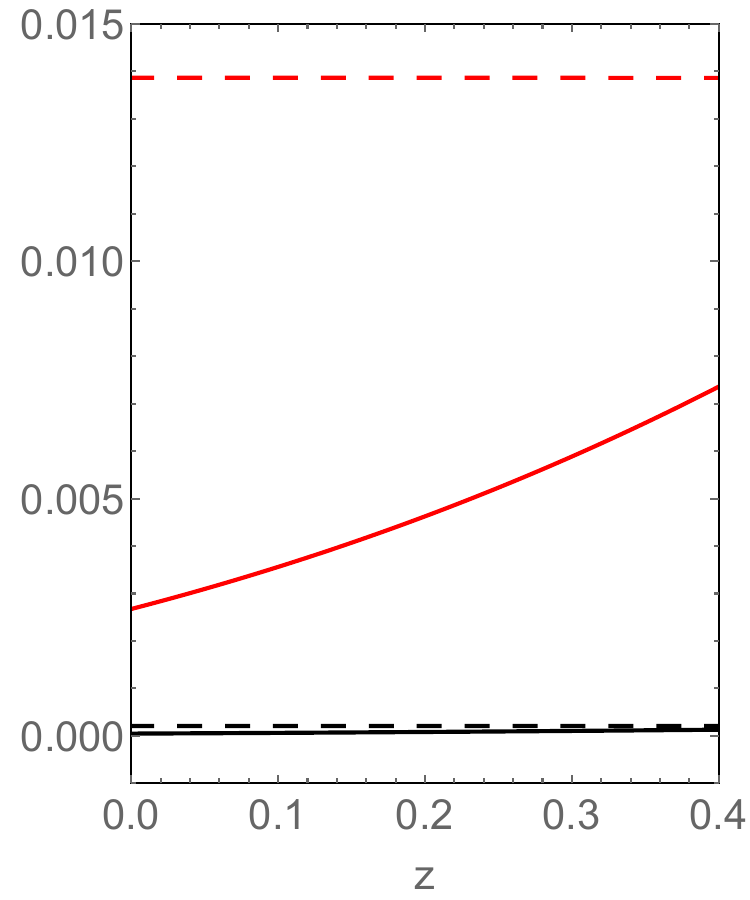}\label{fig:q_const_c1_planck}}
	\quad
	\subfigure[]{%
		\includegraphics[width=8cm,height=6cm]{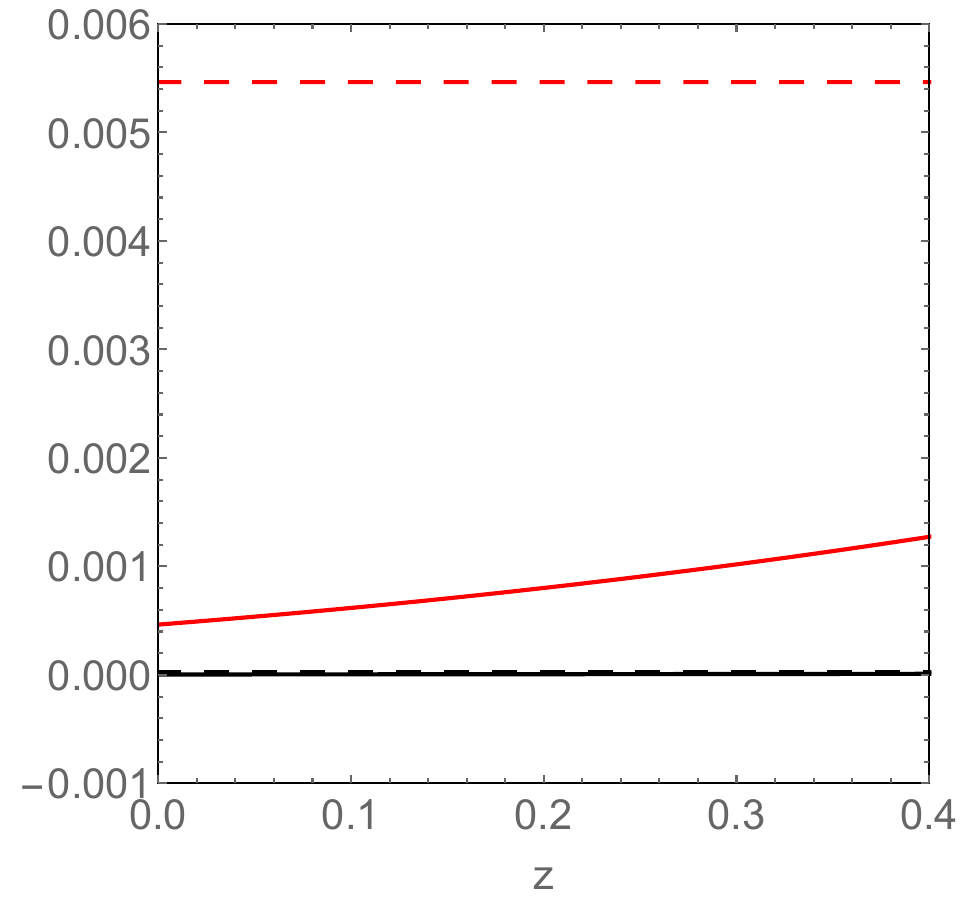}\label{fig:q_const_c1_solar}}
	
	\quad
	\subfigure[]{%
		\includegraphics[width=8cm,height=6cm]{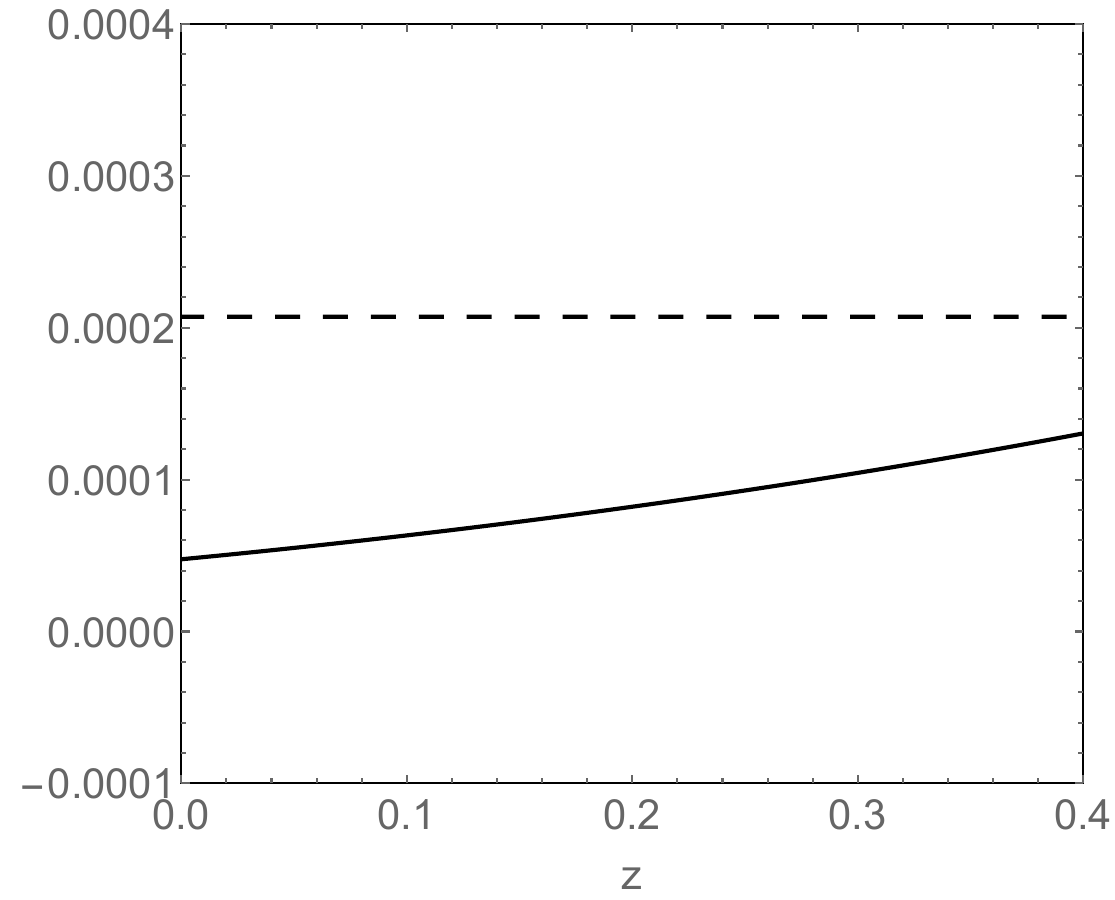}\label{fig:q_const_c1_planck_zoom}}
	\quad
	\subfigure[]{%
		\includegraphics[width=8cm,height=6cm]{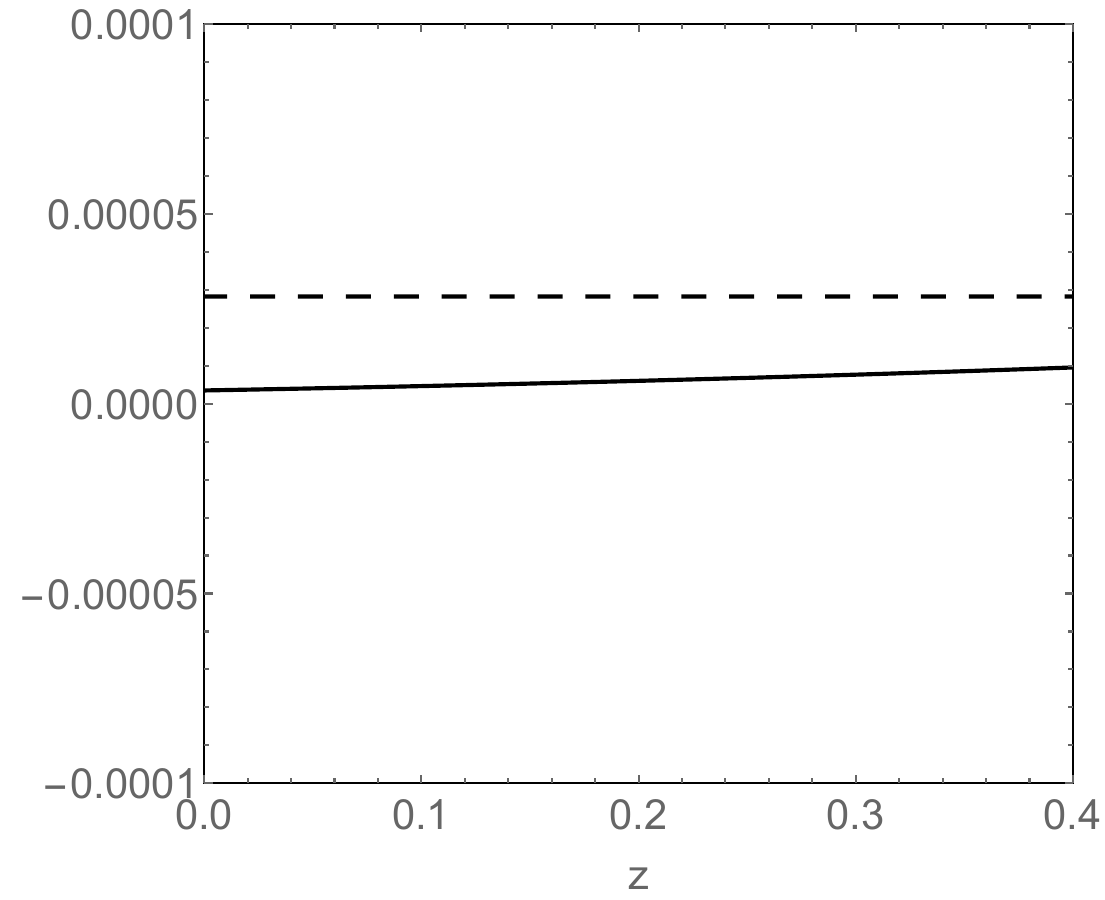}\label{fig:q_const_c1_solar_zoom}}
	
	\caption{In this figure, we depict the left- and right-hand sides of Eq.~(\ref{finalvariation})—which constrains the allowed relative variation of the effective coupling—using solid and dashed curves, respectively. We fix $\alpha = 1$, $\beta H_0^{2n - 2} = -1$ and take $\frac{D}{H_0} = 0.62, 0.01, 0.0015$ and $0.0005$ in panels (a), (b), (c) and (d) respectively; hence these panels scrutinize the bounds inferred from large-scale cosmic and local solar system dynamics, the former being looser. Figs. (e) and (f) are the enlarged versions of some regions in Figs. (c) and (d) respectively. In each plot,  red curves correspond to $n = 0.4$  and black curves correspond to  $n = 1.5$. Our analysis shows that this observational constraint is more easily satisfied towards the present epoch when $n < 1$, consistently with the preliminary assessment based on the structure of the Lagrangian density, which suggests that this branch of the model is more suitable for late-time cosmology \cite{Anagnostopoulos:2022gej}.}
	\label{fig:c1_1}
\end{figure}

	To verify the consistency of the model with observational bounds,  we numerically evaluate the constraint from Eq.~(\ref{for_k_eff_gam1_2}) as a function of redshift $z$. Using $H_0 = 2.1 \times 10^{-18}\, \mathrm{s}^{-1}$ and various values of $D/H_0$, we find that the variation of the effective gravitational coupling remains within acceptable limits up to redshift $z \sim 3.5$. We separately adopt several observationally motivated bounds on $D/H_0$. These include: $D/H_0 \approx 0.62$ from \cite{Chakraborty:2025qlv}; $|D/H_0| \approx 0.0015$ from the Planck-inferred value $|\dot G/G |_0 \approx 10^{-13}$ yr$^{-1}$ at recombination under the assumption of smooth evolution of the effective coupling \cite{Ballardini:2021evv};  $ -4.38 \cdot 10^{-4}  <D/H_0<6.94 \cdot 10^{-4}$ which follows from the reconstructed  $-2.9 \cdot 10^{-14} <\dot G/G< 4.6 \cdot 10^{-14}$ yr$^{-1}$  from solar system dynamics \cite{Pitjeva:2021hnc}; and finally $ -7.55 \cdot 10^{-3}  <D/H_0<0.020$ following from Lunar Laser Ranging Tests which provided $\dot G/G=(4\pm 9) \cdot 10^{-13}$ yr$^{-1}$  \cite{Williams:2004qba}. The latter constitute more stringent conditions, but are also required to hold just towards the present epoch and after the actual formation of the solar system; on the other hand, gravitational wave phenomena do not set tighter bounds  \cite{An:2025cmm}. The results are illustrated in Fig.~\ref{fig:c1_1}, confirming that the model satisfies the bound $\left|\dot{f}_Q/f_Q\right| < D$ throughout the late-time cosmological evolution.
	We specifically consider two representative cases of the exponent $n$, one with $n < 1$ and another with $n > 1$, to explore the differing phenomenological implications. We observe that, toward the present epoch, the constraint is more easily satisfied for the case $n < 1$, with reduced sensitivity to observational uncertainties. This behaviour aligns with the findings of Ref.~\cite{Anagnostopoulos:2022gej}, according to which the relevance of these two different classes of models lies separately in early- or late-time cosmology.

	\section{DYNAMICAL SYSTEM  ANALYSIS  FOR  $\Gamma_2$ }\label{sec:Gamma2}
	
	We now turn to the dynamical system analysis in the connection $\Gamma_2$. In contrast to the  connection  $\Gamma_1$, where the system could be described using just one dynamical variable  due to the simplified geometric structure, the connection $\Gamma_2$ dynamical system formulation requires all five dynamical variables defined in \eqref{dynvar_def}, i.e.
	
	\begin{equation}
		x_1 = -\frac{f}{6H^2 f_Q}, \quad x_2 = \frac{Q}{6H^2}, \quad x_3 = -\frac{\dot{f}_Q}{H f_Q}, \quad x_4 = \frac{\gamma}{2H}, \quad \Omega = \frac{\rho_m}{3H^2 f_Q}\,.
		\label{dynvar_gamma2}
	\end{equation}
	
	Note, here the role of $\gamma$ is a geometric quantity characterising the different connection branches. The above dynamical variables are connected by the Friedmann constraint  \eqref{fried-rev}, which can  be written as
	\begin{equation}
		\Omega = 1 - x_1 - x_2 - x_3 x_4\,.
		\label{friedmann_gamma2}
	\end{equation}
	Hence, the dynamical system for the $\Gamma_2$ connection is, in general, a four-dimensional system given by
	\begin{subequations}\label{eq:ds_2}
		\begin{align}\label{eqx1G2}
			x_1' &= \frac{x_2 x_3}{m} + x_1 \left(x_3 - \frac{2\dot{H}}{H^2} \right), \\
			\label{eqx2G2}
			x_2' &= -\frac{x_2 x_3}{m} - 2x_2 \frac{\dot{H}}{H^2}, \\
			\label{eqx3G2}
			x_3' &= -x_3 \left(\frac{\dot{H}}{H^2} - x_3 + 3\right), \\
			x_4' &= 1 + x_2 - x_4 \left(3 + \frac{\dot{H}}{H^2} \right)\,.
		\end{align}
	\end{subequations}
	
	Recall  $m$ is the function of $Q$ defined in \eqref{defofm} and the derivative of the Hubble parameter is obtained from the Raychaudhuri equation \eqref{raych-rev} as
	
	\begin{equation}
		\frac{\dot{H}}{H^2} = -\frac{3}{2} + \frac{3}{2} x_1 + \frac{3}{2} x_2 + x_3 - \frac{3}{2} x_3 x_4.
		\label{raych_gamma2}
	\end{equation}
	
	To connect the dynamical variables with the underlying geometry of $f(Q)$ in a model-independent way, we consider the auxiliary function $r(Q)$ defined in \eqref{invertibility}. This  auxiliary function can be  expressed as the ratio of phase space variables as:
	\begin{align}
	\frac{x_2}{x_1} = -r(Q).
	\end{align}
	
	This relation enables level curves of constant $Q$ to be defined in the $(x_1, x_2)$ plane, making it possible to parametrise theory space without committing to a specific $f(Q)$. When combined with the auxiliary function $m(Q)$, which can be written as $m(Q(x_1, x_2))$ if invertibility holds, the above dynamical system becomes autonomous, and we can study the behaviour around fixed points generically.
	
	From a physical standpoint, $r(Q)$ characterises the deviation of the theory from the case of pure GR without a cosmological constant, for which $r(Q) = 1$ identically. More generally, models of the form $f(Q) = Q + \text{const.}$ still reproduce Einstein gravity but yield $r(Q) \ne 1$ unless the constant vanishes. Thus, $r(Q)$ provides a useful diagnostic for deviations from GR-like dynamics and plays a central role in analysing the general viability of $f(Q)$ models in the $\Gamma_2$ connection.
	
	The  deceleration parameter $q$ can also be expressed as
	\begin{equation}
		\label{qG2}
		q = \frac{1}{2} - \frac{3}{2} x_1 - \frac{3}{2} x_2 - x_3 + \frac{3}{2} x_3 x_4.
	\end{equation}
	
	For the above dynamical system to be physically viable, we shall impose the conditions $f_Q > 0$ and $\rho_m \geq 0$, which imply
	\begin{equation}\label{phys1_c2}
		\Omega=1-x_1-x_2-x_3 x_4 \geq 0\,.
	\end{equation}
	Hence, the physically allowed four-dimensional phase space for the fixed points of the above dynamical system is given by
	\begin{equation}
		\label{regionG2}
		\left\lbrace (x_1,x_2, x_3,x_4) \in \mathbb{R}^4 \mid x_1+x_2+x_3 x_4 \leq 1\right\rbrace.
	\end{equation}

	Passing by, we note a mathematically intriguing characteristic. The dynamical system for $\Gamma_2$ \eqref{eq:ds_2} reduces to that of $\Gamma_1$ \eqref{eq:dx1_c} in the limit $\gamma\to0$, which makes $x_4\to0$ and $x_2=-1$. This is, however, a purely mathematical characteristic of the nature of the field equations, having nothing to do with physics, as for $\Gamma_2$, the dynamical connection function $\gamma(t)\neq0$ by definition.

	\subsection{Generic considerations for $\Gamma_2$}
	\label{subsec:Gamma2_generic}
	
	We now explore several generic features of the dynamical system \eqref{eq:ds_2} that hold independently of the specific form of $f(Q)$, mirroring the analysis performed for the  connection $\Gamma_1$.  
	\begin{enumerate}
		\item \textbf{GR Limit:} 
		
		The GR limit is recovered when $f_{QQ} \to 0$, which implies $m(Q) \to 0$. Naïvely, one might expect the system to become singular due to the appearance of $m$ in the denominator in equations such as \eqref{eq:ds_2}. However, this is not the case. In fact, $m$ always appears in the combination
		\begin{equation}
			\frac{x_2 x_3}{m} = -\frac{\dot{Q}}{6H^3}\,,
		\end{equation}
		which is regular and independent of $f_{QQ}$. Furthermore, since $f_{QQ} \to 0$ also implies $x_3 \to 0$, all dynamical variables behave smoothly in the GR limit. Therefore, the system remains completely regular\footnote{provided that $H \neq 0$. This assumption is naturally satisfied for any expanding or contracting cosmological solution, and excludes only the measure-zero set of static universes where $H = 0$. } in the GR limit.
		
		\item \textbf{The invariant submanifold $\mathcal{S}_{x_3}$:}
		
		It is clear that $x_3' = 0$ when $x_3 = 0$, from the structure of the dynamical system, hence	
		\begin{equation}
			\label{defSx3}
			\mathcal{S}_{x_3} := \left\{ (x_1, x_2, x_3, x_4) \in \mathbb{R}^4\mid x_3 = 0 \right\}\,,
		\end{equation}
		is an invariant submanifold. By definition, $x_3 = - \frac{\dot{Q}f_{QQ}}{Hf_Q} = 0$. Assuming a finite Hubble parameter, this condition can be satisfied either if $\dot{Q} = 0$ or $f_{QQ}=0$. However, the second case corresponds to $f(Q)$ becoming linear in $Q$, which is not generically the case. Instead, we interpret $\mathcal{S}_{x_3}$ as the submanifold where $\dot{Q} = 0$, i.e., $Q$ is dynamically frozen. As we will show below, this gives rise to a cosmological solution that is \emph{kinematically} equivalent to the General Relativistic $\Lambda$CDM solution.
		
		Interestingly, it appears that the system is integrable on $\mathcal{S}_{x_3}$. On $\mathcal{S}_{x_3}$, the dynamical system reduces to:
		\begin{subequations}\label{dinreduced}
			\begin{eqnarray}
				x_1' &=& -2x_1\frac{H'}{H} = 3x_1(1-x_1-x_2)\,,
				\\
				x_2' &=& -2x_2\frac{H'}{H} = 3x_2(1-x_1-x_2)\,,
				\\
				x_4' &=& 1 + x_2 - x_4\left(3 + \frac{H'}{H}\right) = 1 + x_2 - \frac{3(1 + x_1 + x_2)x_4}{2}\,.
			\end{eqnarray}
		\end{subequations}
		The first two equations decouple and are invariant under the exchange $x_1 \leftrightarrow x_2$, displaying a $\mathbb{Z}_2$ symmetry. They admit the solution
		\begin{equation}\label{x1x2}
			x_1 = \frac{c_1}{H^2}, \quad x_2 = \frac{c_2}{H^2}, \quad \frac{x_2}{x_1} = \frac{c_2}{c_1} = -\bar{r},
		\end{equation}
		where $c_1$, $c_2$, and $-\bar{r}$ are integration constants. 
		We note on $\mathcal{S}_{x_3}$, where $Q$ is dynamically frozen, $\bar{r}$ denotes the value of the auxiliary variable $r(Q)$, which reduces to a constant. Therefore, the parameter $\bar{r}$ effectively captures the on-shell (background-level) constancy of $r(Q)$ in this regime. The above result is consistent with the fact the $\dot{Q}=0$ for cosmological solutions belonging to $\mathcal{S}_{x_3}$.\footnote{At this point one may be tempted to use the relation $x_2/x_1 = -\bar{r}$ to reconstruct $f(Q) \propto (-Q)^{\bar{r}}$ on $\mathcal{S}_{x_3}$, but this would be a wrong conclusion. The reason is that $\dot{Q} = 0$ everywhere on $\mathcal{S}_{x_3}$, so $Q$ is constant. As such, $x_2/x_1$ is constant not because of a particular functional form of $f(Q)$ but because $Q$ does not evolve. This represents an extreme subcase of the reconstruction framework, where the functional form of $f(Q)$ becomes locally unidentifiable. This limitation is also consistent with findings from our earlier analysis in Ref.\cite{Chakraborty:2025qlv}, where imposing $\dot{Q} = 0$ led to degeneracy in the functional form of $f(Q)$ and sometimes to divergent $f_Q$. We note that in Section. \ref{subsec:Gamma1_generic} we reconstructed $f(Q)=-2\Lambda +(-Q)^{1+m}$ from the constancy of $m$ and not that of $Q$.}


		
		In fact, using \eqref{x1x2} into \eqref{dinreduced}, one can explicitly solve
		
		\be\label{x1solsx3}
		x_1 = -\frac{x_2}{\bar{r}} = \frac{1}{1-\bar{r}+\left(\frac{1}{x_0}-1+\bar{r}\right)e^{-3N}}\,, \text{where} ~x_0=x(N=0),
		\ee
		which, from (\ref{qG2}) implies the deceleration parameter evolves as
		\begin{equation}\label{qx3G2}
			q(N) = \frac{1}{2}\left[1 - \frac{3(1-\bar{r})x_0}{(1-\bar{r})x_0 + \left(1 - (1-\bar{r})x_0\right)e^{-3N}} \right]\,,\quad \text{or}\,, \quad q(z) = \frac{1}{2}\left[1 - \frac{3(1-\bar{r})x_0}{(1-\bar{r})x_0 + \left(1 - (1-\bar{r})x_0\right)(1+z)^3} \right],
		\end{equation}
		where we have set $z=0$ to correspond to $N=0$. From this, one can solve for the Hubble parameter evolution
		\begin{equation}\label{hx3G2}
			h^2(N) = [1-(1-\bar{r})x_0]e^{-3N} + (1-\bar{r})x_0\,, \quad \text{or}\,, \quad h^2(z) = [1-(1-\bar{r})x_0](1+z)^3 + (1-\bar{r})x_0\,,
		\end{equation}
		consistently recovering $x_1,x_2 \propto 1/H^2$ within (\ref{x1solsx3}) as from the definition of the dynamical variable (\ref{dynvar_gamma2}) for $Q=const.$. From the Friedmann constraint (\ref{friedmann_gamma2}) on $\mathcal{S}_{x_3}$, one can calculate the evolution of the dynamical variable $\Omega(N)$:
		\begin{equation}
			\label{Omegax3G2}
			\Omega(N) = 1 - \frac{(1-\bar{r})x_0}{(1-\bar{r})x_0 + \left(1 - (1-\bar{r})x_0\right)e^{-3N}}\,,\quad \text{or}\,, \quad \Omega(z) = 1 - \frac{(1-\bar{r})x_0}{(1-\bar{r})x_0 + \left(1 - (1-\bar{r})x_0\right)(1+z)^3}\,.
		\end{equation}
		From \eqref{qx3G2} and \eqref{Omegax3G2}, one can identify $(1-\bar{r})x_0 = \frac{1}{3}(1-2q_0) = 1-\Omega_0$ (subscript `$0$' denotes values at $N=0$), using which one can write the expression of the Hubble parameter \eqref{hx3G2} in a very familiar form:
		\begin{equation}\label{hx3G2_1}
			h^2(z) = \frac{2}{3}(1+q_0)(1+z)^3 + \frac{1}{3}(1-2q_0) = \Omega_0(1+z)^3 + (1-\Omega_0)\,.
		\end{equation}
		Amazingly, the evolution of the Hubble parameter for the cosmological solutions residing on $\mathcal{S}_{x_3}$ is the same as that of the General Relativistic $\Lambda$CDM model. 
		
		In hindsight, the above result is not unexpected, since we can also achieve it  from Eq.~\eqref{fried-rev} by setting $\dot{Q} = 0$:
		\be\label{hx3G2_2}
		h^{2} = - \frac{1}{6f_Q H_0^2}\left(f - Qf_Q\right) + \frac{\rho_{m}}{3f_Q H_0^2} 
		= \left(1 -\bar{r}\right)\frac{c_1}{H_0^2} + \Omega_0(1+z)^3\,.
		\ee
		Alternatively, using the expression for $q$ from \eqref{qG2}, one can calculate that the cosmographic jerk parameter on $\mathcal{S}_{x_3}$ is $j=2q^2+q-q'=1$, which is the kinematical condition behind a $\Lambda$CDM-like cosmic evolution of the form \eqref{hx3G2_1}. Although tempting, one should \emph{not} interpret the results in Eq.\eqref{hx3G2_1} or \eqref{hx3G2_2} as that any theory reduces to GR on $\mathcal{S}_{x_3}$. Since $\dot{Q}=0$, the effective gravitational coupling $1/f_Q$ becomes constant on $\mathcal{S}_{x_3}$, although not necessarily equal to the GR value of unity; we will indeed soon see from the inspection of the first integral \eqref{solintegral} that the auxliary variable $m$ is reduced to a constant on the invariant submanifold $\mathcal{S}_{x_3}$, but not necessarily to the GR value $m=0$. That the underlying theory is not GR is also evident from the fact that in Eq.\eqref{hx3G2_1} or Eq.\eqref{hx3G2_2}, the appearing $\Omega_0$  is not precisely the standard density abundance parameter $\frac{\kappa\rho_{m0}}{3H_0^2}$, but rather $\frac{\kappa\rho_{m0}}{3H_0^2 f_Q}$. It is due to the condition $\dot{Q}=0$ that the \emph{kinematics} of the cosmological evolution on $\mathcal{S}_{x_3}$ mimics that of $\Lambda$CDM. Hence, while the constant value of  $Q$ acts as an effective observational uncertainty on the  Hubble parameter $H_0$ at the level of the kinematics of $x_2$, the constant value of the gravitational coupling dresses as the present-day density of matter $\rho_{m0}$. In this sense, the $\Lambda$CDM-like behaviour arises \emph{on-shell} (i.e., at the level of the background cosmological solution), even though the theory remains distinct from GR at the level of action or field equations. Therefore, $\mathcal{S}_{x_3}$ can be interpreted as a submanifold hosting such emergent $\Lambda$CDM-like cosmologies.
		
		
		\subsubsection*{First Integral and Flow Towards \texorpdfstring{$\mathcal{S}_{x_3}$}{Sx3}}
		
		We now look for  a first integral $\mathcal{I}(x_1,x_2,x_4)$ that remains constant along any trajectory confined to the submanifold. It will help us to compute the evolution of $x_4(N)$ on that invariant submanifold $\mathcal{S}_{x_3}$.
		Such a function satisfies the condition:
		
		\begin{equation}
			\label{defintegral}
			\frac{d\mathcal{I}}{dN} = \frac{\partial \mathcal{I}}{\partial x_1} x_1'
			+ \frac{\partial \mathcal{I}}{\partial x_2} x_2'
			+ \frac{\partial \mathcal{I}}{\partial x_4} x_4' = 0,
		\end{equation}
		where $x_1', x_2', x_4'$ are given by the reduced system on $\mathcal{S}_{x_3}$ [cf. Eq.~\eqref{dinreduced}]. Substituting the expressions yields the following PDE:
		\begin{equation}
			\label{defintegral_explicit}
			3x_1(1 - x_1 - x_2) \frac{\partial \mathcal{I}}{\partial x_1}
			+ 3x_2(1 - x_1 - x_2) \frac{\partial \mathcal{I}}{\partial x_2}
			+ \left[1 + x_2 - \frac{3(1 + x_1 + x_2)x_4}{2} \right] \frac{\partial \mathcal{I}}{\partial x_4} = 0\,.
		\end{equation}
		
		Solving this equation, we get a first integral:
		\beq 
		\label{solintegral}
		\mathcal{I}(x_1, x_2, x_4) = \mathcal{F} \left( -r(Q),\, \frac{\sqrt{x_1}}{3(y-1)y^{3/2}} \left[ (y-1)x_1\, \mathrm{arctanh}(\sqrt{y}) + \left[(3x_4 - 1)x_1 + (3x_4 - 2)x_2\right] \sqrt{y} \right] \right), \quad y \equiv x_1 + x_2,
		\eeq
		where $\mathcal{F}$ is an arbitrary function of its two arguments. Note the variables $x_1$, $x_2$, and $x_4$ appear only in the specific combinations shown, ensuring that $\mathcal{I}$ satisfies the partial differential equation \eqref{defintegral}.
		
		Thus, $\mathcal{I}(x_1, x_2, x_4)$ represents a first integral of the reduced dynamical system on the invariant submanifold $\mathcal{S}_{x_3}$—that is, a conserved quantity that remains constant along each orbit within this subspace.\footnote{Here, we are not claiming that a first integral exists for the full four-dimensional dynamical system. We are only presenting and verifying a first integral for the dynamics restricted to the invariant submanifold $\mathcal{S}_{x_3}$.}
		
		The first argument of $\mathcal{F}$, namely $-r(Q)$, itself defines a particular first integral: it recovers the condition $x_2 = -\bar{r} x_1$  obtained earlier. It reveals that the auxiliary variable $m$ (defined in Eq.~\eqref{defofm}) remains constant along orbits on $\mathcal{S}_{x_3}$.\footnote{We anticipate that this behavior will be consistent with the specific example discussed later in Eq.~\eqref{m_dyn_var_c2}, where $m$ becomes functionally dependent on $x_1$ and $x_2$.}
		
		This identification marks the zeroth step in a broader hierarchy of auxiliary variables $m_i$ introduced in Eq.~\eqref{mip}, each of which becomes constant when $x_3 = 0$.\footnote{For the specific power-law model $f(Q) = \alpha Q + \beta (-Q)^n$ considered later, the derivatives $m_i'$ can be computed explicitly using Eqs.~\eqref{m_dyn_var_c2} and \eqref{dinreduced}.}
		
		Furthermore, by treating the second argument of $\mathcal{F}$ in \eqref{solintegral} as an independent first integral, one can explicitly solve for the time evolution of $x_4(N)$ as
		\begin{equation}\label{x4_Sx3_G2}
			x_4(N) = \frac{(1-\tilde y) [D+{\rm arctanh}(\sqrt{\tilde y})]+(1-2\bar{r})\sqrt{\tilde y}}{3  \sqrt{\tilde y}}\,, \qquad \tilde{y} = (1 - \bar{r})x_1,
		\end{equation}
		where $D$ is an orbit-dependent integration constant, and it is understood that $x_1=x_1(N)$. To ensure the physical real-valuedness of $x_4(N)$, it is necessary to impose $x_1 \geq 0$, which fixes $c_1 \geq 0$ from (\ref{x1x2}), and $\tilde y = (1 -\bar{r})x_1 \geq 0$,  which implies $\bar{r} \geq -1$. This condition provides an additional consistency condition on the parameter space.
		
		Finally, by recalling the definition of the dynamical variable $x_4$ given in (\ref{dynvar_gamma2}), by using $N=-\ln(1+z)$, and using the inferred $\Lambda$CDM-mimiking Hubble history (\ref{hx3G2}), we obtain the redshift evolution of the dynamical connection $\gamma=\gamma(z)$. Reconstruction of the cosmological evolution of the dynamical connection is notoriously difficult, with literature either resorting to numerical techniques \cite{Yang:2024tkw} or by assuming it from the outset \cite{De:2023xua}; here, we have analytically obtained that its evolution on the the invariant submanifold ${\mathcal S}_{x_3}$ should be a specific subcase of the one we reconstructed in \cite{Chakraborty:2025qlv} because the same $\Lambda$CDM-mimiking evolution of the scale factor applies in the compatibility equation,  but for a constant non-metricity scalar: our present result  opens the path towards a comparison against astrophysical datasets for understanding which $\Lambda$CDM-mimiking orbit is the most realistic, whether one lying in ${\mathcal S}_{x_3}$ or one outside it (indeed some $f(Q)$ theories admit matter dominated and de Sitter epochs also outside such an invariant submanifold, and the heteroclinic trajectory connecting them may be $\Lambda$CDM-mimicking; see Table \ref{tab:C2_3} below for an explicit realisation).
		
		In summary, this study of the first integral serves as a non-trivial consistency check of our earlier results for the dynamics on the invariant submanifold ${\mathcal S}_{x_3}$. More importantly, it allows us to extract an exact solution for the time evolution of $x_4$, which would be difficult to obtain directly from its evolution equation, providing a valuable starting point for the astrophysical assessment of such an invariant submanifold via the inferred redshift behaviour of the dynamical connection. 
		It shows the utility of identifying first integrals in simplifying the dynamics. Moreover, the analysis deepens our understanding of the auxiliary variable $m$. On ${\mathcal S}_{x_3}$, $m$ behaves as a conserved quantity, remaining constant along a trajectory belonging to ${\mathcal S}_{x_3}$, as expected from the constancy of the non-metricity scalar itself. While the former argument in (\ref{solintegral}) is nothing but the auxiliary variable, the latter instead plays the role of the compatibility condition of the chosen branch from which we could obtain the dynamical connection. 
		
		Next, we analyze the flow \emph{towards} the invariant submanifold $\mathcal{S}_{x_3}$. We can write the evolution equation for $x_3$  as
		\begin{equation}
			\label{S_x3_stab}
			x_3' = -x_3(2 - q - x_3)\,.
		\end{equation}
		
		Linearizing Eq.~\eqref{S_x3_stab} around the equilibrium point $x_3 = 0$,
		we get
		\begin{equation}
			\delta x_3' = -(2 - q)\delta x_3\,.
		\end{equation}
		Since $q \leq 1$ for a viable cosmological model, this equilibrium point is stable against small perturbations for all values of $x_1$ and $x_2$ in a viable cosmic evolution. This implies that ${\mathcal S}_{x_3}$ is an attractor.
		
		We emphasise that all results in this section are derived without assuming any specific form of $f(Q)$. The invariant submanifold $\mathcal{S}_{x_3}$ plays a pivotal role in the global dynamics: both the matter-dominated and de Sitter fixed points (see Table~\ref{tab:C2_1}) lie on $\mathcal{S}_{x_3}$, so the heteroclinic orbit connecting them is fully contained within this submanifold. It is worth noting that this orbit also exhibits $\Lambda$CDM-like kinematics. One can conclude that the $\Lambda$CDM-like cosmic evolution emerges generically at late times for any $f(Q)$ within the framework of the nontrivial gauge $\Gamma_2$, even if the theory itself does not generically tend to GR\footnote{However, there is no apparent proof that this would be the only possible $\Lambda$CDM-mimicking trajectory, or that all the $\Lambda$CDM-mimicking trajectories must lie on $\mathcal{S}_{x_3}$.}. This conclusion would be validated in the particular example considered below (see Fig.\ref{fig:c2_1} in particular). 
		
		\item {\bf Existence of a vacuum invariant submanifold $\mathcal{S}_\Omega$:}
		
		The vacuum condition $\Omega = 0$ defines an invariant submanifold of the phase space:
		\begin{equation}
			\mathcal{S}_\Omega := \left\{ (x_1, x_2, x_3, x_4) \in \mathbb{R}^4 \mid x_1 + x_2 + x_3 x_4 = 1 \right\}.
		\end{equation}
		In order to assess its stability, we  now examine the evolution equation for $\Omega$:
		\begin{equation} \label{S_Om_stab}
			\Omega' = -3\Omega\left(1 + \frac{1}{3}x_3 - 2x_3 x_4 - \Omega\right).
		\end{equation}
		It shows that $\mathcal{S}_\Omega$ is \emph{stable or unstable depending on whether} 
		\begin{align}
					1 + \frac{1}{3}x_3 - 2x_3 x_4 > 0 \quad \text{or} \quad < 0,
		\end{align}
		respectively. This condition depends dynamically on both $x_3$ and $x_4$, reflecting the geometric influence of the  connection.
		
		\item {\bf Invariant submanifold $\mathcal{S}_{x_2}$:}
		
		A \emph{partially general} invariant submanifold, which exists for \emph{many, but not all} the class of $f(Q)$ models, is given by
		\begin{equation}
			\mathcal{S}_{x_2} := \left\{ (x_1, x_2, x_3, x_4) \in \mathbb{R}^4 \mid  x_2 = 0 \right\}\,.
		\end{equation}
		This invariant submanifold exists for the class of models for which $\lim_{x_2\to0}\frac{x_2}{m(x_1,x_2)}=0$. As we will see in the subsection \ref{sectG2example}, this does exist for the particular example $f(Q)$ we consider. This invariant submanifold contains all those solutions for which, locally, $Q=0$ (on the assumption of a finite Hubble parameter). The existence of this invariant submanifold implies that the sign of $Q$ is preserved during the cosmic evolution, since $x_2\gtreqless0$ implies $Q\gtreqless0$. Preserving a positive sign of $Q$ is specifically important for some $f(Q)$ theories based on the logarithm or general powers of such a non-metricity scalar for preventing them from losing their meanings \cite{Goncalves:2024sem}. 
		
		\item \textbf{Generic fixed points:}
		
		Two generic 1-parameter families of non-isolated fixed points and one isolated fixed point can be obtained from the dynamical system \eqref{eq:ds_2}, under mild conditions on the form of $f(Q)$. They are summarised in Table \ref{tab:C2_1}.
		\begin{table}[H]
			\centering
				\renewcommand\arraystretch{2}
			\begin{tabular}{|c|c|c|c|c|}
				\hline 
				Fixed point& $(x_1,x_2,x_3,x_4)$ & Existence condition & $q$& Cosmological solution \\ 
				\hline 
				$\mathcal{Q}_1 $ & $\left(2-3x_4,3x_4-1,0,x_4\right)$ & Always & $-1$ & $a(t)=a_0 e^{H_0t}$(de-Sitter)  \\ 
				\hline
				$\mathcal{Q}_2 $ & $\left(0,0,0,\frac{2}{3} \right)$ & Always & $\frac{1}{2}$ &  $a=a_0 t^{\frac{2}{3}}$  \\ 
				\hline
				$\mathcal{Q}_3 $ & $\left(0,0,\frac{1}{x_4},x_4\right)$ & $\lim_{\{x_1,x_2\}\to\{0,0\}}m(x_1,x_2)\neq0$ & $2-\frac{1}{x_4}$ &  $a(t)=a_0 t^\frac{x_4}{3x_4-1}$  \\ 
				\hline
			\end{tabular} 
			\caption{Fixed points of system \eqref{eq:ds_2}. } \label{tab:C2_1}
		\end{table}
		\begin{itemize}
			\item \textbf{Family of vacuum de Sitter fixed points ($\mathcal{Q}_1$):} 
			This generic family of fixed points lies at the intersection of the invariant submanifolds $\mathcal{S}_{x_3}$ and $\mathcal{S}_{\Omega}$. Since $q = -1$ and $x_3 = 0$, it follows from the evolution equations \eqref{S_x3_stab} and \eqref{S_Om_stab} that these points are stable. Hence, $\mathcal{Q}_1$ constitutes a robust candidate for a generic attractor of the system.
			
			\item \textbf{GR matter-dominated fixed point ($\mathcal{Q}_2$):} 
			This generic fixed point lies on $\mathcal{S}_{x_3}$ and represents a decelerated dust-dominated solution with $q = 1/2$ and $\Omega = 1$, mimicking standard GR  with minimal geometric influence.
			
			\item \textbf{Family of vacuum fixed points ($\mathcal{Q}_3$):}
			The family $\mathcal{Q}3$ corresponds to the hyperbola $x_3 x_4 = 1$, which lies entirely within the vacuum submanifold $\mathcal{S}_\Omega$. This family of fixed points exists under the mild condition $\displaystyle \lim_{(x_1,x_2)\to (0,0)} m(x_1,x_2) \neq 0$. The existence condition ensures that the dynamical system remains regular at the fixed points and avoids division by zero in expressions like $\frac{x_2}{m(x_1,x_2)}$. Mathematically, these fixed points yield accelerated expansion for $x_4 < 1/2$ and radiation-like behaviour for $x_4 = 1$.
		\end{itemize}
		One may be inquisitive about why the fixed points $\mathcal{Q}_1$ and $\mathcal{Q}_2$ do not require the existence condition $\displaystyle \lim_{(x_1,x_2)\to (0,0)} m(x_1,x_2) \neq 0$ unlike $\mathcal{Q}_3$. It is a very valid and intricate question worth addressing. The quantity $m(x_1,x_2)$ always appear in the dynamical system in the combination $\frac{x_2 x_3}{m}=-\frac{\dot{Q}}{6H^3}$. For the fixed points that lie on $\mathcal{S}_{x_3}$, such as $\mathcal{Q}_1$ and $\mathcal{Q}_2$, $\dot{Q}=0$ identically irrespective of $m(x_1,x_2)$, i.e. irrespective of the given theory $f(Q)$. Therefore, these fixed points \emph{always} exist generically. The same cannot be said about $\mathcal{Q}_3$, which does \emph{not} lie on $\mathcal{S}_{x_3}$.
		
		Considering the family of fixed points $\mathcal{Q}_3$, tight constraints on the variation of the effective gravitational coupling typically require a small value of $x_3$, which in turn demands a large value of $x_4$ due to the constraint $x_3 x_4 = 1$. In such cases, the resulting expansion is characterised by $a(t) \propto t^{1/3}$, corresponding to an effective equation of state $\omega_{\rm eff} \approx 1$. This is the well-known Zel'dovich model \cite{Zeldovich:1972zz}, associated with stiff matter, often interpreted as arising from massless scalar fields.
		
		More importantly, since $\Omega = 0$ at $\mathcal{Q}_3$, this stiff-fluid-like behaviour is not supported by real matter, but instead emerges as a purely geometric effect inherent to the modified gravity sector. As a result, while this family is mathematically allowed, it has very less  physical importance, specifically  in the context of late-time cosmology.
		Moreover, the above analysis shows that any $f(Q)$ model admitting a finite value of $\frac{x_2}{m}$ at a fixed point may support viable cosmological evolution within this framework. In what follows, we examine this possibility by considering the dynamics of a power-law $f(Q)$ model in the context of the $\Gamma_2$ connection.
	\end{enumerate}
	
	We emphasise that the list of fixed points and invariant submanifolds presented in this section is not necessarily exhaustive. Depending on the particular form of $f(Q)$, and therefore the particular form of $m(x_1,x_2)$, more fixed points or broader families of fixed points, as well as new invariant submanifolds, may appear. 
	
	\subsection{Example: $f(Q) = \alpha Q + \beta (-Q)^n$, $\alpha>0,\,\beta\neq0,\,n\neq0,1$}
	\label{sectG2example}
	
	Again, we now consider the power-law model:
	\begin{equation}\label{e.g_fQ_c2}
		f(Q)=\alpha Q+\beta (-Q)^n\,,
	\end{equation}
	with $\alpha>0,\,\beta\neq0,\,n\neq0,1$. Although exact solutions to the field equations in $f(Q)$ gravity are rare, notable progress has been made by either reducing their order \cite[Eqs.~(26),(29)]{Yang:2024tkw} or decoupling them \cite[Sect.~VIII]{Chakraborty:2025qlv}. In this context, the model above offers some analytical tractability. For instance, for the selected theory, the second-order equation \eqref{Q-dot_rev} becomes
	\beq
	Q\ddot Q +\left[(n-2)\dot Q + 3\frac{\dot a}{a}Q  \right]\dot Q=0 \,,
	\eeq
	which can be reduced in order and integrated to yield
	\beq 
	Q(t)=\left[ (n-1) \left( C_1 \int \frac{dt}{a^3(t)}+C_2 \right)\right]^{1/(n-1)} \,,
	\eeq
	where $C_{1,2}$ are integration constants. This form suggests that, following \cite[Eq.~(14)]{Yang:2024tkw}, one can decouple the evolution of $a(t)$ into a single differential equation. Hence, we should resort to our dynamical system formulation to extract further information on this $f(Q)$ theory. 
	
	First of all,  for this model, we obtain the relation between $m$ and dynamical variables $x_1, x_2$ as 
	\begin{equation}\label{m_dyn_var_c2}
		m=n \left(1+\frac{x_1}{x_2}\right)\,.
	\end{equation}
	So the dynamical system \eqref{eq:ds_2} becomes
	\begin{subequations}\label{eq:ds_2_example}
		\begin{eqnarray}
			x_1' &=&  {\frac {{ x_3}\,{x_2}^2}{n\left( x_2+x_1\right)} } + x_1 [-x_3 + 3(1 - x_1 - x_2 + x_3 x_4)]\,,
			\label{eq:c2_ds1_ex1}
			\\
			x_2' &=& -{\frac {{ x_3}\,{x_2}^2}{n\left( x_2+x_1\right)} } + x_2 [-2x_3 + 3(1 - x_1 - x_2 + x_3 x_4)]\,,
			\\
			x_3' &=&-\frac{3x_3}{2}\, \left(1+x_1+x_2 -{x_3}\,{x_4} \right)\,,
			\label{eq:c2_ds1_ex3}      \\
			x_4' &=&1+x_2-\frac{3x_4}{2}\, \left( 1+ x_1+x_2 \right)-x_3 x_4+\frac{3}{2} x_3 x_4^2\,. \label{eq:c2_ds5_ex1}     
		\end{eqnarray}
	\end{subequations}
	
	\subsubsection*{Physically viable region of the phase space}
	
	From the definition of auxiliary quantity $m$ and \eqref{m_dyn_var_c2}, we obtain $Q$ in terms of $\Omega$ as
	\begin{equation}\label{eq:Q_Gamma2_example}
		(-Q)^{n-1}=\frac{\alpha (x_1+x_2)}{\beta ( nx_1+x_2)}\,.
	\end{equation}
	As a result, the physical condition 
	\begin{equation}
		f_Q>0 \,,
	\end{equation}
	can be expressed in terms of dynamical variables $x_1, x_2$ as 
	\begin{equation}\label{phys2_c2}
		\frac{ x_2 (1-n)}{nx_1+x_2}>0 \,.
	\end{equation}
	From \eqref{phys1_c2} and  \eqref{phys2_c2}, the physical phase space of fixed points of the dynamical system \eqref{eq:c2_ds1_ex1}-\eqref{eq:c2_ds5_ex1} is given by
	\begin{equation}\label{psp_c2}
		\left\lbrace (x_1,x_2,x_3,x_4) \in \mathbb{R}^4 \mid x_1+x_2+x_3 x_4 \leq 1\,, \frac{ x_2 (1-n)}{nx_1+x_2}>0  \right\rbrace\,.
	\end{equation}
	
	\subsubsection*{Regularizating the dynamical system}
	
	It is worth noting that the dynamical system \eqref{eq:ds_2_example} is not regular on the 3-dimensional submanifold $x_1 + x_2 = 0$. 
	This configuration is not a fundamental pathology of the theory, but an artefact of our particular choice of the dynamical variables. We can remove the singularity by redefining the time parameter $N$, and to handle this, we consider two disjoint regions of the 4-dimensional phase space defined by $ x_1 + x_2 > 0$ and $x_1 + x_2 < 0$  respectively. Within each region, we can perform a redefinition of the time variable $ N \to \tilde{N}$ as:
	
	\begin{equation}
		\frac{dN}{d\tilde{N}} = |x_1 + x_2|\,.
	\end{equation}
	
	We note that the dynamical system becomes regular in terms of $\tilde{N}$. Further,  the surface $x_1 + x_2 = 0$ divides the phase space into two disjoint sectors, across which the sign of $Q$ cannot change. Hence, each sector defines a physically consistent branch of the theory. Concerning this redefined time variable, the dynamical system \eqref{eq:ds_2_example} transforms into:
	\begin{subequations}\label{eq:ds_2_example_redefined}
		\begin{eqnarray}
			\epsilon x_1' &=&  \frac {{ x_3}\,{x_2}^2}{n} + x_1(x_1 + x_2)[-x_3 + 3(1 - x_1 - x_2 + x_3 x_4)]\,,
			\\
			\epsilon x_2' &=& -\frac {{ x_3}\,{x_2}^2}{n} + x_2(x_1 + x_2)[-2x_3 + 3(1 - x_1 - x_2 + x_3 x_4)]\,,
			\\
			\epsilon x_3' &=&-\frac{3x_3}{2}\left( x_2+x_1\right)\left(1+x_1+x_2 -{x_3}\,{x_4} \right)\,,
			\\
			\epsilon x_4' &=& \left( x_2+x_1\right)\left[1+x_2-\frac{3x_4}{2}\, \left( 1+ x_1+x_2 \right)-x_3 x_4+\frac{3}{2} x_3 x_4^2\right]\,,    
		\end{eqnarray}
	\end{subequations}
	where $\epsilon=+1$ in the region with $(x_1+x_2)>0$ and $\epsilon=-1$ in the region with $(x_1+x_2)<0$. 
	
	The phase space given by the dynamical system \eqref{eq:ds_2_example_redefined} is topologically equivalent to that given by the original system \eqref{eq:ds_2_example}, but it has the added advantage of being regular everywhere in the phase space. So, we will henceforth work with the regularised dynamical system \eqref{eq:ds_2_example_redefined}.

	\subsubsection*{Invariant submanifolds}
	
	\begin{itemize}
		
		\item The two generic invariant submanifolds $\mathcal{S}_{x_3}$ and $\mathcal{S}_{\Omega}$ are present in the dynamical system \eqref{eq:ds_2_example_redefined}. Their stability conditions follow from Eqs.~\eqref{S_x3_stab} and \eqref{S_Om_stab}, respectively.
		
		\item One can also confirm the existence of the invariant submanifold $\mathcal{S}_{x_2}$. It exists in this model because $\lim_{x_2 \to 0} \frac{x_2}{m(x_1,x_2)} \to 0$ for the particular example under consideration.   
		
		To investigate the stability of this invariant submanifold, we write the $x_2'$-equation in \eqref{eq:ds_2_example_redefined} as
		\begin{equation}
			\epsilon x_2' = g(x_1,x_2,x_3,x_4) \equiv - x_2 \left[\frac{x_3 x_2}{n} - (x_1 + x_2)(3 -2x_3 - 3x_1 - 3x_2 + 3x_3 x_4)\right]\,.
		\end{equation}
		The stability is governed by
		\begin{equation}
			\left(\frac{\partial g}{\partial x_2}\right)\bigg\vert_{x_2=0} = x_1(3 - 2x_3 - 3x_1 + 3x_3 x_4).
		\end{equation}
		Hence, in the region $(x_1+x_2)>0$, the submanifold $\mathcal{S}_{x_2}$ is stable or unstable depending on whether
		\begin{equation}
			x_1(3 - 2x_3 - 3x_1 + 3x_3 x_4) < 0 \quad \text{or} \quad x_1(3 - 2x_3 - 3x_1 + 3x_3 x_4) > 0\,.
		\end{equation}
		The opposite conditions hold for the region $(x_1+x_2)<0$.
		
		\item On top of $\mathcal{S}_{x_3},\mathcal{S}_{\Omega},\mathcal{S}_{x_2}$, the regularisation of the dynamical system reveals the existence of another invariant submanifold in the cosmological phase space of the theory under consideration
		\begin{equation}
			\mathcal{S}_{x_1 + x_2} := \left\{ (x_1, x_2, x_3, x_4)\in \mathbb{R}^4 \mid x_1 + x_2 = 0 \right\}\,,
		\end{equation}
		separating the phase space into regions with opposite signs of $(x_1+x_2)$. 
		The invariant submanifolds $ \mathcal{S}_{x_2}$, $ \mathcal{S}_{x_3}$, and $\mathcal{S}_{x_1 + x_2} $ are all three-dimensional, and they intersect along the line defined by $ x_1 = x_2 = x_3 = 0 $. From the definition of the dynamical variable, one can obtain
		\begin{equation}
			x_1 + x_2 = \frac{(n-1)\beta(-Q)^n}{6H^2}\,.
		\end{equation}
		From this expression, it becomes clear that in the limit $\beta \to 0$, or more generally whenever $|\frac{\beta(-Q)^n}{H^2}| \ll 1$, we have $(x_1 + x_2)\to 0$. Hence, the submanifold $\mathcal{S}_{x_1 + x_2}$ can be interpreted as capturing the regime of the theory where the contribution from the non-linear term in the function $f(Q)$ becomes negligible, i.e., where the correction $ \beta (-Q)^n$ is dynamically subdominant. This interpretation aligns with the understanding of the quantity $(x_1+x_2)$ that we presented in section \ref{sec:DSA_intro}, namely, in the absence of a cosmological constant, $(x_1+x_2)$ characterises deviation from GR.
		
		To investigate the stability of $\mathcal{S}_{x_1+x_2}$,  we transform coordinates via in the $x_1-x_2$ plane: 
		\begin{equation}
			(x_1,x_2)\rightarrow(y,z):
			\begin{cases}
				y = x_1 + x_2
				\\
				z = x_1 - x_2
			\end{cases}\,,
		\end{equation}
		In terms of the new variables $(y, z, x_3, x_4)$,  the dynamical system \eqref{eq:ds_2_example_redefined} becomes:
		\begin{subequations}\label{eq:ds_2_example_yz}
			\begin{eqnarray}
				\epsilon y' &=&  h(y,z,x_3,x_4) \equiv 3y^2(1 - y + x_3 x_4) - \frac{3}{2}x_3 y^2 + \frac{1}{2}x_3 y z \,,
				\\
				\epsilon z' &=& \frac{1}{2n}x_3(y-z)^2 + 3yz(1 - y + x_3 x_4) - x_3 y z + \frac{1}{2}x_3 y (y-z)\,,
				\\
				\epsilon x_3' &=& -\frac{3x_3}{2}y\left(1 + y - x_3 x_4 \right)\,,
				\\
				\epsilon x_4' &=& y \left[1 + \frac{1}{2}(y-z) - \frac{3}{2}x_4(1+y) - x_3x_4 + \frac{3}{2}x_3 x_4^2\right]\,.   
			\end{eqnarray}
		\end{subequations}
		The stability of the submanifold $\mathcal{S}_{x_1+x_2}$ is determined by the behaviour of the system under small perturbations in $y$ near $y = 0$. The other equations describe evolution within the submanifold and do not influence whether the system escapes from or is attracted to it. Therefore, we focus on the partial derivative of the $y'$ equation and the stability of $\mathcal{S}_{x_1+x_2}$ is given by the sign of the quantity :
		\begin{equation}
			\left(\frac{\partial h}{\partial y}\right)\bigg\vert_{y=0} = \frac{1}{2}x_3 z = \frac{1}{2}(x_1 - x_2)x_3.
	\end{equation}
		In the region $(x_1+x_2)>0$, the invariant submanifold $\mathcal{S}_{x_1+x_2}$  is stable or unstable depending on whether
		\begin{equation}
			(x_1 - x_2)x_3 < 0 \quad \text{or} \quad (x_1 - x_2)x_3 > 0\,.
		\end{equation}
		Opposite conditions hold for the region $(x_1+x_2)<0$.

	\end{itemize}
	We summarise the stability analysis results of the invariant submanifolds in Table \ref{tab:inv_sub_Gamma2_ex}.
	\begin{table}[H]
		\centering
			\renewcommand\arraystretch{1.5}
		\begin{tabular}{|c|c|c|}
			\hline 
			Invariant submanifolds & Definition & Stability \\ 
			\hline 
			$\mathcal{S}_{\Omega}$ & $\Omega=0\Leftrightarrow x_1 + x_2 + x_3 x_4 = 1$ & Stable for $1+\frac{1}{3}x_3-2x_3x_4>0$ \\ 
			&  & Unstable for $1+\frac{1}{3}x_3-2x_3x_4<0$ \\ 
			\hline
			$\mathcal{S}_{x_2}$ & $x_2=0$ & Stable for $x_1(3 - 2x_3 - 3x_1 + 3x_3 x_4) < 0$ \\ 
			&  & Unstable for $x_1(3 - 2x_3 - 3x_1 + 3x_3 x_4) > 0$ \\ 
			\hline
			$\mathcal{S}_{x_3}$ & $x_3=0$ & Always stable \\ 
			\hline
			$\mathcal{S}_{x_1+ x_2}$ & $x_1+x_2=0$ & Stable for $(x_1 - x_2)x_3 < 0$ \\ 
			&  & Unstable for $(x_1 - x_2)x_3 > 0$ \\ 
			\hline
		\end{tabular} 
		\caption{Invariant submanifolds in the phase space of $f(Q)=\alpha Q+\beta(-Q)^n$ theory for the connection branch $\Gamma_2$. The stability conditions of the invariant submanifolds are for the region $(x_1+x_2)>0$. The stability conditions will be reversed for the region $(x_1+x_2)<0$.} 
		\label{tab:inv_sub_Gamma2_ex}
	\end{table}

	\begin{figure}[H]
		\centering
		\subfigure[]{%
			\includegraphics[width=8cm,height=6cm]{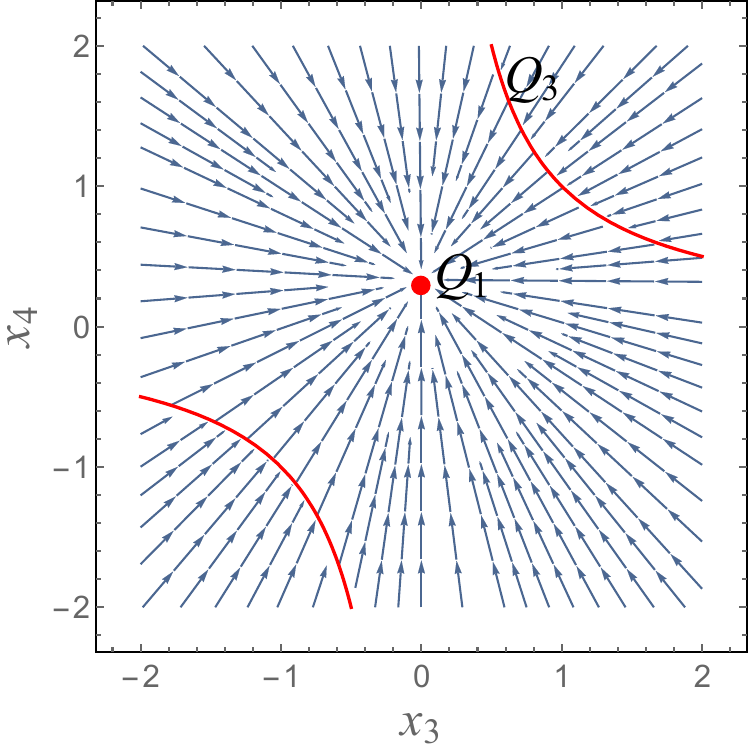}\label{fig:intersection_S_Om_Sx2}}
		\quad
		\subfigure[]{%
			\includegraphics[width=8cm,height=6cm]{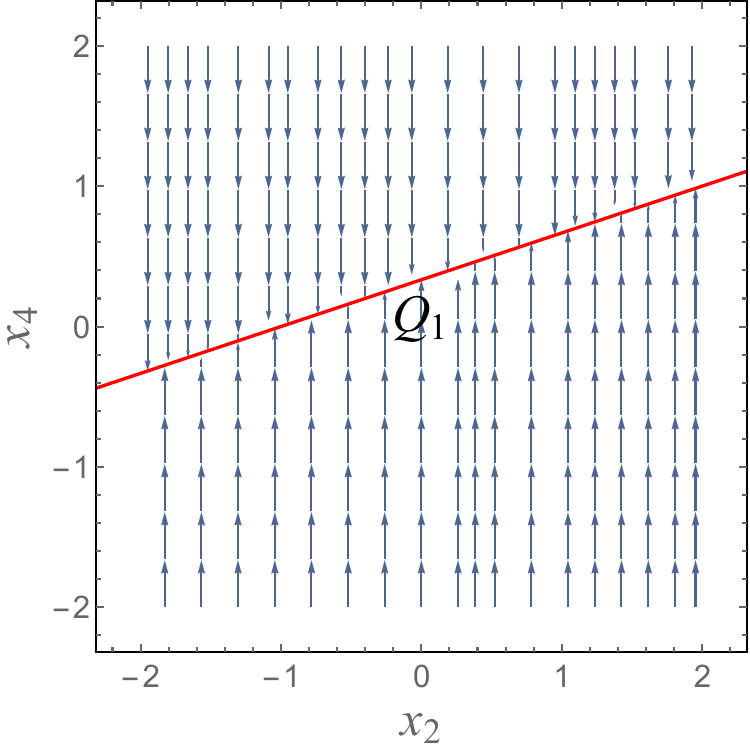}\label{fig:intersection_S_Om_Sx3}}
		\quad
		\subfigure[]{%
			\includegraphics[width=8cm,height=6cm]{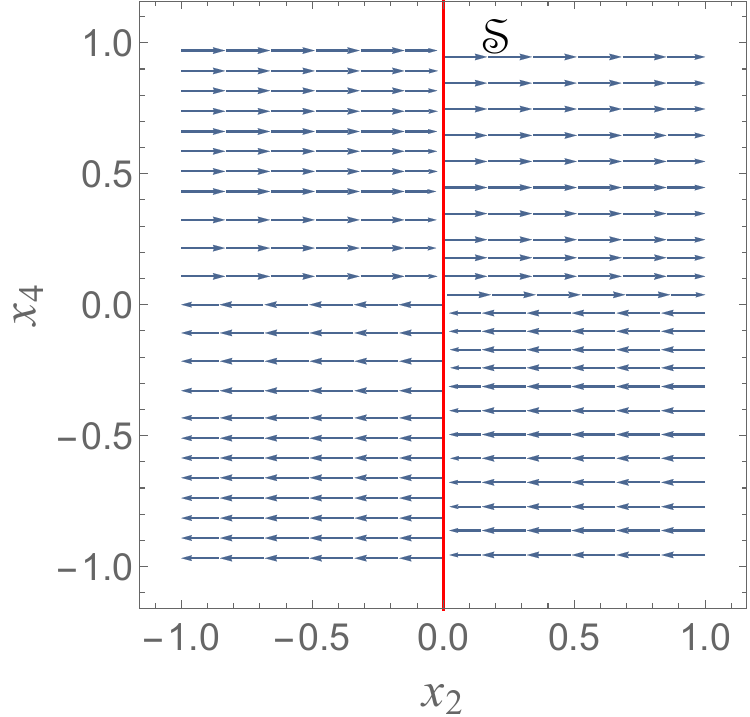}\label{fig:intersection_S_Om_Sx1x2}}
		\caption{Projection of phase trajectories \eqref{eq:ds_2_example_redefined} on the intersection of pairs of invariant submanifolds, shown in the following panels: \\  \textbf{(a)}, We intersect  $\mathcal{S}_{\Omega}$ with $\mathcal{S}_{x_2}$ obtaining a point $\left(0,\frac{1}{3}\right)$ in red, which belongs to the $\mathcal{Q}_1$ in table \ref{tab:C2_1} as an attractor node, while a curve $\mathcal{Q}_3$ (in red) is a saddle;\\     \textbf{(b)} We intersect $\mathcal{S}_{\Omega}$ with $\mathcal{S}_{x_3}$ and display also the  line $x_2=3 x_4-1$ (in red) further confirming the attractive nature, as a node, of $\mathcal{Q}_1$; this numerical analysis in both panels corroborates the analytical discussion on the stability of the set of points $\mathcal{Q}_1$;\\    \textbf{(c)} Finally,  we intersect $\mathcal{S}_{\Omega}$ with $\mathcal{S}_{x_1+x_2}$ and choose $n=-2,$ and confirm that $\mathcal{S}$  (in red) is a saddle. The qualitative behaviors of the trajectories  depicted here are  independent of the parameter $n$.}
		\label{fig:c2_inv_subm_int}
	\end{figure}

	\subsubsection*{Fixed points}	
	
	Let us now turn to the structure of the fixed points of the system \eqref{eq:ds_2_example_redefined}. The existence of the de Sitter fixed point family $\mathcal{Q}_1 $ follows directly from the original dynamical system \eqref{eq:ds_2}. However, the status of $\mathcal{Q}_2$   and $\mathcal{Q}_3 $  is less clear, since the limit $\displaystyle\lim_{\{x_1, x_2\} \to \{0, 0\}} \frac{x_2 x_3}{m(x_1, x_2)} $ becomes indeterminate. Notably,  $\mathcal{Q}_1$  lies entirely in the region $x_1 + x_2 > 0$, which is relevant for late-time cosmology. Hence, we now focus on this particular disjoint sector, where the redefined system fully captures the flow \eqref{eq:ds_2_example_redefined} with $ \epsilon = +1 $. In this region, besides $\mathcal{Q}_1$, we identify another two broader families of fixed points, both of which are a 2-parameter family:
	\begin{itemize}
		
		\item $\mathcal{S}:(x_1,x_2,x_3,x_4)=(0,0,x_3,x_4)$, which contains both $ \mathcal{Q}_2 $ and $\mathcal{Q}_3 $ as special cases. Notice that this is the $x_3-x_4$ plane. The entire 2-dimensional $x_3-x_4$ plane is a set of fixed points. This set of non-isolated fixed points lies entirely at the intersection of two 3-dimensional invariant submanifolds, namely $\mathcal{S}_{x_2}$ and $\mathcal{S}_{x_1+x_2}$.

		\item $\mathcal{T}:(x_1, x_2, x_3, x_4) = (x_1, -x_1, 0, x_4)$, which is another 2-dimensional submanifold that lies entirely at the intersection of two 3-dimensional invariant submanifolds $\mathcal{S}_{x_3}$ and $\mathcal{S}_{x_1+x_2}$. 
		\begin{table}[H]
			\centering
				\renewcommand\arraystretch{1.6}
			\begin{tabular}{|c|c|c|c|c|c|}
				\hline 
				Family of fixed points& $(x_1,x_2,x_3,x_4)$ & Location & $q$ & $\Omega $ &Stability\\ 
				\hline 
				$\mathcal{S} $ & $\left(0,0,x_3,x_4\right)$ & $\mathcal{S}_{x_2} \cap \mathcal{S}_{x_1+x_2}$ & $\frac{1}{2}+\frac{3}{2}x_3x_4-x_3$ & $1-x_3 x_4$ & Saddle (numerically) \\ 
				\hline
				$\mathcal{T}$ & $\left(x_1,-x_1,0,x_4 \right)$ & $\mathcal{S}_{x_3}\cap \mathcal{S}_{x_1+x_2}$& $\frac{1}{2}$ &1 & Undetermined \\ 
				\hline
				$\mathcal{Q}_1 $ & $\left(2-3x_4,3x_4-1,0,x_4\right)$ & $\mathcal{S}_{x_3} \cap \mathcal{S}_\Omega$& $-1$ &  0  & Attractor\\ 
				\hline
			\end{tabular} 
			\vspace{0.5cm}
			\begin{tabular}{|c|c|c|c|c|c|}
				\hline 
				Fixed point& $(x_1,x_2,x_3,x_4)$ & Belongs to & $q$& $\Omega$ & Stability \\ 
				\hline 
				$\mathcal{Q}_2 $ & $\left(0,0,0,\frac{2}{3} \right)$ & $\mathcal{S}\cap \mathcal{T}$ & $\frac{1}{2}$ &  1 & Saddle (numerically)\\ 
				\hline
				$\mathcal{Q}_3 $ & $\left(0,0,\frac{1}{x_4},x_4\right)$ & $\mathcal{S}$ & $2-\frac{1}{x_4}$ &  0 & Saddle (numerically) \\ 
				\hline
			\end{tabular} 
			\caption{{\it Upper table:} Broad families of fixed points of system \eqref{eq:ds_2_example_redefined}. {\it Lower table:} Special set of fixed points of system \eqref{eq:ds_2_example_redefined} lying on the broad families of fixed points presented in the upper table } \label{tab:C2_3}
		\end{table}

	\end{itemize}
	The nature of the broader 2-parameter family of fixed points and specific fixed points contained in such a family of fixed points is summarised in Table \ref{tab:C2_3}. Since all eigenvalues at the fixed point families $\mathcal{S}$ and $\mathcal{T}$ vanish, linear stability theory fails. Further, since there are no negative eigenvalues, there is no reduction in the dynamics near the centre manifold in comparison to the original system. Hence, centre manifold reduction analysis is also inconclusive. Therefore, a more sophisticated mathematical technique is required to analyse the stability behaviour, which is beyond the scope of the present work. 
	
	Nonetheless, in Fig. \ref{fig:c2_inv_subm_int}, we present the projection of the phase trajectories of the system \eqref{eq:ds_2_example_redefined} on the plane of intersection of pairs of invariant submanifolds such as $\mathcal{S}_{\Omega}$ with $\mathcal{S}_{x_2}$, $\mathcal{S}_{x_3}$ and $\mathcal{S}_{x_1+x_2}$, separately, which numerically confirms the attractive nature nature of a family of de Sitter points $\mathcal{Q}_1$. Specifically, our numerical analysis shows that this cosmic epoch is a node regardless of the assumed theory with the dynamics of vacuum space being richer than in GR.

\begin{figure}[H]
		\centering
		\includegraphics[width=7cm,height=6cm]{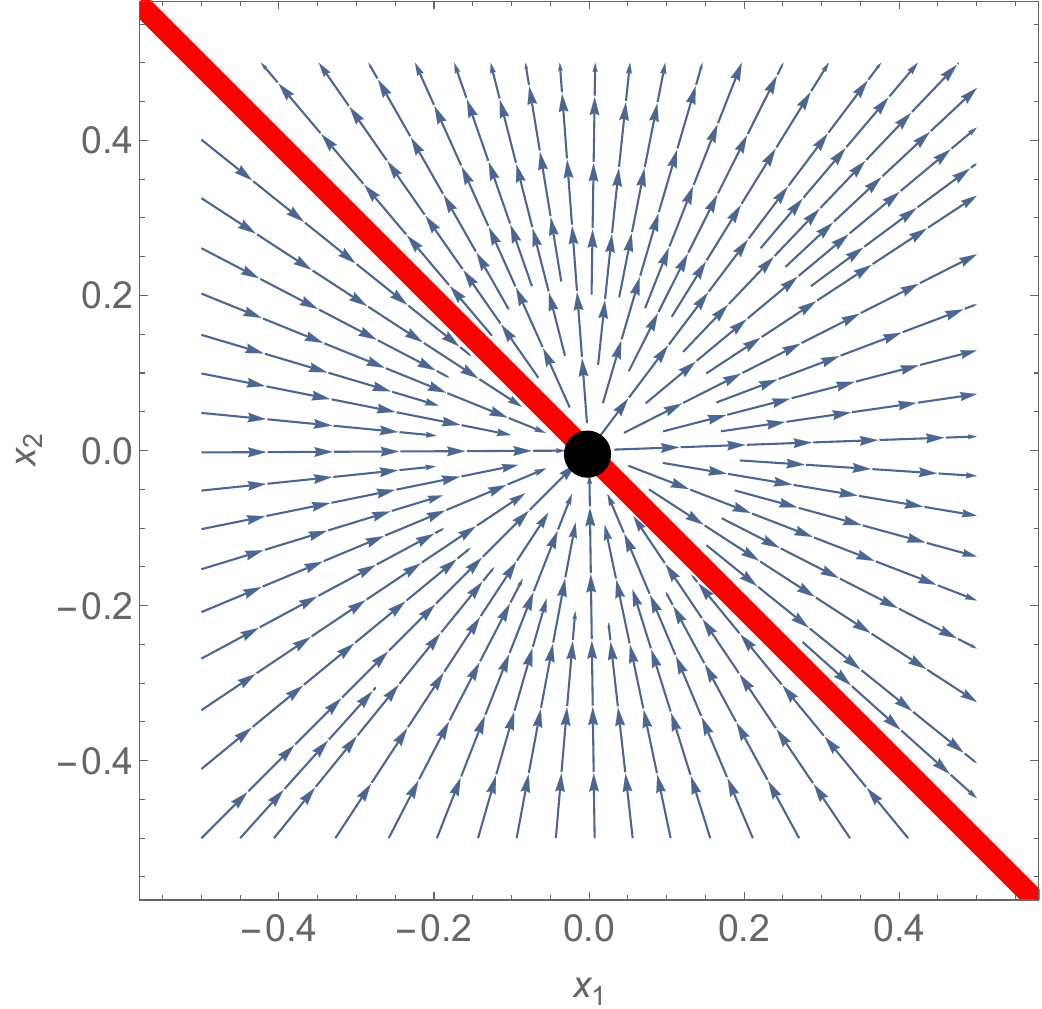}
		\caption{The projection of phase trajectories \eqref{eq:ds_2_example_redefined} on the $(x_1,x_2,x_3,x_4)=\left(x_1,x_2,0,\frac{2}{3}\right)$-plane which shows that $(0,0)$ is saddle and hence a point $\mathcal{Q}_2$ is saddle. Here we have taken $n=0.5$. The diagonal red line corresponds to the invariant submanifold $\mathcal{S}_{x_1+x_2}$. One must be mindful that the above picture is just a \emph{projection} of the complete higher-dimensional phase space. Note that although no physical trajectory actually crosses the invariant submanifold $\mathcal{S}_{x_1+x_2}$, it may seem to do so in the projected phase portrait due to dimensional reduction.} 
		\label{fig:cmt_S}
	\end{figure}

\begin{figure}[H]
		\centering
		\subfigure[]{%
			\includegraphics[width=8cm,height=6cm]{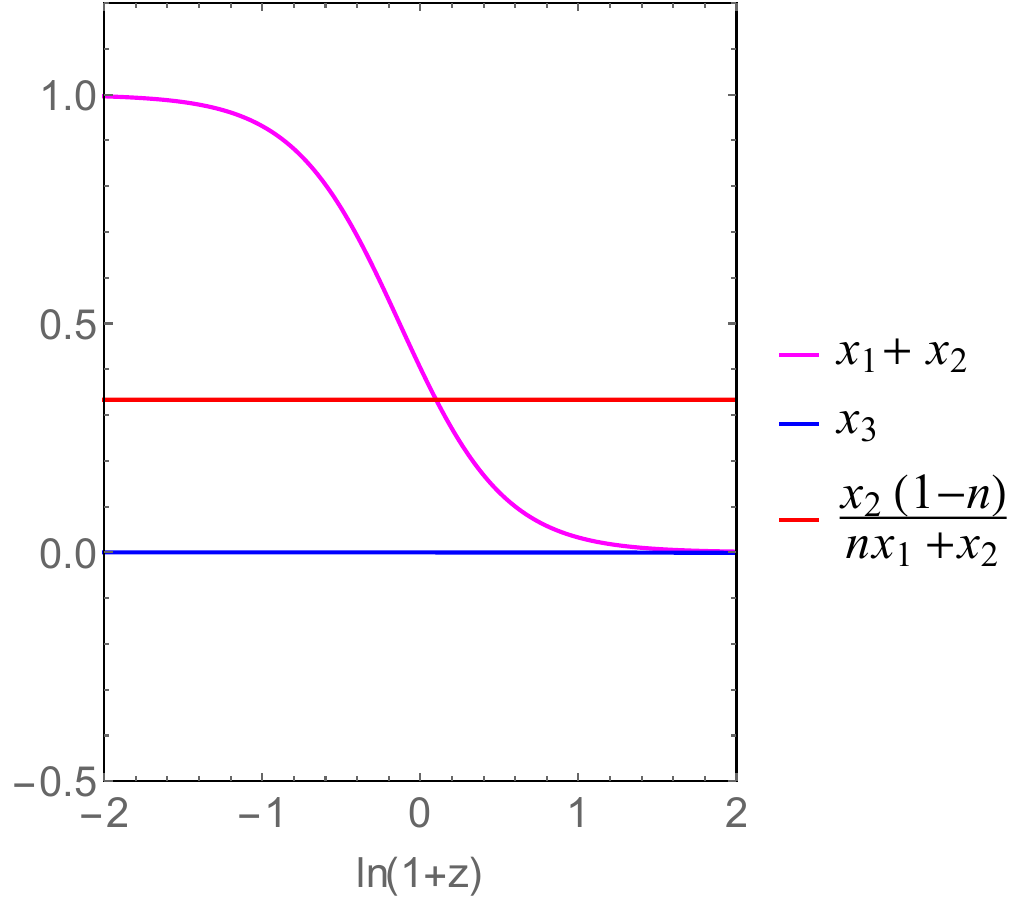}\label{fig:var_c2}}
		\quad
		\subfigure[]{%
			\includegraphics[width=8cm,height=6cm]{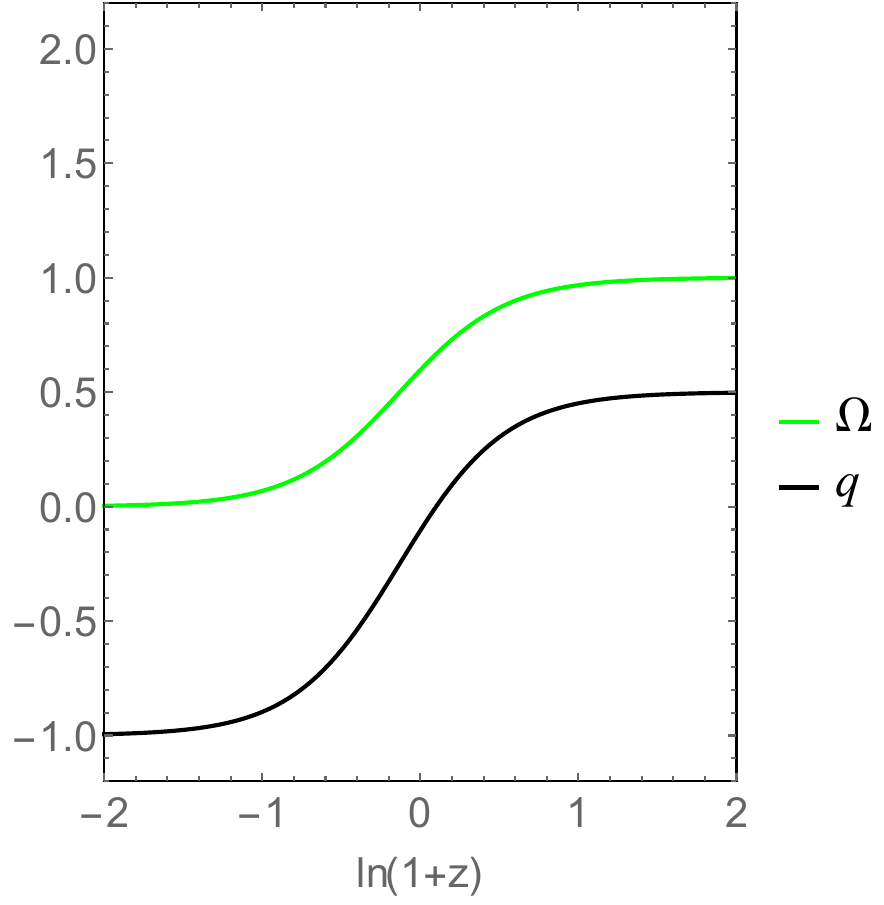}\label{fig:para_c2}}
		\caption{(a) Evolution of the dynamical variables $x_1+ x_2, x_3$ and the quantity $\frac{x_2(1-n)}{nx_1+x_2}$, whose positive value corresponds to $f_Q>0$, (b) Evolution of energy density parameter $\Omega$ and  the deceleration parameter $q$ along a typical trajectory passing through the vicinity of $\mathcal{Q}_2$. Here, $f(Q) = \alpha Q + \beta (-Q)^n$, $n = 0.5$. The initial conditions are taken to be close to the matter-dominated fixed point $\mathcal{Q}_2$ at $z=6$: $x_1(z=6)=0.001, x_2(z=6)=0.001, x_3(z=6)=-0.001, x_4(z=6)=0.5$. The plots are made within a range of $z$ high enough so that one can see the transition between the matter-dominated saddle $\mathcal{Q}_2$ and the de-Sitter attractor $\mathcal{Q}_1$. In figure (a), notice that $x_3=0$ throughout the evolution, implying that the entire transition takes place within $\mathcal{S}_{x_3}$. The positivity of the quantity $\frac{x_2(1-n)}{nx_1+x_2}$ implies preservation of the condition $f_Q>0$ throughout the evolution, while its constancy is an artefact of $\dot{Q}=0$ on $\mathcal{S}_{x_3}$. However, also note that $(x_1+x_2)$ transitions from $0$ to $1$, which means that even if $Q$, and thus $f_Q$, remains constant, the theory of gravity does not remain GR. This difference is because the Hubble-normalised contribution of the nonlinear term in the action, $\frac{(n-1)\beta(-Q)^n}{H^2}$, gains prominence along the course of the evolution.}
		\label{fig:c2_1}
	\end{figure}

	Interestingly, both sets  $\mathcal{S}$ and $\mathcal{T}$ are not disjoint, but intersect at the line $\mathcal{L}:(x_1, x_2, x_3, x_4) = (0,0,0,x_4)$ with the fixed point $\mathcal{Q}_2$ falls on this line. We note that the set $\mathcal{Q}_2:(x_1, x_2, x_3, x_4) = \left(0,0,0,\frac{2}{3}\right)$ is an interesting one from the cosmological point of view as it can be a candidate for the matter-dominated Universe. Therefore, even though analytically we could not extract the stability nature of the set,  we try to assess its stability from the 2-dimensional projection of the phase portrait as shown in Fig. \ref{fig:cmt_S}. From the 2D projections in the $(x_1,x_2)$ slice at $x_3=0$, $x_4=2/3$,  we observe saddle-type flow patterns at the origin. Importantly, the presence of a saddle structure in a lower-dimensional projection guarantees that the fixed point cannot be fully stable or unstable in the full four-dimensional phase space. The reason is that a trajectory leaving the fixed point in the projection corresponds to a genuinely unstable direction, which persists in the full system.
	Therefore, the observed saddle behavior in the projected plane is sufficient to robustly classify the fixed point $\mathcal{Q}_2$ as a \emph{saddle} in the full system on the $(x_1,x_2,x_3,x_4)=\left(x_1,x_2,0,\frac{2}{3}\right)$ plane in its vicinity.  The saddle nature of point $\mathcal{Q}_2$ is more desirable as the Universe should evolve from the decelerated matter-dominated epoch towards an accelerated epoch. Fig.\ref{fig:c2_1} also shows the time variation of several physically interesting quantities along a heteroclinic trajectory connecting the matter-dominated saddle $\mathcal{Q}_2$ and the de Sitter attractor $\mathcal{Q}_1$.

	\subsubsection*{Variation of the effective gravitational coupling}
	
	As previously discussed in the $\Gamma_1$ analysis,  the time variation of the effective gravitational coupling plays a vital role in establishing the physical viability of $f(Q)$ gravity. We describe the time variation of this coupling using the quantity $x_3 = - \frac{\dot{Q} f_{QQ}}{H f_Q}$, so that,
	\begin{align}
		(\ln f_Q)' = -x_3.
	\end{align}
	Hence, the condition $x_3 \to 0$ implies that $f_Q$ approaches a constant, and so does the effective gravitational coupling. In other words, the theory asymptotically approaches a regime where gravity stabilises.
	
	At the fixed points $\mathcal{Q}_1$ and $\mathcal{Q}_2$, we have $x_3 = 0$ exactly, ensuring stability of the gravitational coupling. For the vacuum fixed point $\mathcal{Q}_3$, this condition requires a sufficiently large value of $x_4$ to suppress $x_3$ and thus control the variation of $f_Q$. In a neighbourhood of all three fixed points, $x_3'=0$ by construction.
	
	To probe this behaviour further, we analyse the stability of $x_3$ at second order. Starting from the $x_3'$ equation in the full system \eqref{eq:c2_ds1_ex3}, we compute its derivative:
	\begin{equation}
		(x_3')' = \frac{27 x_3}{4} \left\{ \left(x_4 - \frac{2}{9} \right)x_3^2 x_4 
		+ \left[ \frac{2(1 + x_1 + 3 x_2)}{9} - 2\left(x_1 + x_2 + \frac{2}{3} \right)x_4 \right] x_3 
		+ (x_1 + x_2)^2 + \frac{1}{3} \right\}.
	\end{equation}
	
	Near each of the fixed points $\mathcal{Q}_1$, $\mathcal{Q}_2$, and $\mathcal{Q}_3$, the first derivative vanishes ($x_3' = 0$ by definition). The second derivative becomes proportional to $x_3$, confirming that the variation of $G_{\text{eff}}$ is suppressed also at second order in a neighbourhood of these points.
	Furthermore, when we include the physical condition from the phase space constraint \eqref{regionG2} into the formula for the deceleration parameter \eqref{qG2}, we get the following bound:

	\begin{equation}
		q < 2 - x_3 - 3(x_1 + x_2).
	\end{equation}
	The above bound, in turn, provides an upper limit on the logarithmic derivative of the effective coupling:
	\begin{equation}
		(\ln |x_3|)' < -3(x_1 + x_2).
	\end{equation}
	
	This inequality is valid near the invariant submanifold $\mathcal{S}_{x_3}$, whereas Eq.~\eqref{S_x3_stab} governs the flow directly on it. As the system approaches   $\mathcal{S}_{x_3}$, the dynamics of $x_3$ become increasingly suppressed. This observation further supports the interpretation of $\mathcal{S}_{x_3}$ as a generic late-time attractor where the evolution of the gravitational coupling effectively freezes out.

	These results -consistent across both $\Gamma_1$ and $\Gamma_2$ branches- emphasise the role of $x_3$ as a geometric measure of the effective gravitational dynamics. In the $\Gamma_2$ case, the simultaneous presence of a matter epoch and a stable de Sitter phase, both lying on $\mathcal{S}_{x_3}$, ensures that physically viable cosmological histories also automatically exhibit controlled variation of $G_{\text{eff}}$.

	We emphasise that, unlike the $\Gamma_{1}$ case illustrated in Fig.~\ref{fig:c1_1}, we could not produce an explicit figure showing the time evolution of $G_{\text{eff}}$ in the present $\Gamma_2$ scenario. The reason is that the non-metricity scalar $Q$ here depends not only on the Hubble rate $H$, but also on the dynamical connection and its derivative, making the relation between $Q$ and cosmological observables significantly more intricate. Nevertheless, we could analytically constrain the logarithmic derivative of $x_3$, which encodes the variation of the effective gravitational coupling.

	\section{DYNAMICAL SYSTEM  ANALYSIS  FOR $\Gamma_3$}\label{sec:Gamma3}
	
	We now turn to the third and final connection branch, $\Gamma_3$, to analyse its dynamical system structure. For convenience, we recall the set of five dynamical variables defined in \eqref{dynvar_def}
	\begin{equation}
		x_1 = -\frac{f}{6H^2 f_Q}, \quad x_2 = \frac{Q}{6H^2}, \quad x_3 = -\frac{\dot{f}_Q}{H f_Q}, \quad x_4 = \frac{\gamma}{2H}, \quad \Omega = \frac{\rho_m}{3H^2 f_Q}\,.
		\label{dynvar_gamma3}
	\end{equation}
	Using these  variables, the Friedmann equation \eqref{fried-3_redef_rev} can be wriiten as:
	
	\begin{equation}
		\Omega=1-x_1-x_2+x_3x_4\,.
	\end{equation}
	Again with the help of dynamical variables \eqref{dynvar_gamma3}, the cosmological equations \eqref{Q-3_redef_rev}- \eqref{Qdot-3_redef_rev} can be recast as the following dynamical system:
	\begin{subequations}\label{eq:ds_3}
		\begin{eqnarray}
			x_1' &=& \frac{x_2 x_3}{m}+x_1 \left(x_3- \frac{2\dot{H}}{H^2}\right)\,, \label{eq:c3_ds1}
			\\
			x_2' &=&- \frac{x_2 x_3}{m}- 2x_2\frac{\dot{H}}{H^2}\,,
			\\
			x_3' &=& - x_3 \left[\frac{\dot{H}}{H^2} - x_3 - 1 +\frac{2}{x_4}(1 + x_2)\right]\,,
			\\
			x_4' &=&1+x_2-x_4 \left(3+\frac{\dot{H}}{H^2}\right)\,.\label{eq:c3_ds4}      
		\end{eqnarray}
	\end{subequations}
	We can then substitute $\dot{H}/H^2$ from the Raychaudhuri equation \eqref{raych-3_redef_rev}: 
	\beq
	\label{HdotG3}
	\frac{\dot{H}}{H^2}=-\frac{3}{2}+\frac{3}{2} x_1+\frac{3}{2} x_2+x_3-\frac{1}{2} x_3 x_4\,,
	\eeq
	so that the deceleration parameter $q$ is given by
	\begin{equation}
		\label{qG3}
		q = \frac{1}{2} - \frac{3}{2} x_1 - \frac{3}{2} x_2 - x_3 + \frac{1}{2} x_3 x_4\,.
	\end{equation}
	
	Note that $\gamma$ is nonzero by definition for the connection branch $\Gamma_3$ \cite{Guzman:2024cwa}, so that $x_4\neq0$ as long as the Hubble parameter is finite. Even though $x_4$ appears in the denominator in some of the equations, it doesn't lead to any irregularity in the dynamical system.

	The conditions $f_Q > 0$ and $\rho_m \geq 0$ imply
	\begin{equation}\label{phys1_c3}
		\Omega=1-x_1-x_2+x_3 x_4 \geq 0\,,
	\end{equation}
	so that the physically allowed four-dimensional phase space for the fixed points of the above dynamical system is given by
	\begin{equation}
		\label{regionG3}
		\lbrace (x_1,x_2, x_3,x_4) \in \mathbb{R}^4 \mid x_1+x_2-x_3 x_4 \leq 1 \rbrace\,.
	\end{equation}

	\subsection{Generic considerations for $\Gamma_3$}
	\label{genericG3}
	
	\begin{enumerate}
		\item {\bf GR limit and invariant submanifold $\mathcal{S}_{x_2}$:} The GR limit and the existence of the invariant submanifold $\mathcal{S}_{x_2}$ are exactly the same as in the case of connection $\Gamma_2$.
		
		\item  {\bf Existence of a vacuum invariant submanifold $\mathcal{S}_\Omega$:}
		
		For this connection, we define the vacuum invariant submanifold of the phase space  as
		\begin{equation}
			\mathcal{S}_\Omega := \left\{ (x_1, x_2, x_3, x_4) \in \mathbb{R}^4 \mid x_1 + x_2 + x_3 x_4 = 1 \right\}.
		\end{equation} 
		The evolution equation of $\Omega$ is given by
		\begin{equation} \label{S_Om_stab_c3}
			\Omega' = -3\Omega\left(1 + \frac{1}{3}x_3 + \frac{2}{3}x_3x_4 - \Omega\right)\,,
		\end{equation}
		which implies that $\mathcal{S}_\Omega$ is \emph{stable} if
		\begin{equation}\label{stab_S_Om_3}
			1+\frac{1}{2}x_3+\frac{2}{3}x_3 x_4 > 0,
		\end{equation}
		otherwise, it is unstable. As in the case of $\Gamma_2$, the stability of $\mathcal{S}_\Omega$ depends on the geometric influence of the  connection through  $x_3$ and $x_4$.  
		
		\item {\bf Invariant submanifold $\mathcal{S}_{x_3}$:} The existence of the invariant submanifold $\mathcal{S}_{x_3}$ characterises a generic dynamical feature stemming from the connection branch $\Gamma_3$ as well. The kinematics on this invariant submanifold turns out to be exactly the same as discussed for the $\Gamma_2$ scenario, because the dynamical equations and the Friedmann constraint \emph{on} this invariant submanifold read the same as those in the case of $\Gamma_2$ once we enforce the condition $x_3=0$. As a result, the evolution on this submanifold exactly mimics the $\Lambda$CDM solution, regardless of which  gauge ($\Gamma_2$ or $\Gamma_3$) is chosen. It is noteworthy that the dynamics \emph{on} this invariant submanifold, which exists for both $\Gamma_2$ and $\Gamma_3$, cannot distinguish between these two connection branches at least at the background level. However, the dynamics \emph{towards} this invariant submanifold differ between the two connection branches, as we discuss below.
		
	\end{enumerate}

	\subsubsection*{Flow towards $\mathcal{S}_{x_3}$}
	We now analyze the flow \emph{towards} the invariant submanifold $\mathcal{S}_{x_3}$.  The evolution equation for $x_3$ can be rewritten as
	\begin{equation}
		\label{S_x3_stab_c3}
		x_3' =x_3 \left(2+q+x_3-\frac{2x_2}{x_4}-\frac{2}{x_4}\right)= x_3 \left(\frac{5}{2}- \frac{3}{2} x_1 - \frac{3}{2} x_2+ \frac{1}{2} x_3 x_4-\frac{2x_2}{x_4}-\frac{2}{x_4}\right)\,. 
	\end{equation}
	
	Therefore, ${\mathcal S}_{x_3}$ is an attractor only if $ (x_1 + x_2)+\frac{4}{3x_4}(x_2+1)>\frac{5}{3}  $. As in the case of $\Gamma_2$, the invariant submanifold $\mathcal{S}_{x_3}$ is crucial as both the matter-dominated and de Sitter fixed points (see Table~\ref{tab:C2_1}) and hence the heteroclinic orbit connecting them  lie on $\mathcal{S}_{x_3}$. Hence, the $\Lambda$CDM-like cosmic evolution  also exists generically within the framework of  $\Gamma_3$ connection branch.\\

	\noindent {\bf Generic fixed point:}   Besides the vacuum de Sitter fixed points $\mathcal{R}_1\equiv(2-3x_4,3x_4-1,0,x_4)$ and GR matter-dominated fixed point $\mathcal{R}_2\equiv(0,0,0,2/3)$, the system admits a third  isolated  fixed point $\mathcal{R}_3\equiv(0,0,-1,1)$ which exhibits a radiation-like solution with $q=1$ and $\Omega=0$. It exists when the quantity $\frac{x_2}{m}$ vanishes. Since $\Omega = 0$, this  radiation-like  behavior is purely geometric, with no matter contribution. It is worth noting that the de Sitter set $\mathcal{R}_1$ is always stable, making it a natural candidate for a generic attractor, similar to $\mathcal{Q}_1$ in the $\Gamma_2$ case. As before, any $f(Q)$ model with a finite value of $\frac{x_2}{m}$ at the fixed points yields a viable cosmological evolution. In Table \ref{tab:inv_sub_Gamma3_ex}, we summarize the generic comparative features between the connection $\Gamma_2$ and $\Gamma_3$.
	In what follows, we explore these features concretely for a power-law $f(Q)$ model. Note that depending on the form of $f(Q)$, additional fixed points or invariant submanifolds may arise.

	\subsection{Example:  $f(Q) = \alpha Q + \beta (-Q)^n$}
	
	Again here we consider a particular example  
	\begin{equation} \label{e.g_fQ_c2}
		f(Q)=\alpha Q+\beta (-Q)^n\,.
	\end{equation}
	As in Sect.~\ref{sectG2example}, where the same functional form of $f(Q)$ was considered in the $\Gamma_2$ connection, the corresponding equation \eqref{Q-dot_rev} for the $\Gamma_3$ case becomes:

	\beq
	Q  \ddot Q +\left[(n-2) \dot Q + \left( 2\frac{\dot \gamma}{\gamma}+\frac{\dot a}{a}\right)Q  \right]\dot Q=0\,, 
	\eeq
	which can be one-order integrated into
	\beq 
	Q(t)=\left[ (n-1) \left( C_1 \int \frac{dt}{a(t)\gamma^2(t)}+C_2 \right)\right]^{1/(n-1)} \,,
	\eeq
	with $C_{1,2}$ two arbitrary constants of integration; the evolution for the scale factor is now coupled to a factor $\gamma^2$ as noticed in Sect.\ref{sec:background}.
	
	Resorting to dynamical system techniques, we obtain the relation between $m$ and dynamical variables $x_1, x_2$ as 
	\begin{equation}\label{m_dyn_var_c3}
		m=n \left(1+\frac{x_1}{x_2}\right)\,.
	\end{equation}
	
	So, the dynamical system \eqref{eq:c3_ds1}-\eqref{eq:c3_ds4} becomes 
	\begin{subequations} \label{eq:ds_3_example}
		\begin{eqnarray}
			x_1' &=&  {\frac {{ x_3}\,{x_2}^2}{n\left( x_2+x_1\right)} }+x_1 \left(3-3x_1-3\,{x_2}-\,{x_3}+\,{x_3}\,{x_4}\right),
			\label{eq:c3_ds1_ex1}
			\\
			x_2' &=& -{\frac {{ x_3}\,{x_2}^2}{n\left( x_2+x_1\right)} }+x_2 \left(3-3\,x_2 -3x_1 -2\,{x_3}+\,{x_3}\,{x_4}\right)\,,
			\\
			x_3' &=&-\frac{x_3}{2x_4}\, \left(4+4x_2-5 x_4+ 3{x_1}\,{x_4}+3 x_2 x_4-x_3 x_4^2 \right) 
			\,,
			\\
			x_4' &=&1+x_2-\frac{x_4}{2} \left(3\, \left( 1+ x_1+x_2 \right)-2\,x_3 + x_3 x_4\right)
			\,.\label{eq:c3_ds5_ex1}     
		\end{eqnarray}
	\end{subequations}

	From conditions $\Omega \geq 0$ and \eqref{phys2_c2}, the physical phase space of fixed points of the dynamical system \eqref{eq:c3_ds1_ex1}-\eqref{eq:c3_ds5_ex1} is given by
	\begin{equation}\label{psp_c3}
		\left\lbrace (x_1,x_2,x_3,x_4) \in \mathbb{R}^4 \mid x_1+x_2-x_3 x_4 \leq 1\,, \frac{x_2 (1-n)}{nx_1+x_2}>0  \right\rbrace\,.
	\end{equation}
	While a regularized version of the system \eqref{eq:ds_3_example} can be constructed as in the $\Gamma_2$ case, we refrain from presenting it explicitly for the sake of brevity.
	
	\begin{figure}[H]
		\centering
		\subfigure[]{%
			\includegraphics[width=8cm,height=6cm]{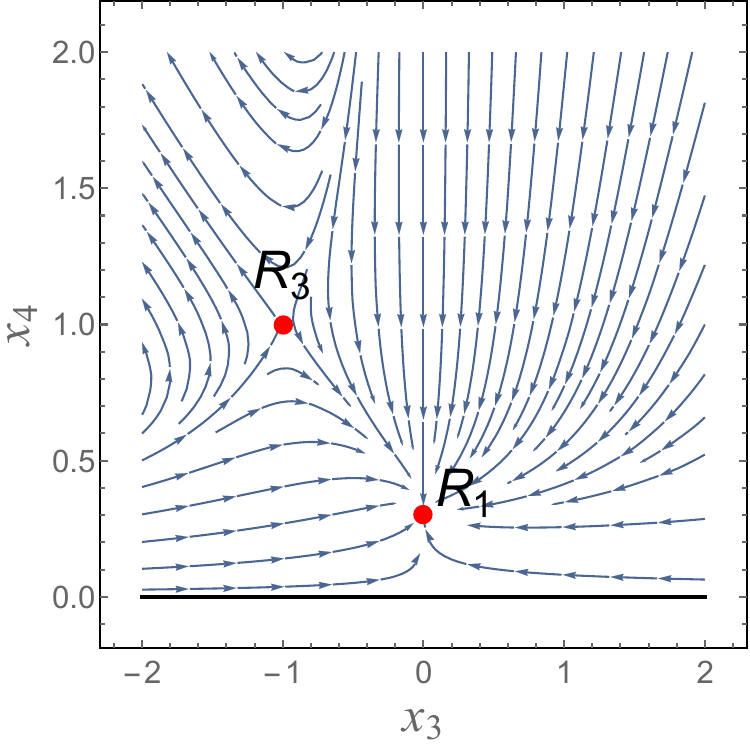}\label{fig:c3_intersection_S_Om_Sx2}}
		\quad
		\subfigure[]{%
			\includegraphics[width=8cm,height=6cm]{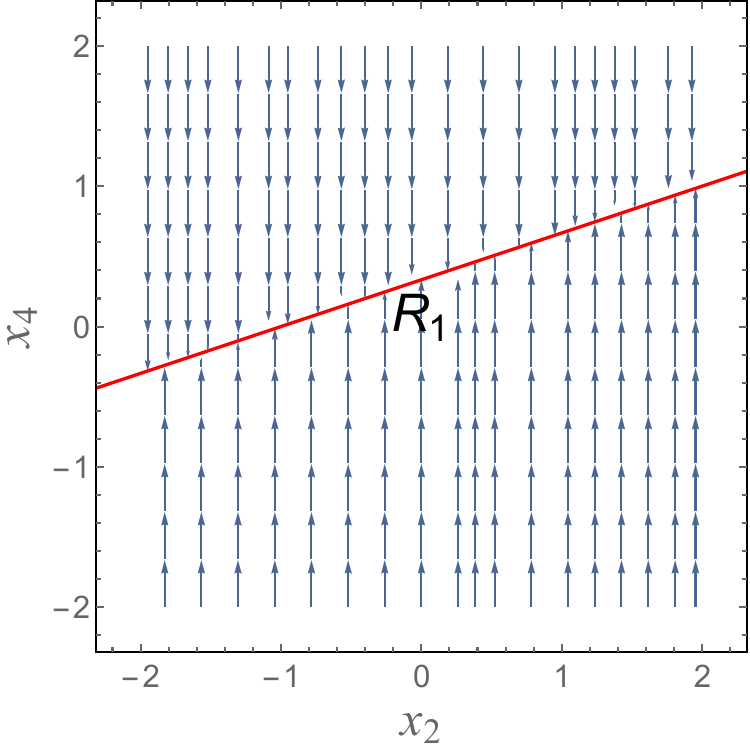}\label{fig:c3_intersection_S_Om_Sx3}}
		\quad
		\subfigure[]{%
			\includegraphics[width=8cm,height=6cm]{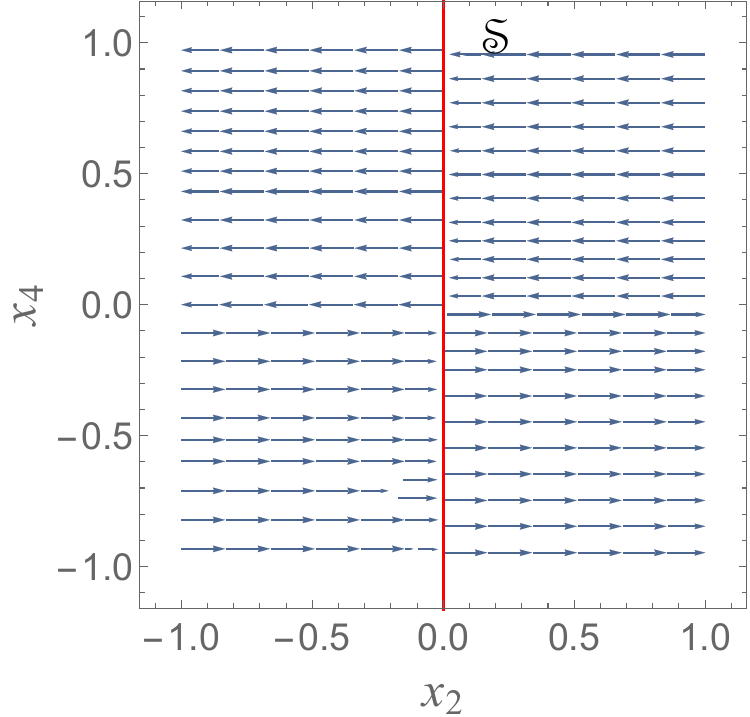}\label{fig:c3_intersection_S_Om_Sx1x2}}
		\caption{Projection of phase trajectories \eqref{eq:ds_3_example} on the intersection of pairs of invariant submanifolds, shown in the following panels:  \\ \textbf{(a)} We intersect  $\mathcal{S}_{\Omega}$ with $\mathcal{S}_{x_2}$ and obtain that point $\left(0,\frac{1}{3}\right)$, which  belongs to the family $\mathcal{R}_1$, is an attractor, while the point $\mathcal{R}_3=(-1,1)$   is a saddle.\\ \textbf{(b)}  We intersect  $\mathcal{S}_{\Omega}$ with $\mathcal{S}_{x_3}$ and display also the line $x_2=3 x_4-1$ (in red) confirming that $\mathcal{R}_1$ is an attractor as node. \\ \textbf{(c)} Finally, we intersect  $\mathcal{S}_{\Omega}$ with $\mathcal{S}_{x_1+x_2}$ choosing $n=-2$ and numerically confirm that $\mathcal{S}$ (in red) is a saddle. The depicted qualitative behaviors of the trajectories in all plots are independent of the values of $n$.}
		\label{fig:c3_inv_manf_int}
	\end{figure}

	\begin{table}[H]
		\centering
			\renewcommand\arraystretch{1.5}
		\begin{tabular}{|c|c|c|}
			\hline 
			Features & Stability condition in $\Gamma_2$ case &  Stability condition in  $\Gamma_3$ case \\ 
			\hline 
			$ \mathcal{S}_{x_1+x_2}$ &	See Table \ref{tab:inv_sub_Gamma2_ex} & Same as $\Gamma_2$ \\ 
			\hline
			$\mathcal{S}_{\Omega}$&  See Table \ref{tab:inv_sub_Gamma2_ex} & $1+\frac{1}{3} x_3+2 x_3 x_4>0$ \\
			\hline
			$ \mathcal{S}_{x_3}$ &  See Table \ref{tab:inv_sub_Gamma2_ex} & Same as $\Gamma_2$ \\ \hline
			$\mathcal{S}_{x_2}$ &  See Table \ref{tab:inv_sub_Gamma2_ex}& Stable for $x_1(3 - 2x_3 - 3x_1 + x_3 x_4) < 0$ \\ 
			&    & Unstable for $x_1(3 - 2x_3 - 3x_1 + x_3 x_4) > 0$ \\ 
			\hline
			Family of points ($\mathcal{S}, \mathcal{T}$) & See Table \ref{tab:C2_3} & Same as $\Gamma_2$\\  
			\hline
			deSitter point & Always stable (see Table \ref{tab:C2_3}, set $\mathcal{Q}_1$) & Always stable  (set $\mathcal{R}_1$)\\        
			\hline
			Matter dominated point & Saddle (see Table \ref{tab:C2_3}, point $\mathcal{Q}_2$) & Saddle (point $\mathcal{R}_2$)\\        
			\hline
			Vacuum fixed point & Saddle (see Table \ref{tab:C2_3}, set $\mathcal{Q}_3$) & Saddle (Point $\mathcal{R}_3$)\\        
			\hline
		\end{tabular} 
		\caption{Key similarities and differences in the stability behavior of invariant submanifolds and fixed points of the phase space of $f(Q)=\alpha Q+\beta (-Q)^n$ theory between $\Gamma_2$ and $\Gamma_3$ dynamical systems. The stability conditions of the invariant submanifolds are for the region $(x_1+x_2)>0$. For the region $(x_1+x_2)<0$, the stability conditions will be reversed. } 
		\label{tab:inv_sub_Gamma3_ex}
	\end{table}
	
\begin{figure}[H]
		\centering
		\subfigure[]{%
			\includegraphics[width=8cm,height=6cm]{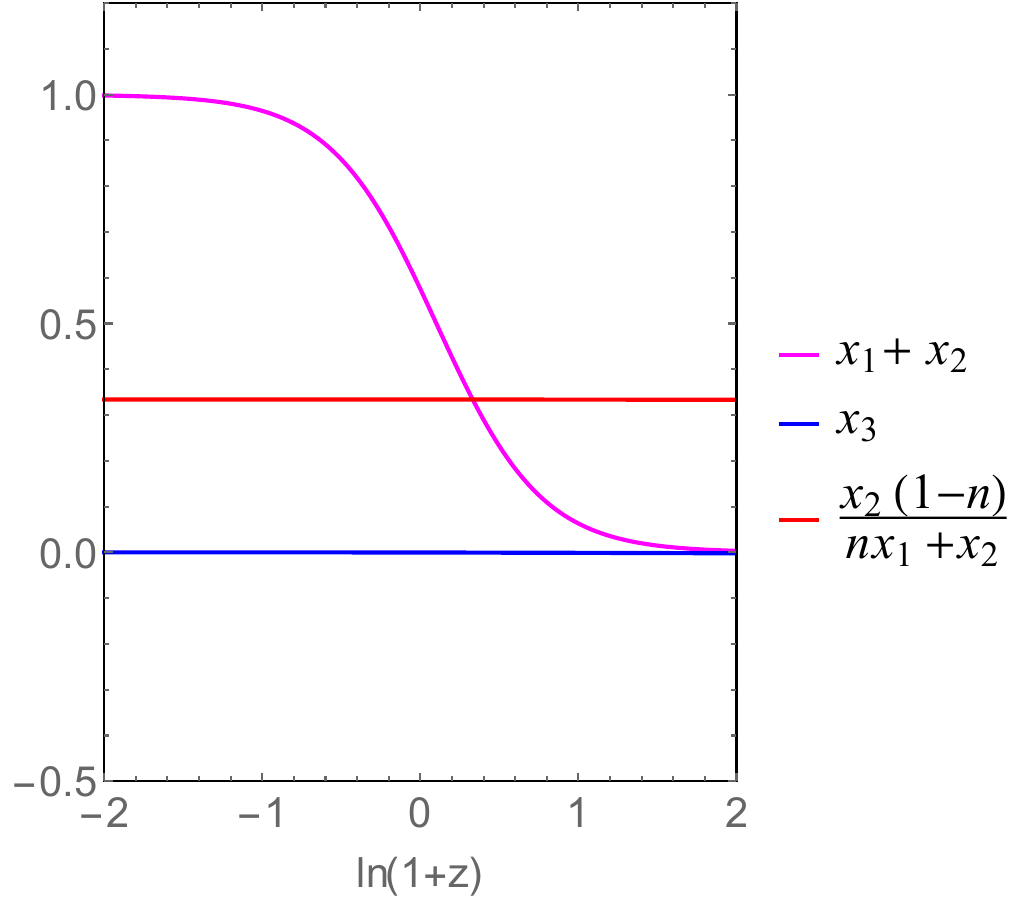}\label{fig:var_c3}}
		\quad
		\subfigure[]{%
			\includegraphics[width=8cm,height=6cm]{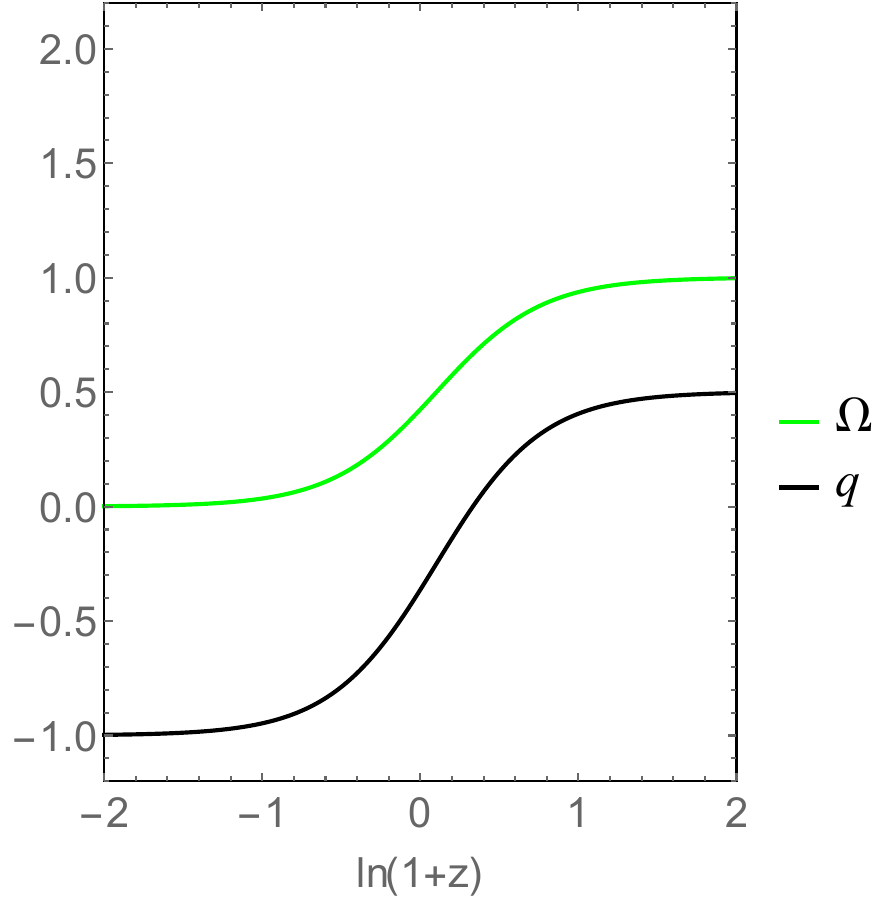}\label{fig:para_c3}}
		\caption{(a) Evolution of the dynamical variables $x_1+ x_2, x_3$ and the quantity $\frac{x_2(1-n)}{nx_1+x_2}$, whose positive value corresponds to $f_Q>0$, (b) Evolution of energy density parameter $\Omega$ and  the deceleration parameter $q$ along a typical trajectory passing through the vicinity of $\mathcal{Q}_2$. Here, $f(Q) = \alpha Q + \beta (-Q)^n$, $n = 0.5$. The initial conditions are taken to be close to the matter dominated fixed point $\mathcal{R}_2$ at $z=6$: $x_1(z=6)=0.002, x_2(z=6)=0.002, x_3(z=6)=-0.002, x_4(z=6)=0.6$.  The positivity of the quantity $\frac{x_2(1-n)}{nx_1+x_2}$ implies preservation of the condition $f_Q>0$ throughout the evolution.}
		\label{fig:c3_1}
	\end{figure}

	In Table \ref{tab:inv_sub_Gamma3_ex}, we present and compare the invariant submanifolds and the fixed points of the regularised system with those of connection $\Gamma_2$.  Interestingly,  as in $\Gamma_2$, the system besides the generic fixed point has two broader families of fixed points (same as $\mathcal{S}$ and $\mathcal{T}$ of $\Gamma_2$) and one isolated fixed point $\mathcal{R}_4 \left(\frac{2(4n-3)}{(2n+1)(n-1)},\frac{2n(3-4n)}{(2n+1)(n-1)},\frac{10(n-1)}{(2n+1)},\frac{1-2n}{2(n-1)}\right)$. We note that point  $\mathcal{R}_4$ corresponds to a vacuum dominated point and hence belongs to submanifold $\mathcal{S}_\Omega$. It is physical if $-\frac{1}{2}<n<0$ or $n>\frac{3}{4}$. It can also correspond to an accelerated universe for $n<-\frac{1}{2}$ or $n>2$. The stability and physical nature of the remaining family of fixed points ($\mathcal{S}, \mathcal{T}, \mathcal{R}_1, \mathcal{R}_2, \mathcal{R}_3$) are the same as in the connection $\Gamma_2$. As in the case of $\Gamma_2$, in Fig. \ref{fig:c3_inv_manf_int} we present the projection of the phase trajectories of the system \eqref{eq:ds_3_example} on the planes of intersection of pairs of invariant submanifolds such as  $\mathcal{S}_{\Omega}$ with $\mathcal{S}_{x_2}$, $\mathcal{S}_{x_3}$ and $\mathcal{S}_{x_1+x_2}$, separately; the same considerations about the de Sitter epoch as in $\Gamma_2$ apply. Fig. \ref{fig:c3_1} also illustrates the time variation of several physically interesting quantities along a heteroclinic trajectory connecting the matter-dominated saddle $\mathcal{R}_2$ and the de-Sitter attractor $\mathcal{R}_1$.

	\section{Towards the analysis of broader classes of $f(Q)$ theories}\label{app: broader_class}
	
	Many $f(Q)$ theories of cosmological interest fail to yield a closed autonomous system because the auxiliary function $ m(Q)$ cannot always be expressed solely in terms of dimensionless dynamical variables. The dynamical system framework presented in this paper—formulated as Eq.~\eqref{eq:dx1_c} for $\Gamma_1$, \eqref{eq:ds_2} for $\Gamma_2$, and \eqref{eq:ds_3} for $\Gamma_3$—relies critically on $m(Q)$ being either constant or expressible in terms of $r(Q)=-\frac{x_2}{x_1}$, the latter being the case of our particular example $f(Q) = \alpha Q + \beta(-Q)^n$.  
	
	When this condition fails, one can no longer construct a closed autonomous system unless $m(Q)$ is promoted to a new dynamical variable and its evolution equation $m'$ is computable from $m$ and the other variables. This naturally motivates a hierarchy of \textit{shape parameters} $m_i$, where each $m_i$ encodes higher-order logarithmic derivatives of $f(Q)$.
	
	In this section, we develop an extended framework that incorporates this $m_i$ hierarchy and enables the dynamical analysis of a broader class of $f(Q)$ theories.
	
	To find a set of first order differential equations that can be closed under some assumption, let us define the hierarchy of the auxiliary quantities
	\begin{equation} \label{def-mi}
		m_i \equiv Q \frac{d^{i+1}f/dQ^{i+1}}{d^{i}f/dQ^{i}} = \frac{d\ln\left(d^{i}f/dQ^{i}\right)}{d\ln Q}\,,
	\end{equation}
	so that $r=m_0,\,m=m_1$. Differentiating Eq.~\eqref{def-mi} with respect to time and using Eq.~\eqref{dynvar_def} to express $\dot{Q}$ in terms of $x_3$, one arrives at a recursion relation 
	\begin{equation} \label{mip}
		m_i' = - \left(1 - m_i + m_{i+1}\right) \frac{m_i x_3}{m_1}\,.
	\end{equation}
	This relation tracks how the structure of 
	$f(Q)$ changes at higher orders and plays a central role in building the extended dynamical system. 
	For the case $\Gamma_1$, the variable $x_3$ can be rewritten as: 
	\begin{align}
			x_3 = -\frac{f_{QQ} \dot{Q}}{H f_Q} = \frac{12 f_{QQ} \dot{H}}{f_Q} = - 2 m_1 \frac{\dot{H}}{H^2} = 2m_1(1+q)\,.
			\end{align}
	Using the identity in Eq.~\eqref{trade}, we  eliminate $x_3$ in favor of $\Omega$, yielding a recursive evolution equation for the $m_i$:
	
	\begin{equation} \label{mip1}
		m_i' = -3\Omega m_i \left(\frac{1 - m_i + m_{i+1}}{1 + 2 m_1}\right)\,.
	\end{equation}

	The set of equations in Eq.~\eqref{mip} or Eq.~\eqref{mip1} can be closed at the $i$-th level of the hierarchy when the sequence of shape parameters $m_i$ admits a truncation. This occurs if either $m_{i+1}$ is a constant or can be written as a function of $m_i$ and other dynamical variables. In such cases, one can promote $m_1$ (or higher $m_i$) to a dynamical variable and construct an extended autonomous system. This fact reveals that even when $m=m_1$ cannot be expressed purely in terms of $r=m_0$ for a given $f(Q)$, one can still analyze the theory dynamically by building a systematic extension of the standard framework. We illustrate this hierarchy technique with two particular examples below. For the sake of simplicity we confine ourselves to the case of $\Gamma_1$, but the technique is general and can be applied for all the three connection branches.


	\subsubsection*{Example 1: Constant $m_{i+1}$ at some level}

	Let us first consider an example for which $m_{i+1}$ is constant at some order $i$. Consider the theory:
	\begin{equation}
		f(Q) = -2\Lambda + \alpha Q + \beta(-Q)^n ~~{\rm with}~ \Lambda,\alpha,\beta\neq0,\,n\neq0,1\,.
	\end{equation}
	For this $f(Q)$, one cannot in general express $m_1=m$ uniquely in terms of $m_0=r=-\frac{x_2}{x_1}$, so that the the standard dynamical system formulation fails. However, for this case one has
	\begin{equation}
		m_2 = \frac{Q f_{QQQ}}{f_{QQ}} = n-2\,,
	\end{equation}
	so that if one extends the standard dynamical system framework by including only one more differential equation, namely the dynamical equation for $m_1$, one ends up with an autonomous system. For the case of $\Gamma_1$, the autonomous system is
	\begin{subequations}
		\begin{eqnarray}
			\Omega' &=& 3\Omega \left[\Omega\left(\frac{1 + m_1}{1 + 2m_1}\right) - 1\right]\,,
			\label{omp-mi}
			\\
			m_1' &=& -3 m_1 \Omega\left(\frac{n - m_1 - 1}{1 + 2m_1}\right)\,.
			\label{m1p-mi}
		\end{eqnarray}
	\end{subequations}
	The system presents the fixed points
	\begin{table}[H]
		\begin{center}
				\renewcommand\arraystretch{1.3}
			\begin{tabular}{|c|c|}
				\hline
				Fixed points & $(\Omega,m_1)$ \\ 
				\hline
				$\mathcal{A}$ & $(0,m_1)$ \\
				\hline
				$\mathcal{B}$& $(1,0)$ \\
				\hline
				$\mathcal{C}$& $\left(2-\frac{1}{n},n-1\right)$ \\
				\hline
			\end{tabular}
		\end{center}
		\caption{Fixed points of the theory $f(Q) = -2\Lambda + \alpha Q + \beta(-Q)^n$, $(\Lambda,\alpha,\beta\neq0,\,n\neq0,1)$.}
	\end{table}
	$\mathcal{A}$ is a line of De-Sitter fixed points and the $\mathcal{B}$ is the standard General Relativistic matter-dominated fixed point ($m_1=0\Rightarrow f_{QQ}=0$).
	
	\subsubsection*{Example 2: Non-constant $m_{i+1}$ }
	
	Let us now consider a theory where the auxiliary functions $m_i$ do not become constant at any finite order, but a truncation of the hierarchy is still possible because we can express  $m_{i+1}$ in terms of $m_i$. Consider:
	\begin{equation}\label{f-exp1}
		f(Q) = \alpha (-Q)^n + e^{\beta Q}\,.
	\end{equation}
	For this model, \( m_1 = Q f_{QQ} / f_Q \) cannot be expressed in terms of other dynamical variables like \( r = -x_2 / x_1 \), making the standard dynamical system approach inapplicable. However, for \( i \geq n+1 \), one finds
	\begin{equation}
		m_i = \beta Q \quad \text{and hence} \quad m_{i+1} = m_i\,.
	\end{equation}
	This means that the hierarchy of shape parameters can be truncated beyond level \( i = n+1 \). Promoting \( m_1, m_2, \ldots, m_{n+1} \) to dynamical variables, one obtains a closed autonomous system:
	\begin{subequations}
		\begin{align}
			\Omega' &= 3\Omega \left[\Omega\left(\frac{1 + m_1}{1 + 2m_1}\right) - 1\right]\,, \\
			m_1' &= -3 m_1 \Omega \left(\frac{1 - m_1 + m_2}{1 + 2 m_1}\right)\,, \\
			&\ \vdots \nonumber \\
			m_{n+1}' &= -3 m_{n+1} \Omega \left(\frac{1}{1 + 2 m_1}\right)\,.
		\end{align}
	\end{subequations}
	The system presents, among all possible fixed points, definitely the following two
	
	\begin{table}[H]
		\centering
		\begin{tabular}{|c|c|}
			\hline
			Fixed points & $(\Omega, m_1, m_2, \ldots, m_{n+1})$ \\
			\hline
			$\mathcal{A}$ & $(0, m_1, m_2, \ldots, m_{n+1})$ \\
			\hline
			$\mathcal{B}$ & $(1, 0, 0, \ldots, 0)$ \\
			\hline
		\end{tabular}
		\caption{Fixed points of the theory $f(Q) = \alpha (-Q)^n + e^{\beta Q}$.}
	\end{table}
	
	The fixed point $\mathcal{A}$ represents a de Sitter phase and exists for arbitrary values of the shape parameters $m_i$, while $\mathcal{B}$ corresponds to a General Relativistic matter-dominated point, where $ f_{QQ} = 0 $ . This fixed point exists because setting all $ m_i = 0$  ensures that $m_i' = 0 $ for all  $i$, satisfying the consistency condition for truncating the hierarchy.

	In general, for the $\Gamma_1$ connection, if the $m_i$-hierarchy closes at the $n$-th level, the generic structure of the dynamical system takes the form:
	\begin{subequations}
		\begin{eqnarray}
			\Omega' &=& 3\Omega \left[\Omega\left(\frac{1 + m_1}{1 + 2m_1}\right) - 1\right]\,,
			\\
			m_i' &=& -3\Omega m_i \left(\frac{1 - m_i + m_{i+1}}{1 + 2 m_1}\right)\,, \qquad\qquad (i=1,2,......n)\,.
		\end{eqnarray}
	\end{subequations}
	
	The above generic structure for the $\Gamma_1$ dynamical system reveals the existence of a generic family of de Sitter fixed points $\{\Omega, \mathbf{m_i}\} = \{0, \mathbf{m_i}\}$, except in the special case $m_1 = -\frac{1}{2}$, which corresponds to a degenerate model excluded from our analysis in Sec.~\ref{subsec:Gamma1_generic}. However, a standard (General Relativistic) matter-dominated fixed point $\{\Omega, \mathbf{m_i}\} = \{1, \mathbf{0}\}$ is not guaranteed. Such a point exists only if the condition $\mathbf{m_i} = 0$ also implies $\mathbf{m_i}' = 0$. If this fails, then the standard matter-dominated point is absent. The requirement that $\mathbf{m_i} = 0$ leads to $\mathbf{m_i}' = 0$ thus imposes structural constraints on viable $f(Q)$ theories. The following example makes this explicit.

	\subsubsection*{Example 3: $f(Q) = \alpha Q^2 e^{\beta Q}$ — Absence of matter dominated point}

	For this model, the first shape parameter evaluates to
	\be
	m_1 = \frac{2 + 4 \beta Q + \beta^2 Q^2}{2 + \beta Q}\,,
	\ee
	which can be inverted to express $Q$ in terms of $m_1$ as
	\be
	Q = \frac{-4 \beta + \beta m_1 \pm \beta \sqrt{8 + m_1^2}}{2 \beta^2}\,.
	\ee
	Computing $m_2$ and writing $Q$ in the expression for $m_2$ in terms of $m_1$ using the above equation, we get
	\be
	m_2 = -\frac{1}{2 m_1} (2 + m_1)\left(4 - m_1 \mp \sqrt{8 + m_1^2}\right)\,.
	\ee
	
	Thus,  we get the following the autonomous system:
	\begin{subequations}
		\begin{align}
			\Omega' &= 3\Omega \left[\Omega\left(\frac{1 + m_1}{1 + 2m_1}\right) - 1\right]\,, \\
			m_1' &= \frac{3 m_1 \Omega}{1 + 2 m_1}\left(-1 + m_1 + \frac{(2 + m_1) (4 - m_1 \pm \sqrt{8 + m_1^2})}{2 m_1}\right)\,.
			\label{m1p-exp2}
		\end{align}
	\end{subequations}
	
	Note that both the $+$' and $-$' solution branches in Eq.~\eqref{m1p-exp2} are mathematically valid for any choice of model parameters. However, to ensure physical viability of the resulting fixed points, the parameters must be chosen such that the scalar $Q$ is negative, so that fixed points lie within the domain.

	From the above system, we get  the  following two fixed points:
	\begin{table}[H]
		\begin{center}
				\renewcommand\arraystretch{1.3}
			\begin{tabular}{|c|c|}
				\hline
				{\bf Fixed Points} & $(\Omega, m_1)$ \\ 
				\hline
				$\mathcal{A}$ & $(0, 1)$ \\
				\hline
				$\mathcal{B}$ & $\left(\frac{3}{2}, 1\right)$ \\
				\hline
			\end{tabular}
		\end{center}
		\caption{Fixed points of the theory $f(Q) = \alpha Q^2 e^{\beta Q}$. The fixed point $\mathcal{B}$ exists only when the minus sign is chosen in Eq.~\eqref{m1p-exp2}. In the $+$' branch, both $\mathcal{A}$ and $\mathcal{B}$ correspond to $Q = 0$ and are thus unphysical. The point $\mathcal{B}$ becomes physically meaningful in the `$-$' branch only when $\beta > 0$.}. 
	\end{table}
	
	The system features a de Sitter line $\Omega = 0$, but a standard GR matter-dominated fixed point $\{\Omega, m_1\} = \{1, 0\}$ is absent. This failure arises because $m_1 = 0$ does not imply $m_1' = 0$, violating the above structural condition. This violation suggests that the matter-dominated fixed point $(\Omega, m_1) = (1, 0)$ can disappear if the power-law $f(Q)$ is multiplied by an exponential term. Hence,  we can rule out such $f(Q)$ theories from the perspective of cosmological viability. 
	
	In summary, the $m_i$-hierarchy method systematically extends the dynamical systems approach to a broader class of $f(Q)$ models, including those that previously lacked a closed autonomous formulation. By promoting higher-order shape parameters to dynamical variables, this framework reveals structural conditions for the existence of GR-like matter phases. It serves as a diagnostic for identifying cosmologically viable $f(Q)$ theories. This criterion applies to many phenomenologically relevant models, including those studied in \cite{Anagnostopoulos:2022gej}.


	\section{Summary and Discussion}
	\label{Sect:Conclusion}
	
	In this work, we have developed a unified dynamical systems framework to analyse homogeneous and isotropic cosmology in $f(Q)$ gravity, incorporating all three independent affine connection branches. This framework is based on a standard set of Hubble-normalised phase space variables and an auxiliary function $m(Q)$, allowing a model-independent and connection-agnostic treatment of the dynamics. This unification permits direct comparison across the three connection choices for any given $f(Q)$ model, provided the system can be closed (see below for more details), which is impossible using connection-specific formulations.

	The dynamical system can be closed if the function 
	\begin{align}
		r(Q) \equiv \frac{Qf_Q}{f} = -\frac{x_2}{x_1}\,,
	\end{align}
	can be inverted to express \( Q \) as a function of the ratio $x_2/x_1 $. This invertibility condition is essential for casting the system entirely in terms of the dynamical variables.
	
	While this condition holds for a wide range of phenomenologically interesting models—such as power-law and exponential forms—it may fail for more complex $f(Q)$ functions. In such cases, our framework can be extended by promoting the auxiliary function  to a dynamical variable and tracking its evolution through a recursive hierarchy of shape parameters . This hierarchy, developed in Section~\ref{app: broader_class}, enables a generalized dynamical treatment beyond the class of directly invertible models.
	We find that the dimensionality of the system depends on the connection branch. For the connection branch $\Gamma_1$,, the system reduces to a single autonomous equation, consistent with previous studies \cite{Boehmer:2022wln, Boehmer:2023knj}. In contrast,  both  branches $\Gamma_2$ and $\Gamma_3$ yield 4-dimensional dynamical systems, consistent with Ref.~\cite{Shabani:2023nvm} which analyzed  $\Gamma_2$ specifically. A part of our work extends this to all branches simultaneously. Interestingly, the dimensionality of the phase space is the same as in \cite{Guzman:2024cwa}, confirming that it is independent of the particular choice of dynamical variables. We note that claims of a 5- or 6-dimensional phase space occasionally found in the literature arise from treating some variables as independent when in fact they are related by the Friedmann constraint.
	
Unlike earlier studies (e.g., \cite{Paliathanasis:2023hqq}) that employed tailored variable sets for each connection, our formulation introduces a universal variable set applicable to all cases. This choice not only enables a comparative analysis between connection branches but also reveals new features of the phase space structure of $f(Q)$ cosmology that would otherwise remain obscured.
	
The unified choice of dynamical variables enables a model-independent comparison across the three connection branches. In what follows, we summarise the key cosmological insights from our general analysis. We can find details in the discussions regarding the generic considerations about each connection branch in the main text.

	\begin{itemize}
		\item \textbf{Regularity in the GR limit:} The dynamical system remains well-defined and regular in the GR limit $f_{QQ}\to 0$ for all three connection branches. When higher-order corrections vanish with this property, we can ensure smooth continuity with standard GR.

		\item \textbf{Existence of de Sitter attractors:} A viable late-time cosmology typically requires a de Sitter future attractor. Our analysis reveals that while the existence and stability of such attractors depend on specific conditions in the connection branch (\( \Gamma_1 \)), they arise generically and robustly in the  connections  $\Gamma_{2,3}$ (see Table~\ref{tab:ds_attractor} for details).

		\begin{table}[H]
			\begin{center}
					\renewcommand\arraystretch{1.5}
				\begin{tabular}{|c|c|c|}
					\hline
					Connection branch & Existence condition & Attractor condition \\
					\hline
					$\Gamma_1$ & $\Omega \left(\frac{1+m(Q)}{1+2m(Q)} \right) -1 \sim -\frac{C}{\Omega^p}, \,\,\, p<1 \,\,\, \text{near} \,\,\, \Omega \gtrsim 0$ & $m'(0)\cdot 0^2+ (2m(0) + 1)^2 > 0$ \\
					\hline
					$\Gamma_{2,3}$ & Always & Always an attractor \\
					\hline
				\end{tabular}
			\end{center}
			\caption{Existence and stability conditions for de Sitter fixed points across connection branches.}
			\label{tab:ds_attractor}
		\end{table}

		\item \textbf{Existence of a matter-dominated fixed point:} We know that a viable cosmological model must reproduce a standard matter-dominated epoch, where $f(Q)$ effectively reduces to STEGR. For the connection branch $\Gamma_1$, the existence of such a fixed point requires a model-dependent condition. In contrast, the existence is structurally generic for the connection branches $\Gamma_{2,3}$. The stability, however, remains model-dependent across all branches. The results are summarised in Table~\ref{tab:matter_dom}. For the specific model $f(Q)=\alpha Q + \beta(-Q)^n$ considered in this work, we find that the matter point is a repeller for $\Gamma_1$ when $n<1$, while it becomes a saddle for $\Gamma_{2,3}$.


		\begin{table}[H]
			\begin{center}
					\renewcommand\arraystretch{1.5}
				\begin{tabular}{|c|c|c|}
					\hline
					Connection branch & Existence condition &  Nature of Stability   \\
					\hline
					$\Gamma_1$  &  The equation $ \Omega  =  \frac{2m(\Omega) + 1}{m(\Omega) + 1}$  has a positive root & Model-dependent \\
					\hline
					$\Gamma_{2,3}$ & Always & Model-dependent \\
					\hline
				\end{tabular}
			\end{center}
			
			\caption{ Existence and stability conditions of matter-dominated fixed points across connection branches.}
			\label{tab:matter_dom}
		\end{table}

		\item \textbf{Stability of the vacuum submanifold:} We now turn to the vacuum dynamics, which naturally defines an invariant submanifold in phase space. In the 1-dimensional system associated with $\Gamma_1$, this submanifold coincides with the de Sitter solution, whose existence and stability follow from the conditions we summarised in Table~\ref{tab:ds_attractor}.

		In contrast, for the 4-dimensional systems corresponding to the  branches $\Gamma_{2}$ and $\Gamma_{3}$, the vacuum evolution spans a 3-dimensional invariant submanifold denoted by $\mathcal{S}_\Omega$ (with $\Omega = 0$). This submanifold always exists for $\Gamma_{2,3}$, regardless of the form of $f(Q)$. The condition for its local stability reduces to
		\begin{equation}
			1 + \frac{1}{2}x_3 \pm \frac{2}{3}x_3x_4 > 0 \qquad \text{($-$ sign for $\Gamma_2$, $+$ for $\Gamma_3$)}\,,
		\end{equation}
		which depends on the values of $x_3$ and $x_4$ along $\mathcal{S}_\Omega$. Since $x_3$ is related to the variation of the effective gravitational coupling via $x_3 = -(1+z) \frac{\kappa_{\rm eff}'(z)}{\kappa_{\rm eff}(z)}$ and astrophysical observations tightly restrict the variation of the effective gravitational constant for scalar-tensor theories like $f(Q)$-gravity, which constrains $x_3$ to be very small. We have shown that this condition is generically satisfied in both $\Gamma_2$ and $\Gamma_3$, thereby ensuring that the vacuum submanifold functions as a late-time attractor.
		
		Stability of the cosmological vacuum solution is a natural requirement for any viable late-time cosmological model, since the contribution from nonrelativistic matter dilutes as the Universe expands. Remarkably, the astrophysical constraint on the variation of the effective gravitational constant ensures, for the $\Gamma_2, \Gamma_3$ gauges of $f(Q)$, the stability of the cosmological vacuum solution. This stability contrasts with many other modified gravity theories, where vacuum stability often requires additional restrictions on model parameters, or may even be absent altogether \cite[Sect.8]{Bahamonde:2017ize}. Notably, Horndeski gravity exhibits a self-tuning mechanism toward a de Sitter vacuum, which is analogous in spirit to the attractor behaviour found in our analysis \cite{Martin-Moruno:2015bda}.
		

		\item \textbf{Non-existence of a global past or  future attractor for $\Gamma_{2,3}$:}
		Due to the existence of multiple invariant submanifolds,  the global connectivity of the phase space is broken. As a result, the dynamics becomes partitioned into distinct sectors, each exhibiting its own asymptotic behaviour. This structural feature generically precludes the existence of a single global past or future attractor in the full phase space corresponding to the these connection branches, in contrast to simpler dynamical systems.
		
		\item \textbf{$\Lambda$CDM-mimicking solutions:} 
		We have identified a distinguished invariant submanifold, $\mathcal{S}_{x_3}$, within the  connection branches $\Gamma_{2,3}$, corresponding to the condition $x_3 = 0$. This submanifold reflects a dynamical freezing of the nonmetricity scalar $Q$, leading to cosmic trajectories that are \emph{kinematically}, but not \emph{dynamically}, equivalent to $\Lambda$CDM. Kinematical equivalence implies that the trajectories \emph{on} this submanifold correspond to a cosmic evolution of the form 
		
		\begin{equation}
			h^2(z) = \frac{2}{3}(1+q_0)(1+z)^3 + \frac{1}{3}(1-2q_0)\,,
		\end{equation}
		
		but dynamical inequivalence implies that the underlying theory is not essentially STEGR, matching $\Lambda$CDM at the background level for any $f(Q)$ model that admits this submanifold.

		However, the underlying dynamics is not GR: despite $Q$ being constant, the nonlinear terms in the Lagrangian can grow dynamically important. This is clearly illustrated in the example $f(Q) = \alpha Q + \beta (-Q)^n$ (see Figs.~\ref{fig:c2_1}, \ref{fig:c3_1}), where the trajectories remain $\Lambda$CDM-like while the effective contribution from the $(-Q)^n$ term increases with time. This assumption on the gravitational paradigm should be contrasted with the class of GR $\Lambda$CDM-mimicking solutions discovered in \cite{Chakraborty:2022evc}.
		
		The existence of $\mathcal{S}_{x_3}$ implies that $\Lambda$CDM-like evolution arises generically in $f(Q)$ gravity formulated in $\Gamma_2$ and $\Gamma_3$, without requiring fine-tuned parameters. For $\Gamma_2$, this submanifold tends to be a global late-time attractor; for $\Gamma_3$, it can act as an attractor, though not generically. This $\Lambda$CDM-like evolution is characterized by bounded dynamical variables $x_{1,2}$ in Eq.(\ref{x1x2}), while the divergence of  $x_4$ in Eq.(\ref{x4_Sx3_G2}) occurs only in the far past when $H \to \infty$ and $x_1 \to 0$ hence constituting a run-away and not a blow-up.
		
		Importantly, the dynamics \emph{on} $\mathcal{S}_{x_3}$ is identical across both branches, rendering the background evolution insensitive to the choice of connection. This degeneracy, however, can be lifted at the level of perturbations: the matter density contrast equation depends on the effective gravitational coupling, which differs between the branches~\cite{BeltranJimenez:2019tme, Khyllep:2021pcu}. Thus, observational probes beyond the background—such as structure growth- are required to discriminate between dynamically distinct  yet kinematically equivalent cosmologies. 
		
		A promising future direction is to investigate both from the theoretical and observational perspective whether other $\Lambda$CDM-mimicking trajectories exist outside $\mathcal{S}_{x_3}$, particularly in reconstructed models like those constructed in our earlier work~\cite{Chakraborty:2025qlv}. Regarding the former, our generic dynamical system formulation will become handy in identifying possible ghosts, as mentioned above, for any of those trajectories. Regarding the latter stand instead, identifying those best aligned with observational data will be crucial if multiple such classes exist. For now, adopting $q_0 \approx -0.55$, we estimate $x_0 \approx 0.70/(1-\bar{r})$ for the $\Lambda$CDM-like solutions on $\mathcal{S}_{x_3}$ from Eq.~\eqref{qx3G2}, and provide the closed-form expression for the dynamical connection: 
		\begin{equation}
			\gamma(z) = \frac{2}{3 \sqrt{c1(1-{\bar r})}} \left[(1-2{\bar r})\sqrt{c1(1-{\bar r})}H_{\Lambda\rm{CDM}}(z)+[H_{\Lambda\rm{CDM}}^2(z)+c_1({\bar r}-1)] \left( D+{\rm arctanh}\left(\frac{\sqrt{c1(1-{\bar r})}}{H_{\Lambda\rm{CDM}}(z)} \right) \right)\right]\,,
		\end{equation}
		from Eq.(\ref{x4_Sx3_G2}) with (\ref{x1x2}), and where $H_{\Lambda\rm{CDM}}(z)$ is  given from Eq.~\eqref{hx3G2_1}. This equation illustrates how the dynamical connection becomes a physically meaningful quantity in $\Gamma_2, \Gamma_3$ formulations— and in some cases it is of interest to reconstruct the connection function itself—see, for example,~\cite{Yang:2024tkw}.
		
		These results suggest that $f(Q)$ gravity offers a novel route to $\Lambda$CDM-like cosmology driven by geometry rather than dark energy, with observable differences emerging only at the perturbation level.
		
\end{itemize} 
	
	To better understand these general findings, we have also investigated the concrete model $f(Q)=\alpha Q+\beta(-Q)^n$, with $\alpha,\beta \neq 0$ and $n\neq0,1$. We find that within $\Gamma_1$, the model fails to yield a viable cosmology due to the absence of a physical matter-dominated fixed point. In contrast, within $\Gamma_{2,3}$, it gives rise to well-defined heteroclinic trajectories connecting a matter saddle to a de Sitter attractor, yielding a cosmic expansion that closely mimics $\Lambda$CDM — even though the theory asymptotically deviates from GR. These behaviours are vividly captured in the evolution of key variables in Figs.\ref{fig:c2_1} \text{and}~\ref{fig:c3_1}.
	
	The findings highlight the significant advantages of the connection branches $\Gamma_2$ and $\Gamma_3$ over the connection branch $\Gamma_1$ in $f(Q)$ cosmology. For $\Gamma_2$ and $\Gamma_3$, a de-Sitter attractor and a matter-dominated fixed point generically exist within the phase space of any $f(Q)$, provided,  that they lie within the physically viable region of the corresponding phase space.  Another appealing  feature of these branches is the generic presence of $\Lambda$CDM-like solutions on the $x_3=0$ invariant submanifold, a property absent in $\Gamma_1$.
	
	Using the same examples as considered in our paper, Guzman et.al.\cite{Guzman:2024cwa} has found in the context of $\Gamma_2$ and $\Gamma_3$ a sudden singularity ($\dot{H}\to\infty$ whereas $H$ is finite) in the limit $\gamma\to0$, which, in terms of our variables, implies $x_4\to0$. It would have been interesting to investigate whether such a situation is recovered also from the generic phase space analysis as we have presented here, thus ensuring it as a generic feature of the connection branches $\Gamma_2$ and $\Gamma_3$. However, as one can see from the relations \eqref{raych_gamma2} and \eqref{HdotG3}, a situation like a sudden singularity will appear at the infinity of the phase space. Phenomenona like sudden singularity can therefore only be captured by an asymptotic fixed point analysis after compactifying the phase space, something that we have not delved into in this paper. In particular, a sudden singularity in the $\gamma\to0$ limit as obtained in \cite{Guzman:2024cwa}, if present, may show up on the submanifold $x_4\to0$ at the infinity of the phase space.
	
	It is almost impossible for a reader familiar with cosmological phase space analysis in $f(R)$ gravity to miss the striking parallel between the dynamical variables defined in (Eq.\eqref{dynvar_def}) and the standard ones  usually employed in $f(R)$ gravity (see e.g. \cite{Chakraborty:2021mcf}, or \cite[Sec(8.3)]{Bahamonde:2017ize}). This analogy  allows us to draw some generic comparisons between $f(Q)$ ($\Gamma_2$ or $\Gamma_3$) and $f(R)$ gravity. A detailed comparison of the generic phase space structure is out of the scope of this paper, but some of the structural similarities and differences that catch the eyes easily by a first glance at the respective dynamical systems is presented in ~\ref{Appendix:f(R)_comp}. To attract the reader's attention, the main conclusions of the comparative study is tabulated in Table \ref{tab:fQ-fR}.
		\begin{table}[H]
			\centering
			\resizebox{\textwidth}{!}{
					\renewcommand\arraystretch{1.3}
				\begin{tabular}{|c|c|c|}
					\hline
					& $f(R)$ & $f(Q)$ ($\Gamma_2$, $\Gamma_3$)\\
					\hline
					Reduction of phase space dimensionality for monomial theories & Yes & Yes \\
					\hline
					Stability of the vacuum invariant submanifold $\mathcal{S}_{\Omega}$ & Stable & Stable \\
					\hline
					The invariant submanifold $S_{x_2}$ & When exists, $R$ doesn't change sign & When exists, $Q$ doesn't change sign \\
					\hline
					Generic existence of a de-Sitter fixed point & Yes & Yes \\
					\hline
					Generic existence of a matter-dominated fixed point & No & Yes \\
					\hline
					Generic existence of $\Lambda$CDM-like solutions & No & Yes \\
					\hline
				\end{tabular}
			}
			\caption{Comparison of $f(R)$ and $f(Q)$ ($\Gamma_2$ and $\Gamma_3$) phase-space properties.}
			\label{tab:fQ-fR}
			\label{tab:fQ-fR}
		\end{table}

	Our findings suggest that $f(Q)$ gravity formulated in $\Gamma_2, \Gamma_3$ gauges admits structurally robust and observationally viable cosmological behaviours at the background level. In particular, the emergence of generic late-time attractors without strong dependence on model parameters distinguishes these branches from many alternative modified gravity scenarios, where fine-tuning is often essential. While a full confrontation with observational data, including cosmic growth and structure formation, requires a separate perturbation-level analysis, our results provide a solid foundation for such efforts. A natural next step is to extend our dynamical framework to study linear and nonlinear perturbations in $\Gamma_2$ and $\Gamma_3$. This study could reveal how seemingly similar cosmologies—identical at the background level—differ in their physical behaviour. In particular, exploring perturbations may offer new insights into open questions in cosmology, such as the $H_0$ tension, by highlighting the role played by the underlying connection.

	\section*{Acknowledgments}

	SC is supported by the Second Century Fund (C2F), Chulalongkorn University, Thailand. JD acknowledges the support of IUCAA,
	Pune (India) through the visiting associateship program. DG is a member of the GNFM working group of Italian INDAM. The authors thank Laur J{\"a}rv and Jackson Levi Said for valuable comments.

	\renewcommand{\thesection}{Appendix}
	\section{ Comparison between the phase space structures of $f(Q)$ and $f(R)$ gravity}
	\label{Appendix:f(R)_comp}

	$f(R)$ gravity has been popular in the late time context for a long time now, and there has been many different dynamical system approaches proposed in this context (see e.g. \cite{Chakraborty:2021mcf}, or \cite[Sec(8.3)]{Bahamonde:2017ize}). This appendix is dedicated towards a comparison from the point of view of the homogeneous and isotropic cosmological phase space between $f(Q)$ gravity and $f(R)$ gravity.  To facilitate the comparison, let us write down the standard dynamical variables and the auxiliary quantity for $f(R)$ gravity in complete parallel to \eqref{dynvar_def} and \eqref{defofm}:
	\begin{equation}
		x_1=-\frac{f}{6H^2 f_R},\,~x_2=\frac{R}{6H^2},\,~x_3=-\frac{\dot{f_R}}{H f_R},\, \Omega=\frac{\rho_m}{3H^2 f_Q}=1-x_1-x_2-x_3\,, 
	\end{equation}
	and
	\begin{equation}
		r(R)\equiv R\frac{f_{R}}{f} = \frac{d \ln f}{d \ln R},\,  m(R) \equiv R \frac{f_{RR}}{f_R} = \frac{d \ln f_R}{d \ln R} = -\frac{x_2}{x_1}\,,
	\end{equation}
	with the success of the formulation dependent on the ability to write $m=m(r)$. The dynamical equations in the spatially flat case (see \cite[Eq.(8.46)]{Bahamonde:2017ize}) can be written in a form completely parallel to \eqref{eq:ds_2} and \eqref{eq:ds_3}
	\begin{subequations}\label{DS_f(R)}
		\begin{eqnarray}
			x_1' &=& \frac{x_2 x_3}{m} + x_1\left(x_3 - \frac{2\dot{H}}{H^2}\right)\,,
			\\
			x_2' &=& -\frac{x_2 x_3}{m} - 2x_2\frac{\dot{H}}{H^2}\,,
			\\
			x_3' &=& -x_3\left(\frac{\dot{H}}{H^2} - x_3 + 2\right) - (1 + 3x_1 + x_2)\,,
		\end{eqnarray}
	\end{subequations}
	with $\frac{\dot{H}}{H^2}=x_2-2$. Being a completely metric theory of gravity, the connection is not independent, and correspondingly there is no additional $x_4'$-equation. Without going into the details, we list below some of the crucial similarities and contrasts between FLRW cosmology in $f(R)$ and $f(Q)$ theory (especially the connection branch $\Gamma_{2,3}$) that can be revealed by a careful first look at the system \eqref{DS_f(R)}: 
	\begin{itemize}
		\item \textbf{Some similarities between $f(R)$ and $f(Q)$ cosmology:}
		\begin{itemize}
			\item Just as in the monomial case $f(Q)\propto(-Q)^n$, for monomial theories of the form $f(R)\propto R^n$ the phase-space dimensionality also reduces by one, owing to the additional constraint $n x_1 + x_2 = 0$.
			
			\item It can be shown that
			\begin{equation}
				\Omega' = -2\Omega\left[\left(\frac{1}{2}-q\right) - \frac{1}{2}x_3\right]\,, \qquad\qquad \left(q=-1-\frac{\dot{H}}{H^2}\right)\,.
			\end{equation}
		The variable $x_3$ can again be related to the variation of the effective gravitational constant, which is constrained to be small by observations. Since the present-day value of the deceleration parameter $q$ is definitely less than $\frac{1}{2}$, the astrophysical bound on the variation of the effective gravitational coupling ensures the stability of the invariant submanifold $\mathcal{S}_{\Omega}$ containing vacuum cosmological solutions, similar to what occurs for the $f(Q)$ theories in the $\Gamma_2$ and $\Gamma_3$ gauges.

			\item For theories with $\lim_{x_2\to0}\frac{x_2}{m(x_1,x_2)}=0$, the invariant submanifold $\mathcal{S}_{x_2}$ exists and contains solutions with vanishing Ricci scalar ($R=0$), in similarity with what happens for $f(Q)$  ($\Gamma_2 $ and $ \Gamma_3$) connection branches. The existence of this invariant submanifold implies that $R$ does not change sign during the cosmic evolution.
			\item Just like all the connection branches of $f(Q)$ theory, $f(R)$ theory also provides a generic de-Sitter fixed point $(x_1,x_2,x_3)=(-1,2,0)$.
		\end{itemize}
		\item \textbf{Some differences between $f(R)$ and $f(Q)$ cosmology:}
		\begin{itemize}
			\item The de-Sitter fixed point $(x_1,x_2,x_3)=(-1,2,0)$ for $f(R)$ gravity implies $x_2=-2x_1$ or $Rf_R-2f=0$, solving which one gets $f(R)=\alpha R^2$. This is the unique $f(R)$ that yields a de-Sitter solution for \emph{all} $R$. On the contrary, the line of de-Sitter fixed points $(x_1,x_2,x_3,x_4)=(2-3x_4,3x_4-1,0,x_4)$ for $f(Q)$ ($\Gamma_2,\,\Gamma_3$ branches) implies $\frac{x_2}{x_1}=-\frac{Qf_Q}{f}=\frac{2-3x_4}{3x_4-1}$. This does not specify a unique $f(Q)$ which yields a de-Sitter solution for \emph{all} $Q$; the reason being the existence of $x_4$ in the relation.
			\item The fixed point $(x_1,x_2,x_3,x_4,\Omega)=\left(0,0,0,\frac{2}{3},1\right)$ is a generic feature of the connection branches $\Gamma_2$ and $\Gamma_3$, which is a standard General Relativistic matter dominated fixed point, with $\frac{\dot{H}}{H^2}=-\frac{3}{2}\Leftrightarrow q=\frac{1}{2}$. On the contrary, a standard matter dominated fixed point is \emph{not} a generic feature of $f(R)$ gravity. In the system \eqref{DS_f(R)}, for $\frac{\dot{H}}{H^2}=-\frac{3}{2}$ one gets $x_2=\frac{1}{2}$, matter-domination implies $\Omega=1$, and a General Relativistic matter dominated phase implies $x_3=0$. This gives $x_1=-\frac{1}{2}$. One can now easily verify that the point $(x_1,x_2,x_3,\Omega)=\left(-\frac{1}{2},\frac{1}{2},0,1\right)$ is not a fixed point of the system \eqref{DS_f(R)}. The requirement of the existence of a matter-dominated fixed point in the phase space enforces additional conditions on $f(R)$; it is \emph{not} a generic feature.
			\item One crucial difference between $f(R)$ and the nontrivial branches $\Gamma_{2,3}$ of the $f(Q)$ theory is that there is no counterpart of the generic invariant submanifold $\mathcal{S}_{x_3}$ for $f(R)$. However, there are structural similarities between the submanifold $x_3=0$ in $f(R)$ and $f(Q)$ gravity.  The reduced dynamical system \emph{on} the submanifold $x_3=0$ for $f(R)$ gravity appears completely equivalent to the nontrivial gauge branches of $f(Q)$. Similar to the $f(Q)$ ($\Gamma_1$) branch, the function $r(x_1,x_2)$ is an integral of motion \emph{on} $x_3=0$;

			\beq r'\vert_{x_3=0} = \left[\frac{\partial r}{\partial x_1}x_1' + \frac{\partial r}{\partial x_2}x_2'\right]_{x_3=0} = 0\,. \eeq

Nonetheless, demanding it to actually be an \emph{invariant} submanifold requires enforcing a further condition $1+3x_1+x_2=0$. Moreover, unlike the $f(Q)$ nontrivial gauge branches, the cosmology \emph{on} the submanifold $x_3=0$ for $f(R)$ gravity does not mimic $\Lambda$CDM. This is best understood by setting $x_3=0$ in the $x_2'$-equation and replacing $x_2=1-q$; the latter gives $q'=2q^2-2$ and consequently the jerk parameter as $j=2q^2+q-q'=q+2\neq1$\footnote{To see some interesting works addressing the question whether $f(R)$ cosmologies can mimic $\Lambda$CDM specifically from the dynamical system point of view, the reader is suggested the references \cite{Fay:2007uy,Chakraborty:2021jku}.}. The generic existence of the invariant submanifold $\mathcal{S}_{x_3}$ containing $\Lambda$CDM-mimicking solutions within its nontrivial $\Gamma_{2,3}$ branches constitutes a distinct advantage of the $f(Q)$ framework compared to $f(R).$
		\end{itemize}
	\end{itemize}
	Of course, the above comparison between $f(Q)$ and $f(R)$ is neither detailed nor exhaustive. A detailed and exhaustive comparison between the $f(Q)$ and $f(R)$ cosmological phase space with parallelly defined dynamical variables, which is beyond the scope of the present work, is reserved as an interesting future research problem.

	\bibliography{refs}
	\bibliographystyle{unsrt}

\end{document}